\PassOptionsToPackage{unicode}{hyperref}
\PassOptionsToPackage{hyphens}{url}
\PassOptionsToPackage{dvipsnames,svgnames,x11names}{xcolor}
\documentclass[
  12pt]{article}

\usepackage{amsmath,amssymb}
\usepackage{iftex}
\ifPDFTeX
  \usepackage[T1]{fontenc}
  \usepackage[utf8]{inputenc}
  \usepackage{textcomp} % provide euro and other symbols
\else % if luatex or xetex
  \usepackage{unicode-math}
  \defaultfontfeatures{Scale=MatchLowercase}
  \defaultfontfeatures[\rmfamily]{Ligatures=TeX,Scale=1}
\fi
\usepackage{lmodern}
\IfFileExists{upquote.sty}{\usepackage{upquote}}{}
\IfFileExists{microtype.sty}{% use microtype if available
  \usepackage[]{microtype}
  \UseMicrotypeSet[protrusion]{basicmath} % disable protrusion for tt fonts
}{}
\makeatletter
\@ifundefined{KOMAClassName}{% if non-KOMA class
  \IfFileExists{parskip.sty}{%
    \usepackage{parskip}
  }{% else
    \setlength{\parindent}{0pt}
    \setlength{\parskip}{6pt plus 2pt minus 1pt}}
}{% if KOMA class
  \KOMAoptions{parskip=half}}
\makeatother
\usepackage{xcolor}
\usepackage{longtable,booktabs,array}
\usepackage{calc} % for calculating minipage widths
\usepackage{graphicx}
\makeatletter
\def\maxwidth{\ifdim\Gin@nat@width>\linewidth\linewidth\else\Gin@nat@width\fi}
\def\maxheight{\ifdim\Gin@nat@height>\textheight\textheight\else\Gin@nat@height\fi}
\makeatother
\setkeys{Gin}{width=\maxwidth,height=\maxheight,keepaspectratio}
\makeatletter
\def\fps@figure{htbp}
\makeatother

\makeatletter
\@ifpackageloaded{caption}{}{\usepackage{caption}}
\@ifpackageloaded{subcaption}{}{\usepackage{subcaption}}
\makeatother

\usepackage{amsthm}
\usepackage{bm}
\usepackage{dsfont}
\usepackage{tikz}
\usetikzlibrary{positioning,arrows.meta}
\usepackage[section]{placeins}
\theoremstyle{plain}
\newtheorem{theorem}{Theorem}
\newtheorem{proposition}{Proposition}
\newtheorem{corollary}{Corollary}
\theoremstyle{definition}

\usepackage[]{natbib}
\usepackage{bookmark}

\IfFileExists{xurl.sty}{\usepackage{xurl}}{} % add URL line breaks if available
\bibpunct{(}{)}{;}{a}{}{,}

\makeatletter
\renewcommand\@seccntformat[1]{\csname the#1\endcsname\csname sbr@dot@#1\endcsname\quad}
\newcommand\sbr@dot@section{.}
\g@addto@macro\@floatboxreset{\spacingset{1}}
\makeatother
\hypersetup{
  pdftitle={Characterizing Estimator-Regime Transitions in Adaptive Enrichment Trials},
  pdfkeywords={confounding; doubly robust estimation; subpopulation identification; synthetic control; targeted maximum likelihood estimation},
  colorlinks=true,
  linkcolor={blue},
  filecolor={Maroon},
  citecolor={Blue},
  urlcolor={Blue}}
\hypersetup{pdfauthor={Xin Luo; Maria Kudela; Margaret Gamalo}}

\newcommand{\R}{\mathbb{R}}
\newcommand{\E}{\mathbb{E}}
\newcommand{\Var}{\operatorname{Var}}

\newcommand{\sgn}{\operatorname{sgn}}
\newcommand{\Ybar}{\bar{Y}}
\newcommand{\rank}{\operatorname{rank}}
\newcommand{\one}{\mathbf{1}}
\newcommand{\Dx}{D_{x}}
\newcommand{\Tpre}{T_{\mathrm{pre}}}
\newcommand{\Dz}{D_{z}}
\newcommand{\ind}{\mathds{1}}
\DeclareMathOperator{\expit}{expit}

\begin{document}

\def\spacingset#1{\renewcommand{\baselinestretch}%
{#1}\small\normalsize} \spacingset{1}

%%%%%%%%%%%%%%%%%%%%%%%%%%%%%%%%%%%%%%%%%%%%%%%%%%%%%%%%%%%%%%%%%%%%%%%%%%%%%%

{
  \title{\bf Characterizing Estimator-Regime Transitions in Adaptive Enrichment Trials}
  \author{Xin Luo\textsuperscript{a}\thanks{Corresponding author. Email:
    xin.luo@pfizer.com.}, Maria Kudela\textsuperscript{b}, and
    Margaret Gamalo\textsuperscript{c}\hspace{.2cm}\\[6pt]
    {\small\itshape \textsuperscript{a}Pfizer AS, Drammensveien 288, 0283 Oslo, Norway}\\
    {\small\itshape \textsuperscript{b}Pfizer Inc., 610 Main Street, Cambridge, MA 02139, USA}\\
    {\small\itshape \textsuperscript{c}Pfizer Inc., 500 Arcola Road, Collegeville, PA 19426, USA}}
  \date{}
  \maketitle
}

\spacingset{1.8} % DON'T change the spacing! (from the abstract on: the journal asks for double
% spacing including the abstract, keywords, and references; the title block above stays single-spaced)
\bigskip
\begin{abstract}
Adaptive enrichment trials that identify responsive subpopulations
allocate a fixed sample across them and estimate each one's effect from
the resulting data. For a synthetic control design with adaptive
recruitment and borrowing of control outcomes, we ask whether a doubly
robust family of targeted maximum likelihood estimators should replace its
sample mean; the choice can be made largely at the design stage.
A crossing condition weighs the bias removed and the efficiency gained by
the doubly robust family against the finite-sample inflation of nuisance
estimation; an equal-error boundary, located by simulation, separates the
families in the plane of sample size and bias. A consistency principle
prices recruiting with the wrong family's variance functional. With three
and eight simulated subpopulations the boundary's sign prediction for the
mean squared error difference holds in $99.6\%$ of $10{,}302$ comparisons
decided at two standard errors. A pre-specified selector commits to the
doubly robust family where the squared anticipated bias exceeds the net
variance term. Recruiting along that family's path costs up to $0.035$ in
the area under the receiver operating characteristic curve;
recruiting with the sample mean functional and routing only the analysis
removes this cost and about nine tenths of its error excess.
\end{abstract}

\noindent%
\vfill

\newpage

% ===========================================================================
\section{INTRODUCTION}
% ===========================================================================
Treatment effects differ across patients, and a treatment with a small average effect may still benefit an identifiable subpopulation. Trials that aim to identify such
subpopulations, called adaptive enrichment or subpopulation identifying
designs, must solve two connected problems: how to allocate a limited
sample across subpopulations and arms, and how to estimate the treatment
effect in each subpopulation from the data that the allocation produces.

The synthetic control design of \citet{huyuk2024}, which we refer to as
the base design, addresses both problems: it recruits and allocates
patients adaptively, and it estimates the control mean of each
subpopulation by a synthetic control, a weighted combination of control
samples from all subpopulations including the target itself, with the
subpopulation outcomes estimated by sample means. We ask whether a
doubly robust estimator \citep{robins1994,bang2005,vanderlaan2006} should
be preferred instead, and under what conditions: it removes the bias of
nonrandomized assignment and gains efficiency when the covariates
predict the outcome \citep{hahn1998}, but it incurs a finite-sample
variance inflation for estimating the nuisance functions, an inflation
that grows with the complexity of the nuisance models and that small
trials feel
\citep{shao2026covariates}, and it interacts with adaptive recruitment. Which effect dominates depends on the design and on the
data generating process, and our goal is to characterize this dependence
and turn it into design guidance.

The three design forces interact. Adaptation directs the budget
toward the subpopulations whose effect signs are least certain and thereby
keeps many subpopulations small, where the variance inflation of the
doubly robust family is largest, and since the allocation is driven by a
variance functional, the
estimator family also changes which data are collected. Borrowing
redistributes confounding bias without removing it when the confounding
geometry is the same across subpopulations, and cannot remove it at all
with nonnegative weights when the alignment has one sign
(Corollary~\ref{cor:noremoval}). Confounding contributes the only
error term that does not decrease with the sample size and differs
between the families, and whether it is visible in pooled operating characteristics depends on whether its direction is the same across subpopulations, while a ranking measure sees only the part of it that differs between the subpopulations of an environment. The regime map of this article shows which term
dominates where.

The main contribution is a characterization of the choice between the
families. We extend the base design with a doubly robust family, derive
an explicit crossing condition that expresses the advantages of
both families in mean squared error per subpopulation, prove that the
boundary between the families is an equal-error boundary, the
zero set of the difference in mean squared error, whose sign, as a
function of the sample size per subpopulation and arm and of the
magnitude of the confounding bias, depends on the configuration only through the
outcome variance explained by the covariates, the residual outcome
variance, and a variance inflation curve that is estimated by simulation
before the trial in the same way as power is, and combine the results
into a
pre-specified selector that commits to one estimator family per trial
from design stage quantities. The second contribution is a recruitment--analysis consistency principle: when the variance
functional that drives recruitment is not the variance of the final
estimator, the achieved worst case  variance exceeds that of a matched design by a closed-form factor. The simulations realize the penalty for the sample mean analysis and reverse it for the doubly robust analysis, whose own path is the worse one in every core scenario (Section~\ref{sec:mismatch-results}). Two supporting elements complete the
design: a degeneracy diagnostic (Section~\ref{sec:estimators}), by
which the hard-constrained borrowing weights collapse, almost surely, to
no borrowing whenever the number of subpopulations does not exceed the
number of matching constraints, a
condition that is arithmetic in the design constants for centroids in
general position, and an optimal transport relaxation of the
borrowing weights (Supplementary Section~S4), which coincides,
with each subpopulation represented by its centroid, with the
pairwise discrepancy penalty of penalized synthetic control
\citep{abadie2021penalized} and keeps borrowing active in the degenerate
regime.

The empirical findings are as follows. On identical patients the variance of the doubly robust analysis as implemented, fallback rule included, matches that of the sample mean up to about $9$ patients per subpopulation and arm, lies a few percent below it up to about $23$, and $14$--$25\%$ below beyond $27$; without the fallback replications the targeted estimator's own inflation under randomization crosses the edge below about seven patients per arm at $K = 3$ and between about $15$ and $20$ at $K = 8$ (Supplementary Section~S5.8). At fixed counts the boundary, evaluated at each subpopulation's true constants and realized allocation with the inflation curve estimated at its confounding strength, predicts the realized sign of the mean squared error difference in $99.6\%$ of the $10{,}302$ comparisons decided at two standard errors ($71\%$ of $14{,}450$ points over sample sizes from $4$ to $64$ per arm and confounding strengths from $0$ to $3$; a constant inflation of $2$ predicts $78\%$), and inside the adaptive design it has the sign of all $60$ decided tiles once each family is placed at the sample size its own recruitment gave it, a comparison that conditions on outcome-dependent counts and, with only two decided tiles adjacent to a change of predicted sign, is consistent with the boundary rather than a sharp test of it; at the design level the doubly robust family has the larger mean squared error in every adaptive configuration but the anchor configuration of Section~\ref{sec:simstudy}, an implementation check with well-separated effects, through its targeted fits on the six to nine patients of starved subpopulations (with the six-patient fallback rule; at a threshold of ten or fourteen patients the deficit is within error in the aligned $K = 3$ scenario and falls by about nine tenths in the randomly oriented $K = 8$ scenario, the two scenarios in which the threshold was varied, Supplementary Section~S5.9). With the one-step augmented inverse probability weighted (AIPW) or a regression-adjusted estimator in place of the targeted one, each recruiting with its own design stage inflation curve, the sample mean's margin in discrimination at $K = 8$ falls from $0.017$ to $0.005$ and $0.006$ under randomization and reverses under aligned confounding; under randomization most of the AIPW arm's change is the steeper recruitment curve rather than the estimator, since the targeted estimator recruited with the AIPW curve reaches $0.008$ (the regression-adjusted arm was not decomposed; Supplementary Section~S5.9). With adaptive
recruitment across eight subpopulations, where most subpopulations stay
small, the sample mean family is ahead in discrimination in seven of the
eight paired contrasts at about $10$ patients per subpopulation and arm,
and at about $26$ with three subpopulations under randomization the two
families are within simulation error in discrimination at the individual
level; the gap in
discrimination at small counts is a recruitment-path effect, since
analyzing the data of the sample mean family with the doubly robust
estimator matches or exceeds the discrimination of the sample mean
family, and uniform recruitment reverses the gap at the same sample size ($1.8$ standard errors on ten environments, within the two-standard-error criterion). Under aligned confounding the doubly robust family removes a false positive rate excess of $6$ to $14$ points, and the point estimates favor it at matched false positive rate by $2$ to $9$ points of true positive rate, within two standard errors in the core scenarios and at $2.0$ standard errors in the $H = 400$ probe, where the pre-specified prediction of no crossing fails in this metric (Supplementary Table~S5); under randomly oriented confounding at $K = 3$ it is ahead there by $9$ points ($3.0$ standard errors), while the misspecified $K = 3$ scenario favors the sample mean by $0.6 \pm 1.8$. The
selector commits to the doubly robust family whenever the squared
anticipated bias band exceeds the net variance term, which at the
simulated strength holds in most confounded trials; where it commits it
removes the bias, and the selector arm evaluated as a matched protocol, which recruits and analyzes with the family it selects, pays for this with that family's deficit in threshold-free discrimination, up to $0.02$ in the area under the receiver operating characteristic curve (AUC) at ten patients per arm and $0.020$ under the misspecified mechanism at twenty-six ($0.032 \pm 0.013$ in the $62$ environments it routes, $51$ of them with a defined AUC), and with its larger mean squared error at both sampled sizes; in the randomized core scenarios it is within about $0.004$ AUC of the better family. The discrimination cost is a cost of the recruitment path: with the sample mean functional driving recruitment and only the analysis routed, the selector is within two standard errors of the better family in AUC in every core scenario ($-0.005$ to $+0.012$ AUC), removes the same bias, and sheds about nine tenths of the matched arm's mean squared error excess; the remainder, $+0.08 \pm 0.02$ and $+0.09 \pm 0.03$ against the sample mean family at $K = 8$ under aligned and random confounding, is the doubly robust analysis's own error on the starved subpopulations of the sample mean path (Supplementary Section~S5.8).

The work is related to adaptive enrichment and biomarker-driven subgroup
designs \citep{freidlin2005,rosenblum2011,simon2013}, their
machine-learning formulations \citep{curth2023adaptive},
response-adaptive randomization \citep{hu2006}, thresholding bandits \citep{locatelli2016},
and data-driven two-stage designs that certify treatment effects
\citep{cortesgomez2025}; ``adaptive enrichment'' is used here in the
broad sense of a design that adapts to and identifies responsive
subpopulations.
Bayesian basket designs also identify subpopulations
while borrowing across them, through a hierarchical prior
\citep{berry2013,chu2018}; the present design borrows through
synthetic control weights on subpopulation aggregates and chooses the
estimator family by regime. The borrowing mechanism builds on the synthetic control methodology
\citep{abadie2003,abadie2010,abadie2021} and its connections to balancing
weights \citep{doudchenko2016}; the augmented synthetic control method
\citep{benmichael2021} corrects the weighted combination for an imperfect
pre-treatment fit, whereas our doubly robust family targets the arm means
of each subpopulation against a fitted propensity and leaves the weights
unchanged, and two recent constructions, targeted-learning borrowing
across pre-specified subgroups in a fixed randomized trial
\citep{qiu2026subgroup} and a targeted update of synthetic control
weights in panel data \citep{wang2026targeted}, involve neither adaptive
recruitment nor the choice of family by regime. The transport relaxation
builds on penalized synthetic control \citep{abadie2021penalized} and
differs from distributional synthetic controls \citep{gunsilius2023}. The
consistency principle has antecedents in the interaction of design and
analysis: interacted regression adjustment can be inefficient under a
stratified design \citep{cytrynbaum2024}, and adaptive Neyman allocation
is formulated for the variance of the estimator used in the analysis
\citep{dai2023clipogd,zhao2023neyman,neopane2025neyman,chen2025sigmoid}. The doubly robust
family draws on augmented inverse probability weighting
\citep{robins1994,robins1995,bang2005}, semiparametric efficiency
\citep{hahn1998,hirano2003}, double machine learning
\citep{chernozhukov2018}, and targeted maximum likelihood estimation
\citep{vanderlaan2006,gruber2009,vanderlaan2011,schuler2017}; the
directional consistency result relates to transportability arguments
\citep{pearl2014,bareinboim2016,dahabreh2020}, and the design questions
connect to the estimand framework \citep{ich2019} and to regulatory
guidance on adaptive designs \citep{fda2019}.

Section~\ref{sec:setup} formulates the problem. Section~\ref{sec:estimators}
defines the estimator families and the comparators and states the
borrowing-collapse proposition.
Section~\ref{sec:theory} states the theoretical results and
the selector. Sections~\ref{sec:simstudy} and~\ref{sec:results} describe
the simulation study and its results, and Section~\ref{sec:discussion}
gives recommendations and limitations; the proofs, the implementation
details, and the full results of every scenario are in the Supplementary
Material.

% ===========================================================================
\section{PROBLEM SETUP}
\label{sec:setup}
% ===========================================================================
We consider $K$ subpopulations indexed by $k \in [K] = \{1,\dots,K\}$. A
patient $j$ recruited from subpopulation $k$ has a covariate vector
$x_{kj} \in \R^{\Dx}$ and a latent loading $z_{kj} \in \R^{\Dz}$,
both drawn around subpopulation-specific centers, the covariate centroid
$x_{k}$, which is known to the design, and the latent loading $z_{k}$,
which is not. In the potential
outcome framework \citep{rubin1974}, each patient has potential outcomes
$Y_{kj}(0)$ and $Y_{kj}(1)$ under control and treatment. We observe a
pre-treatment outcome trajectory of length $\Tpre$ and a single
post-treatment outcome. We let $\Ybar_{k,a}$ denote the running sample
mean of the post-treatment outcomes in arm $a \in \{0,1\}$ of
subpopulation $k$ and $\Ybar_{k,\neg T}$ the running mean of the
pre-treatment observations. The treatment effect in subpopulation $k$ is
$r_{k} = \E[Y_{k}(1) - Y_{k}(0)]$, and the target of inference is the sign pattern
$\{k : r_{k} > 0\}$, the set of subpopulations with a positive effect.

After a forced balanced burn-in of one patient per subpopulation and
arm, patients arrive sequentially over $H$ adaptive episodes, so the total
sample size is $2K + H$. In each episode the design selects a
subpopulation to recruit from and, in the base design, an arm.
Under the adaptive recruitment policy the design first identifies
the least certain subpopulation, the one with the smallest sensitivity
index $|\hat r_{k}|/\sqrt{V_{k}}$ (effect estimate over its standard
error), and then recruits the subpopulation and arm whose additional
patient reduces the variance $V_{k}$ of that subpopulation the most,
which with active borrowing can be a donor subpopulation; under the uniform policy subpopulations and arms are filled by a fixed
balanced schedule.

\label{sec:arm}
Two mechanisms determine the arm of a recruited patient. Under design-controlled assignment the arm is the one chosen by the
recruitment look-ahead just described, a deterministic function of the
accrued record that does not depend on the covariates or outcomes of the
patient; we call this mechanism randomized for short, because it
makes the assignment ignorable as randomization does, although no
randomization device is involved. Under confounded assignment the arm is
drawn as $A_{kj} \sim \mathrm{Bernoulli}(g(x_{kj}))$ with propensity
$g(x) = \expit(\alpha^{\top} x)$, where $\expit(t) = 1/(1 + \exp(-t))$ and
$\alpha \in \R^{\Dx}$ is a propensity direction \citep{rosenbaum1983}. The
confounded mechanism stands for settings in which the arm depends on
patient characteristics, such as hybrid or externally augmented control
arms whose donor observations were not randomized against the treated arm
\citep{zhu2026externally}, so that every borrowed control mean inherits
the selection mechanism of the donors; in the simulations it acts on the target's own arms as well.
The alignment of $\alpha$ with the post-treatment outcome weights
$w_{T,k}$ is central to the analysis of Section~\ref{sec:theory}.

Given effect estimates $\hat r_{k}$, the design declares the set
$\hat{\mathcal{I}} = \{k : \hat r_{k} > 0\}$, evaluated against the truth
by the true positive rate (TPR), the fraction of the subpopulations with
$r_{k} > 0$ that are declared, and the false positive rate (FPR), the
fraction of those with $r_{k} \le 0$ that are declared, both reported as
functions of $H$ and averaged over independent data generating
environments and trial replications.

% ===========================================================================
\section{ESTIMATORS}
\label{sec:estimators}
% ===========================================================================
We compare six estimators arranged in a $2 \times 3$ grid: two families,
family 1 (sample mean) and family 2 (doubly robust), each in three
variants, a subpopulation-level variant (a), an individual-level variant
(b), and a mean-squared-error (MSE) optimal oracle (c). The variant axis
combines a data-resolution step, from the summary-level (a) to the
individual-level (b), with an objective step to the oracle MSE variant
(c); ``subpopulation-level'' and ``summary-level'' are used as synonyms
throughout, and a variant is called implementable when its objective has no
oracle bias term, that is, no bias evaluated at the true outcome weights
and latent loadings as in the (c) variants; the word refers to this bias term only, since every arm uses at least one oracle variance input, the rates $\lambda_{k}$ or $\tilde\lambda_{k}$ (Supplementary Section~S6.3), so the recruitment comparisons are between oracle-informed paths and the estimated-variance probe of Section~\ref{sec:probes} is the evidence on the practical case; Supplementary
Section~S1 collects what is gained and lost by developing
the methodology at each level.

\subsection{Synthetic Control Weighting}
For a target subpopulation $k$, the synthetic control for the control arm
is a weighted combination $\sum_{l} \beta_{l}\,\Ybar_{l,0}$ of the control
means of the subpopulations, with weights $\beta$ chosen so that
subpopulation $k$ is matched on the covariate and pre-treatment
dimensions. The matching constraints are $\Gamma\beta = q$, where
$q = (1, x_{k}^{\top}, \Ybar_{k,\neg T}^{\top})^{\top}$ is the $k$th column
of
\begin{equation}
\Gamma =
\begin{pmatrix}
\one^{\top} \\ X \\ \widehat{Y}_{\neg T}
\end{pmatrix}
\in \R^{m \times K},
\qquad
m = 1 + \Dx + \Tpre,
\end{equation}
with $\one$ the vector of ones, $X = [x_{1}\,\cdots\,x_{K}] \in
\R^{\Dx \times K}$ the matrix of centroids, and $\widehat{Y}_{\neg T} \in
\R^{\Tpre \times K}$ the matrix whose column $l$ is the running
pre-treatment mean $\Ybar_{l,\neg T}$; the weights solve a quadratic
program with a family-specific objective. The estimated effect is
\begin{equation}
\hat r_{k}(\beta) \;=\; \hat\theta_{k}^{(1)} \;-\; \sum_{l} \beta_{l}\,\hat\theta_{l}^{(0)},
\end{equation}
where $\hat\theta_{k}^{(1)}$ and $\hat\theta_{l}^{(0)}$ are the arm-mean
estimates supplied by the chosen family. Both arms are estimated by the
family, so that under confounded assignment its bias behavior applies to
the treated arm as well.

Let $n_{k}^{(a)}$ be the number of patients in subpopulation $k$ and arm
$a$, and $n_{k} = n_{k}^{(0)} + n_{k}^{(1)}$. The working variance
functional of $\hat r_{k}(\beta)$, of the same form for every variant, is
\begin{equation}
V_{k}(\beta)
= \frac{\sigma^{2}_{k,\bullet}}{n_{k}^{(1)}}
+ \sum_{l} \beta_{l}^{2}\,\frac{\sigma^{2}_{l,\bullet}}{n_{l}^{(0)}}
+ \tilde\lambda_{k}\,\sigma^{2}
  \sum_{l} \frac{(\beta_{l} - \ind\{l=k\})^{2}}{n_{l}},
\label{eq:Vbound}
\end{equation}
where $\sigma^{2}_{l,\bullet}$ is the post-treatment outcome variance of
subpopulation $l$ under the chosen family: the marginal variance
$\sigma^{2}_{l,\mathrm{eff}}$ for the sample mean family and the smaller
conditional variance $\sigma^{2}_{l,\mathrm{cond}}$ for the doubly robust
family, and $\sigma^{2}$ is the outcome noise variance
(Section~\ref{sec:dgp}). The three terms are the sampling variance
of the treated arm, the sampling variance of the synthetic control, and a
representation error penalty for departing from the no-borrowing weight
$e_{k}$, the $k$th standard basis vector. The penalty rate
$\tilde\lambda_{k}$ corrects the rate
$\lambda_{k} = \|M_{\neg T}^{+}\mu_{T,k}\|^{2}$ of the base design, where
$M_{\neg T} = [\mu_{1}\,\cdots\,\mu_{\Tpre}]$ is the matrix of the global
pre-treatment latent weights of the factor model of
Section~\ref{sec:dgp} and $M_{\neg T}^{+}$ its Moore--Penrose
pseudoinverse; the correction accounts for the cross-patient noise that
running means cannot remove (Supplementary Section~S6.3
gives the formula). On the subpopulation-level pool, and at $\beta = e_{k}$,~\eqref{eq:Vbound} is the variance of the error of $\hat r_{k}(\beta)$.
Away from $e_{k}$ on the individual-level pool the same patient noise
enters both the running pre-treatment means and the arm means, and the
variance of the error representation with the weights frozen at $\beta$
contains in addition a cross-covariance term,~(S3) of
Supplementary Section~S6.3, for both families; under
confounded assignment the functionals of both families are working
approximations: the family-1 arm means have the arm-conditional covariate variance $E_{k}^{c}(a) \ne E_{k}$ (below $E_{k}$ under the linear-index mechanisms, Proposition~\ref{prop:crossing}; slightly above it in the treated arm under the quadratic one, Supplementary Section~S5.8), and the
expansion behind the family-2 terms has a remainder of the same order as
the terms kept (Supplementary Section~S6.3).
This frozen-weight functional is the working variance of the
individual-level
variants, which minimize it with the term included; it is the variance of
the estimator only at $\beta = e_{k}$, since a weight vector on the
realized constraint set depends on the same pre-treatment noise; the
supplement checks the size of the gap (within $2\%$ at $200$ patients per
arm).
The family-2 variants \textbf{(2b, 2c)} add the covariate sampling terms
discussed after Proposition~\ref{prop:variance} and inflate the
conditional variances by the simulated inflation factor $1 + \kappa(\bar n_{l})$ of
Proposition~\ref{prop:crossing}, so that the recruitment rule uses the
variance that the design realizes, whereas \textbf{(2a)} uses
influence curve variance estimates and no inflation factor.

When the post-treatment weights are heterogeneous the estimator is
biased. Enforcing $z_{k} \approx \sum_{l}\beta_{l} z_{l}$ through the
trajectory matching gives
\begin{equation}
\E[\hat r_{k}(\beta)] - r_{k}
= \sum_{l} \beta_{l}\, d_{kl},
\qquad
d_{kl} = (w_{T,k} - w_{T,l})^{\top} x_{l} + (\mu_{T,k} - \mu_{T,l})^{\top} z_{l}.
\label{eq:repbias}
\end{equation}
This representation bias is the same for both families, since
targeted maximum likelihood removes the confounding bias but not this
term, and it vanishes at $\beta = e_{k}$. The variance-minimizing
variants minimize $V_{k}(\beta)$; the mean-squared-error--optimal variants
minimize $M_{k}(\beta) = (d_{k}^{\top}\beta)^{2} + V_{k}(\beta)$ with
$d_{k} = (d_{k1},\dots,d_{kK})^{\top}$.

The hard constraints also decide whether borrowing is active at all.
\begin{proposition}[Borrowing collapse]
\label{prop:degeneracy}
Let $m = 1 + \Dx + \Tpre$ be the number of hard matching constraints. With
probability one the realized constraint matrix has
$\rank(\Gamma) = \min(K, m)$ (for fixed centroids the condition is
$\rank([\one^{\top}; X]) \ge \min(K, m) - \Tpre$, void for $K \le \Tpre$
and almost sure under the environment law), so the feasible set
$\{\beta : \Gamma\beta = q\}$ is the single point $\beta = e_{k}$ whenever
$K \le m$, in which case the family-specific objective is inactive and all
hard-constrained variants return the same weights, and it is an affine
set of dimension $K - m$ when $K > m$. The noiseless constraint matrix
$\Gamma^{\circ}$, with the running means replaced by their noiseless values
under the factor model of Section~\ref{sec:dgp}, has, for centroids,
loadings, and pre-treatment latent weights in general position, rank
$\rank(\Gamma^{\circ}) = \min(K, 1 + \Dx + \min(\Tpre, \Dz))$, which is
smaller than $\min(K, m)$ when $\Tpre > \Dz$ and $K > 1 + \Dx + \Dz$; the
$\Tpre - \Dz$ surplus trajectory constraints have no counterpart in the
noiseless factor structure and are satisfied only by reproducing the
sampling noise of the running means.
\end{proposition}
\noindent
The proof is in Supplementary Section~S6.9. In the
noiseless factor structure the trajectory rows carry information about
only $\min(\Tpre, \Dz)$ factor directions, but the realized rows are
running
means with their own sampling noise and every one of them binds as a
hard equality. In this study $m = 7$ and $\rank(\Gamma^{\circ}) = \min(K, 5)$:
the weights are fixed at $e_{k}$ for every $K \le 7$, and the $K = 8$
scenarios have a single free direction. At or below $m$ the point
estimators within a family and level coincide on identical data, which is
a diagnostic for the degenerate regime; the noiseless count is reported
alongside for soft-constrained weighting. The collapse differs from
shrinkage, which fades continuously with the estimated heterogeneity: the
constrained weighting is a projection whose feasible set loses dimensions
discretely, so the regime can be decided from $(K, \Dx, \Tpre, \Dz)$
before any patient is enrolled, for centroids in general position (the
solvers confirm it by the realized rank), and even in the collapsed
regime adaptive
recruitment couples the subpopulations through the shared budget
(Supplementary Section~S5.6).

\subsection{The Two Families}
\label{sec:family2}
\emph{Family 1 (sample mean).} The arm means are the running means
$\Ybar_{l,a}$. All three variants use the same matching constraints on
subpopulation aggregates (centroids, running pre-treatment means, counts)
and differ in the data they are evaluated on, in the variance inputs of~\eqref{eq:Vbound}, and in the objective. \textbf{(1a)} is the base design, run on the
subpopulation-level pool in which every patient is placed at the centroid
of its subpopulation, with the homoscedastic $\sigma^{2}$ and the rates
$\lambda_{k}$, which are exact on that pool. \textbf{(1b)} is the same
objective on the individual-level pool with the oracle variance inputs of
that pool, the marginal variances $\sigma_{l,\mathrm{eff}}^{2}$, the
corrected rates $\tilde\lambda_{k}$, and the cross-covariance
term~(S3), all functions of the true outcome weights, the
pre-treatment weight matrices, and the noise scales but not of the latent
loadings $z_{l}$. \textbf{(1c)} adds the squared
representation bias $(d_{k}^{\top}\beta)^{2}$ evaluated at the true
outcome weights and latent loadings; it is a benchmark, not implementable,
and MSE-optimal under randomized assignment only, since its objective
omits the confounding bias.

\emph{Family 2 (doubly robust).} Each arm mean is replaced by a targeted
maximum likelihood estimate (TMLE) \citep{vanderlaan2006,gruber2009}, a
construction closely related to augmented inverse probability weighting
and double machine learning \citep{bang2005,chernozhukov2018}. For each
subpopulation an outcome model $Q(a,x) = \E[Y \mid A = a,\, x_{kj} = x]$
and a propensity model $g(x) = \Pr(A = 1 \mid x_{kj} = x)$ are fitted with
linear and logistic working models, the outcome model is updated arm by
arm along the clever covariates $\ind\{A=1\}/g(x)$ and
$\ind\{A=0\}/(1-g(x))$ by a linear fluctuation on the unscaled outcome,
so that the two arm-specific efficient-score equations hold exactly in
the sample (Supplementary Section~S3; a bounded-outcome fluctuation on the logistic scale \citep{gruber2010} was the family-2 estimator of the preliminary runs on which the pre-specified predictions of Supplementary Table~S5 are based; its range-pinned predictions shifted the estimate at small counts, those adaptive runs implied a randomized threshold beyond $26$ patients per arm, and it was replaced by the linear fluctuation before the runs reported here and is kept as an option; its design stage inflation is two to five times the linear one at $4$--$10$ patients per arm, with a randomized threshold of about $15$ against $7$ on the isotonic fits, Supplementary Section~S5.9), and the targeted plug-in mean over the covariate sample of the subpopulation is reported. Throughout, ``the doubly robust family'' denotes this implementation, the linear-fluctuation targeted estimator with the propensity truncated to $[0.05, 0.95]$ and the fallback to sample means below six patients per subpopulation or two per arm (Section~\ref{sec:metrics-reporting}); the one-step AIPW comparator at the same weights has the smaller inflation and the higher AUC in every core scenario (Table~\ref{tab:contrasts}), Supplementary Section~S5.8 records what its inflation curve would imply for the confounded threshold, and Supplementary Section~S5.9 runs AIPW and the regression-adjusted estimator as recruiting arms with their own design stage curves, which close most of the targeted arm's discrimination gap at $K = 8$ (for the AIPW arm, under randomization largely through the curve and under aligned confounding largely through the estimator; the regression-adjusted arm was not decomposed). The estimator is consistent if
the outcome model is correctly specified, or if the propensity model is
correctly specified and its truncation does not bind; in the core simulations the outcome model is always correct and carries the consistency (Supplementary Section~S5.9 probes a misspecified one). Since the targeting step
conditions on $x_{kj}$, the arm-mean estimate has, under randomized
assignment, asymptotic variance
$\sigma_{k,\mathrm{cond}}^{2}/n_{k}^{(a)} + E_{k}/n_{k}$ with the
conditional variance $\sigma_{k,\mathrm{cond}}^{2} = \sigma^{2}
+ \sigma_{z}^{2}\|\mu_{T,k}\|^{2}$ and $E_{k} = \sigma_{x}^{2}
\|w_{T,k}\|^{2}$ the covariate-explained variance of
Proposition~\ref{prop:variance}, where $\sigma_{x}$ and $\sigma_{z}$ are
the covariate and latent dispersions within a subpopulation
(Section~\ref{sec:dgp}), in place of the sample mean's
$\sigma_{k,\mathrm{eff}}^{2}/n_{k}^{(a)}$, which is the source of the
efficiency edge of Proposition~\ref{prop:variance}. Variants
\textbf{(2a)}, \textbf{(2b)}, and \textbf{(2c)} parallel the sample mean
variants; \textbf{(2a)} uses influence curve variance estimates in the
weighting and the recruitment index, while \textbf{(2b, 2c)} use the
oracle conditional variances multiplied by the simulated inflation
factor, with the
covariate sampling and cross-covariance terms added.

\emph{Comparators.}
\label{sec:comparators}
Two doubly robust comparators use the nuisance models of family 2 at the
weights of \textbf{(2b)} and attribute its finite-sample behavior to its
components. The augmented inverse probability weighted (AIPW, one-step)
estimator \citep{robins1994,bang2005}, \textbf{(3a)} below, adds the
empirical mean of the efficient influence curve correction instead of
performing the targeting update; a difference between the two isolates
the targeting step as implemented. A cross-fitted variant \citep[cf.][]{vanderlaan2011,chernozhukov2018},
\textbf{(3b)}, splits each subpopulation
into two folds stratified by arm, fits the nuisance models on one fold
and evaluates on the other, fitting the fluctuation within each held-out
fold and averaging the two targeted arm means (so that the nuisance
sample and the targeting sample are halved together); its performance
shows whether the finite-sample penalty is due to own-sample overfitting,
in which case cross-fitting helps, or to the cost of halving these
samples, in which case it hurts.

\paragraph{Data resolution.}
The base design is a summary-level method that uses only subpopulation
moments and running means, whereas the doubly robust family and its
comparators require patient-level records. Supplementary
Section~S1 sets out the consequences for identification,
efficiency, borrowing, auditability, operations, and data governance, of
which two recur below: a summary-level analysis cannot remove confounding
without a maintained outcome model, and aggregates can be computed from
records but never the reverse.

% ===========================================================================
\section{THEORETICAL CONSIDERATIONS}
\label{sec:theory}
% ===========================================================================
We state the mechanisms behind the regime map. The regularity
conditions (Assumptions~1--5) and the proofs are
in Supplementary Section~S6. The fixed-count statements
hold conditionally on the arm assignments, for subpopulation sizes fixed
in advance or chosen independently of the outcomes and, under
design-controlled assignment, for an allocation of the arms that uses the
counts only (Assumption~3); the remark after the
assumptions describes what changes under adaptive recruitment.

\subsection{Efficiency Edge of the Doubly Robust Family}
\begin{proposition}[Asymptotic variance]
\label{prop:variance}
Under randomized assignment and correct outcome modeling, with
$E_{k} = \sigma_{x}^{2}\|w_{T,k}\|^{2}$ the part of the post-treatment
outcome variance explained by the covariates:
(i) the doubly robust estimator of the marginal arm mean has asymptotic
variance $\sigma_{k,\mathrm{cond}}^{2}/n_{k}^{(a)} + E_{k}/n_{k}$, which is
no larger than the variance $\sigma_{k,\mathrm{eff}}^{2}/n_{k}^{(a)}$ of
the sample mean, with gap
$E_{k}\,\big(1/n_{k}^{(a)} - 1/n_{k}\big) \ge 0$; and
(ii) for the effect within a subpopulation at the no-borrowing weight
$\beta = e_{k}$, because the treatment effect is homogeneous within the
subpopulation ($Q(1,x) - Q(0,x) \equiv r_{k}$; Section~\ref{sec:dgp}), the
covariate sampling components of the two targeted arm means cancel in the
difference, and the variance gap at the effect level is the full
$E_{k}\,s_{k}$, where $s_{k} = 1/n_{k}^{(1)} + 1/n_{k}^{(0)}$.
\end{proposition}
\noindent
The distinction matters because the estimand is marginal: no regular
estimator of the marginal arm mean can remove the covariate sampling term
$\Var_{x}(Q(a,x)) = E_{k}$ without knowledge of the covariate law, and the
targeted plug-in carries it at rate $1/n_{k}$, so the gap at the arm level
is half of the na\"ive $E_{k}/n_{k}^{(a)}$ at balanced arms, whereas both
targeted arm means average the updated outcome model $Q^{*}(a,\cdot)$
(Supplementary Section~S3) over the same covariate
sample, so that with a constant effect the covariate sampling components
cancel exactly, which is why the crossing condition below contains
$E_{k}\,s_{k}$. Under heterogeneous
effects within a subpopulation the doubly robust side retains
$\Var_{x}(Q(1,x) - Q(0,x))/n_{k}$ and the edge is smaller; at general
borrowing weights the donor means average over different covariate
samples and the frozen-weight functional contains the noncanceling terms
$(\beta_{l} - \ind\{l=k\})^{2} E_{l}/n_{l}$ that the family-2
variants add to~\eqref{eq:Vbound}. Both gaps vanish as
$\sigma_{x} \to 0$.

\subsection{Bias Under Confounded Assignment}
\begin{proposition}[Directional bias]
\label{prop:bias}
Under the confounded mechanism of Section~\ref{sec:arm}, the sample mean
effect estimate at the no-borrowing weight $\beta = e_{k}$ has,
conditionally on the arm assignments and for subpopulation sizes fixed in
advance or chosen independently of the outcomes
(Assumption~3), the exact bias (for the patients whose
arms the mechanism draws)
\begin{equation}
\E[\hat r_{k}(e_{k})] - r_{k}
\;=\; c_{k}\,\big(\alpha^{\top} w_{T,k}\big),
\qquad
c_{k} \;=\; \frac{\sigma_{x}^{2}\,
\E\!\big[g(x_{kj})\big(1 - g(x_{kj})\big)\big]}{\pi_{k}(1 - \pi_{k})}
\;\in\; \big(0,\, \sigma_{x}^{2}\big],
\label{eq:biasconstant}
\end{equation}
where $\pi_{k} = \E[g(x_{kj})]$ is the marginal treated fraction of the
subpopulation. The doubly robust estimator removes this bias to leading
order when the outcome model is correctly specified, or when the
propensity model is correctly specified and its truncation does not bind.
\end{proposition}
\noindent
The mechanism is a covariate shift between the arms: by Stein's identity
\citep{stein1981}, $\E[x_{kj} \mid A_{kj}=1] - \E[x_{kj} \mid A_{kj}=0]
= c_{k}\,\alpha$, and the outcome weights convert the shift into the
outcome bias $c_{k}\,\alpha^{\top} w_{T,k}$, while the latent loadings,
independent of $x_{kj}$, do not shift. In the simulations the forced
burn-in patient of each arm is assigned without regard to the propensity,
which dilutes this bias by a known factor (Supplementary
Section~S3). At general borrowing weights the
confounding component of the bias, which adds to~\eqref{eq:repbias}, is
\begin{equation}
b_{k}(\beta)
= \alpha^{\top}\!\Big[(1-\pi_{k})\,c_{k}\, w_{T,k}
+ \sum_{l} \beta_{l}\, \pi_{l}\, c_{l}\, w_{T,l}\Big],
\label{eq:biasborrowed}
\end{equation}
which under homogeneous $(\pi_{l}, c_{l}, w_{T,l})$ is constant in
$\beta$: borrowing redistributes the confounding bias but cannot remove
it, because the control arm of every donor is selected by the same
mechanism.

\begin{corollary}[Sign-consistent confounding cannot be removed by borrowing]
\label{cor:noremoval}
Let the borrowing weights range over the simplex, as the relaxation of
Supplementary Section~S4 requires, and suppose that $\sgn(\alpha^{\top} w_{T,l})$
is constant in $l$ with $\alpha^{\top} w_{T,l} \neq 0$ for all $l$. Then
\begin{equation}
\inf_{\beta \in \Delta^{K-1}} \big|b_{k}(\beta)\big|
\;\ge\; (1-\pi_{k})\,c_{k}\,\big|\alpha^{\top} w_{T,k}\big|
\;+\; \min_{l}\, \pi_{l}\, c_{l}\,\big|\alpha^{\top} w_{T,l}\big|
\;>\; 0,
\label{eq:noremoval}
\end{equation}
uniformly in the sample sizes of the subpopulations.
\end{corollary}
\noindent
Alignments of different signs across the subpopulations of one
environment allow partial cancellation by reweighting, and on the simplex
full cancellation is possible; the next corollary concerns whether the
bias of each environment is visible after aggregation across
environments.

\begin{corollary}[Visibility of bias correction]
\label{cor:consistency}
To first order in the bias and under a normal approximation to the estimates, for subpopulation sizes fixed in advance or chosen independently of the outcomes (Assumption~3), the bias component of the aggregate advantage of the doubly robust family over the sample mean family under confounded assignment is determined by the directional consistency of the induced bias across environments. It is largest when
the confounding direction is consistent and can be negligible when the
direction is randomly oriented.
\end{corollary}
\noindent
With $\mathrm{se}_{k}$ the standard error of $\hat r_{k}$ and
$b_{k} = c_{k}\,\alpha^{\top} w_{T,k}$, the normal approximation
$\Pr(\hat r_{k} > 0) \approx \Phi((r_{k} + b_{k})/\mathrm{se}_{k})$, with $\Phi$ and $\phi$ the standard normal distribution and density functions, gives,
to first order in $b_{k}$, aggregate shifts
\begin{equation}
\Delta \mathrm{TPR}
\approx \E_{\mathrm{env}}\!\left[
\frac{\phi(r_{k}/\mathrm{se}_{k})}{\mathrm{se}_{k}}\, b_{k}
\;\middle|\; r_{k} > 0\right],
\qquad
\Delta \mathrm{FPR}
\approx \E_{\mathrm{env}}\!\left[
\frac{\phi(r_{k}/\mathrm{se}_{k})}{\mathrm{se}_{k}}\, b_{k}
\;\middle|\; r_{k} \le 0\right].
\label{eq:aggshift}
\end{equation}
If the orientation is symmetric (Assumption~5) both
expectations vanish, so the bias contributes no aggregate shift to first
order and the aggregate contrast between the families reduces to their
variance difference, although every environment has a nonzero bias; if
the orientation is aligned, the shifts accumulate and the correction
becomes visible. This is related to transportability
\citep{bareinboim2016,dahabreh2020}.

\subsection{A Crossing Condition for Estimator Choice}
Let $\bar n_{k} = n_{k}/2$ denote the number of patients per arm of
subpopulation $k$ along a balanced path (written $\bar n^{(a)}$ in the
tables), and let $\kappa_{k}(\bar n_{k}) \ge 0$ denote the finite-sample
variance inflation of the doubly robust estimate relative to
$\sigma_{k,\mathrm{cond}}^{2}\, s_{k}$, the variance of an oracle contrast
with known nuisance functions under randomized assignment and the same
arm counts, so that $1 + \kappa_{k}$ multiplies that variance (the curves
of Section~\ref{sec:crossing-selector} plot the inflation term
$\kappa_{k}\sigma_{k,\mathrm{cond}}^{2}$ against the edge; the term is
unrelated to the variance inflation factor of collinearity diagnostics); it is the variance inflation from estimating the nuisance
functions from the $n_{k} = 2\bar n_{k}$ observations of the
subpopulation and, under confounded assignment, from the overlap. The
inflation is indexed by the per-arm count because the simulation-based
estimate of Section~\ref{sec:crossing-selector} and the selector are. We assume it to be nonincreasing with a limit
$\kappa_{k,\infty} \ge 0$ as $\bar n_{k} \to \infty$: under randomized
assignment with nuisance functions fitted at parametric rates
$\kappa_{k,\infty} = 0$, and under confounded assignment
$\kappa_{k,\infty} = \omega_{k}^{\dagger} - 1$, where
$\omega_{k}^{\dagger} \ge 1$ is the large-sample variance factor of the
doubly robust estimator as implemented, introduced after the proposition.
Under confounded assignment the covariate variance within arm $a$,
$E_{k}^{c}(a) = \Var(w_{T,k}^{\top} x_{kj} \mid A_{kj} = a)$, differs from
the explained variance $E_{k} = \sigma_{x}^{2}\|w_{T,k}\|^{2}$ because the
assignment selects on $\alpha^{\top} x_{kj}$: for Gaussian covariates
$E_{k}^{c}(a) = E_{k} - \gamma_{k,a}\,(\alpha^{\top} w_{T,k})^{2}$ with a
selection coefficient $\gamma_{k,a}$ that is nonnegative whenever the
selection weight $\Pr(A_{kj} = a \mid \alpha^{\top} x_{kj})$ is
log-concave in the index \citep{brascamp1976}, as $\expit$ of a linear
index and its complement are, so that $E_{k}^{c}(a) \le E_{k}$ with
equality under randomization; we call the aligned and randomly oriented
mechanisms, whose index is linear, the linear-index mechanisms.
Under the quadratic index of the misspecified mechanism the treated-arm
selection weight is not log-concave and $\gamma_{k,1} < 0$ in most
subpopulations (Supplementary Section~S5.8). In the
confounded scenarios of Section~\ref{sec:simstudy}
the ratio $E_{k}^{c}(a)/E_{k}$ is $0.8$--$0.9$ in both arms under the
linear-index mechanisms and about $1$ in the treated and $0.5$ in the
control arm under the quadratic index (first ten environments;
Supplementary Section~S5.8). We write $\bar E_{k}^{c}$ for the count-weighted
mean $[E_{k}^{c}(1)/n_{k}^{(1)} + E_{k}^{c}(0)/n_{k}^{(0)}]/s_{k}$.

\begin{proposition}[Crossing condition]
\label{prop:crossing}
At the no-borrowing weight $\beta = e_{k}$, the mean squared errors of the
two families in subpopulation $k$ are
\begin{equation}
\begin{aligned}
\mathrm{MSE}_{1}
&= \big(c_{k}\,\alpha^{\top} w_{T,k}\big)^{2}
+ \sigma_{k,\mathrm{cond}}^{2}\, s_{k}
+ \frac{E_{k}^{c}(1)}{n_{k}^{(1)}} + \frac{E_{k}^{c}(0)}{n_{k}^{(0)}} \\
&= \big(c_{k}\,\alpha^{\top} w_{T,k}\big)^{2}
+ \big(\sigma_{k,\mathrm{cond}}^{2} + \bar E_{k}^{c}\big)\, s_{k}, \\
\mathrm{MSE}_{2}
&= \big(1 + \kappa_{k}(\bar n_{k})\big)\,\sigma_{k,\mathrm{cond}}^{2}\, s_{k},
\end{aligned}
\end{equation}
the first exactly and the second to leading order, so the doubly robust
family is preferred if and only if
\begin{equation}
\underbrace{\big(c_{k}\,\alpha^{\top} w_{T,k}\big)^{2}}_{\text{bias removed}}
\;+\;
\underbrace{\bar E_{k}^{c}\, s_{k}}_{\text{variance edge}}
\;>\;
\underbrace{\kappa_{k}(\bar n_{k})\,\sigma_{k,\mathrm{cond}}^{2}\, s_{k}}_{\substack{\text{finite-sample}\\ \text{inflation term}}} ,
\label{eq:crossing}
\end{equation}
where $\bar E_{k}^{c} = E_{k} = \sigma_{x}^{2}\|w_{T,k}\|^{2}$ under
randomized assignment.
\end{proposition}
\noindent
Under randomized assignment ($\alpha = 0$) the condition compares the
explained-variance edge with the inflation term of nuisance estimation,
which the sample mean family wins at sample sizes small enough for the
inflation term to exceed the edge, the adaptive-recruitment regime at
large $K$; under
confounded assignment the bias term is $\mathcal{O}(1)$ while both
variance terms decrease as $s_{k}$, so the doubly robust family is
preferred for all sufficiently large sample sizes, at a threshold that
decreases with the alignment. The inflation $\kappa_{k}$ is a property of
the nuisance estimators at the sample sizes the design realizes and is
estimated by simulation in Section~\ref{sec:crossing-selector}. For a
linear working model under randomization the inflation has the classical
small-sample form of the analysis of covariance,
$\kappa \approx p/(2\bar n - p - 3)$ for $p$ covariates
\citep{cochran1957,cox1982}, which for $p = \Dx = 2$ gives $0.67$,
$0.18$, $0.13$, and $0.04$ at $\bar n = 4$, $8$, $10$, and $26$, against
the design stage simulated values $0.77$, $0.27$, $0.16$, and $0.04$
for the targeted estimator and $0.64$, $0.26$, $0.14$, and $0.04$ of AIPW. For any $\alpha \neq 0$ the
doubly robust variance carries in addition an overlap factor: in the
model that leaves the outcome regression unrestricted the efficient
variance of the contrast is
$\omega_{k}\,\sigma_{k,\mathrm{cond}}^{2}\, s_{k}$ with
$\omega_{k} = \pi_{k}(1-\pi_{k})\,\E[1/\{g(1-g)\}] \ge 1$, whereas the
estimator as implemented, with the fitted propensity truncated to
$[0.05, 0.95]$ and a fitted linear outcome model, has a product remainder
of first order wherever the truncation binds or the propensity working
model is misspecified (with mean zero under the correct outcome model, so
the bias removal stands); its large-sample factor $\omega_{k}^{\dagger}$ is
therefore bounded below not by $\omega_{k}$ but by the factor
$\omega_{k,\mathrm{lin}} \in [1, \omega_{k}]$ of the least-squares
coefficient of the arm indicator in the linear model (proof of the
proposition) and may lie on either side of $\omega_{k}$; in the simulated configurations it lies below ($\omega_{k,\mathrm{lin}}$ has median $1.5$ against $2.5$--$2.6$ for the targeted estimator and $2.1$ for AIPW, Supplementary Section~S5.8), so in this process, where the linear outcome model is correct, a plain regression adjustment on the covariates would remove the same bias at a lower variance than either doubly robust estimator; the results characterize the propensity-based correction specifically; a regression-adjusted recruiting arm is evaluated as a sensitivity in Supplementary Section~S5.9. We compute $\omega_{k}^{\dagger}$ from the influence curve (a script of the accompanying code) and fold it into $\kappa_{k}$, so that $\kappa_{k,\infty} = \omega_{k}^{\dagger} - 1 \ge
0$. In the confounded scenarios of Section~\ref{sec:simstudy},
$\omega_{k}^{\dagger}$ has median $2.5$--$2.6$ under aligned and randomly
oriented confounding and $1.1$--$1.2$ in the misspecified scenario
(first ten environments; Supplementary Section~S5.8), so
$\kappa_{k,\infty} \approx 1.5$--$1.6$ exceeds
$\bar E_{k}^{c}/\sigma_{k,\mathrm{cond}}^{2} \approx 0.3$ in every
subpopulation under the linear-index mechanisms, where the variance terms
alone never favor the doubly robust family and the crossing is produced
by the bias term, whereas in the misspecified scenario
$\kappa_{k,\infty}$ is of the order of the edge and lies below it in
about half of the subpopulations, where the variance-only boundary is
finite.
The condition compares the
families subpopulation by subpopulation with the design held fixed; the
coupling through the recruitment path is addressed by
Proposition~\ref{prop:mismatch}.

\subsection{The Boundary Between the Two Regimes}
\label{sec:surface}
Proposition~\ref{prop:crossing} gives the difference in mean squared
error between the families as
\begin{equation}
\Delta\mathrm{MSE}_{k}(\bar n, u)
\;=\; \mathrm{MSE}_{2} - \mathrm{MSE}_{1}
\;=\; s(\bar n)\,\big[\kappa_{k}(\bar n)\,\sigma_{k,\mathrm{cond}}^{2} - E_{k}\big]
\;-\; \big(c_{k}\, u\big)^{2},
\label{eq:deltaR}
\end{equation}
where $\bar n$ indexes the number of patients per arm of the
subpopulation along a balanced-growth path, treated as continuous, so
that $s(\bar n)$ is positive, continuous, and strictly decreasing from
$\infty$ to $0$ (at exact balance $s(\bar n) = 2/\bar n$; any strictly
increasing reparameterization, for instance by the subpopulation total
$2\bar n$, leaves every statement below unchanged), $u = \alpha^{\top}
w_{T,k}$ is the signed confounding alignment, and $E_{k}$ stands for the
count-weighted arm-conditional covariate variance $\bar E_{k}^{c}$ of
Proposition~\ref{prop:crossing}, held at its anticipated value (it equals
the explained variance under randomization; its own dependence on $u$ is
treated after the theorem). The doubly robust family has the smaller
mean squared error exactly where $\Delta\mathrm{MSE}_{k} < 0$, and the
boundary between the two regimes is the equal-error boundary
$\mathcal{T}_{k} = \{\Delta\mathrm{MSE}_{k} = 0\}$.

\begin{theorem}[Equal-error boundary]
\label{thm:surface}
Suppose that the variance inflation curve
$\kappa_{k} : (0, \infty) \to [0, \infty)$ is
continuous and strictly decreasing with limit $\kappa_{k,\infty} \ge 0$
as $\bar n \to \infty$, and that at sufficiently small sample sizes the
inflation term exceeds the edge,
$\lim_{\bar n \downarrow 0} \kappa_{k}(\bar n)\,\sigma_{k,\mathrm{cond}}^{2}
> E_{k}$. Then:
\begin{enumerate}
\item[(i)] (\emph{Single crossing}) For every alignment $u$ there is a
boundary sample size $n_{k}^{*}(u) \in (0, \infty]$ such that
$\Delta\mathrm{MSE}_{k}(\bar n, u) > 0$ for all $\bar n < n_{k}^{*}(u)$ and
$\Delta\mathrm{MSE}_{k}(\bar n, u) < 0$ for all $\bar n > n_{k}^{*}(u)$. The boundary
is finite if and only if $|c_{k} u| > 0$ or
$\kappa_{k,\infty}\,\sigma_{k,\mathrm{cond}}^{2} < E_{k}$.
\item[(ii)] (\emph{Partition}) The boundary depends on $u$ only through
$|u|$, so on the design quadrant $(\bar n, |u|)$ the boundary
$\mathcal{T}_{k}$ is the graph of $n_{k}^{*}(\cdot)$ and separates the
quadrant into exactly two connected regions, a sample mean region on the
side of small sample sizes and a doubly robust region on the side of large
sample sizes (when $c_{k} \ne 0$; only $(c_{k}u)^{2}$ enters, and under
the linear index~\eqref{eq:biasconstant} gives $c_{k} > 0$; if $c_{k} = 0$
and the boundary is infinite, the doubly robust region is empty).
\item[(iii)] (\emph{Structural dependence}) The sign of $\Delta\mathrm{MSE}_{k}$
depends on the configuration only through the sample sizes of the
subpopulations (via $s$), the bias magnitude $|c_{k} u|$, the edge
$E_{k}$, the conditional variance $\sigma_{k,\mathrm{cond}}^{2}$, and the
variance inflation curve $\kappa_{k}(\cdot)$. The horizon, the number of
environments,
the number of replications, and the random seeds enter only through which
points $(\bar n, |u|)$ a configuration realizes.
\item[(iv)] (\emph{Comparative statics}) Where finite, $n_{k}^{*}$ is
strictly decreasing in the bias magnitude $|c_{k} u|$ and in the edge
$E_{k}$, and strictly increasing in $\sigma_{k,\mathrm{cond}}^{2}$. When
$\kappa_{k,\infty}\,\sigma_{k,\mathrm{cond}}^{2} < E_{k}$, the boundary
$n_{k}^{*}(0)$ under randomization is the unique root of
$\kappa_{k}(\bar n)\,\sigma_{k,\mathrm{cond}}^{2} = E_{k}$; it is the
threshold $n^{*}$ reported with the variance inflation curves in
Section~\ref{sec:crossing-selector}.
\end{enumerate}
\end{theorem}
\noindent
The terms of~\eqref{eq:deltaR} mark the regimes: where
$s\,\kappa_{k}\,\sigma_{k,\mathrm{cond}}^{2}$ is largest, variance
dominates and the sample mean family is preferred (small sample sizes
under adaptive recruitment); where $(c_{k}u)^{2}$ is largest, bias
dominates and the doubly robust family is preferred, which every
configuration with $c_{k}u \neq 0$ reaches at the finite boundary
$n_{k}^{*}(u)$ since the bias term is $\mathcal{O}(1)$ while the variance
terms are $\mathcal{O}(1/n)$. Whether the biases of the subpopulations
accumulate or cancel in pooled operating characteristics is governed by
the directional consistency of $\sgn(u)$ (Corollary~\ref{cor:consistency}), which is why the confounding
axis carries an orientation label as well as a magnitude. The arm-conditional
edge itself depends on the alignment,
$\bar E_{k}^{c} = E_{k} - \bar\gamma_{k} u^{2}$ with $\bar\gamma_{k}$ the
count-weighted mean of the $\gamma_{k,a}$; the term can be absorbed into
the edge, so that parts (i)--(iii) hold with $E_{k} - \bar\gamma_{k} u^{2}$
in place of $E_{k}$ and the comparative statics in $|u|$ hold wherever
$c_{k}^{2} > \bar\gamma_{k} s(\bar n)$, with $c_{k}$ the between-arm
shift coefficient, which is throughout the sampled range under the
linear-index mechanisms and in all but a minority of subpopulations
under the quadratic index (Supplementary Section~S5.8).

Only one object in~\eqref{eq:deltaR} is estimated by simulation rather than derived, the variance inflation curve $\kappa_{k}(\cdot)$, a property of the chosen nuisance estimators at the sample sizes the design realizes. It is also a property of the assignment mechanism: under confounding the overlap worsens as $\|\alpha\|$ grows and the inflation rises with it (median $\kappa$ at ten patients per arm from $0.2$ at strength $0$ to $3.3$ at strength $3$, Section~\ref{sec:map}), so the theorem holds at a fixed mechanism, its comparative statics in $|u|$ refer to the alignment of $w_{T,k}$ at a fixed strength, a change of strength moves the bias term and the curve together, and Figure~\ref{fig:map} draws one boundary per strength. The raw simulated curves are not monotone at every strength (Supplementary Section~S5.7); the isotonic fit imposes the monotone curve of the hypothesis. Changing the nuisance learner or the outcome model moves
$\kappa_{k}$ and with it the location of $\mathcal{T}_{k}$, but not the
existence of the boundary, its single-crossing structure, or its
comparative statics, which follow from the
$\mathcal{O}(1)$-versus-$\mathcal{O}(1/n)$ asymmetry for any admissible
variance inflation curve: the boundary is defined by the theory, and
the simulations locate it by estimating $\kappa_{k}$. A simulated curve
is an isotonic fit, nonincreasing rather than strictly decreasing and
held constant outside the simulated range,
so the boundary is reported as the operational threshold
$n^{*} = \inf\{\bar n : \kappa_{k}(\bar n)\,\sigma_{k,\mathrm{cond}}^{2}
\le E_{k}\}$, infinite when the set is empty (Supplementary
Section~S6.1 records the weaker conditions under which
(i) still holds). At or below the rank threshold $m$ of
Proposition~\ref{prop:degeneracy} the weights are fixed at $e_{k}$
and~\eqref{eq:deltaR} applies subpopulation by subpopulation; with active
borrowing the two families use different weights and the representation,
covariate sampling, and cross-covariance terms enter, and we do not
extend the statement to that case. The recruitment rule is a channel
that the crossing condition does not describe, since the sensitivity
index follows the running
estimates of each family, so the two families collect different data even
where their mean squared errors per subpopulation are equal; this coupling is stylized
by Proposition~\ref{prop:mismatch} and measured in
Section~\ref{sec:mismatch-results}.

\subsection{Recruitment--Analysis Consistency}
\begin{proposition}[Mismatch penalty]
\label{prop:mismatch}
Let a control budget be allocated by the minimax-variance rule
$n_{k}^{(0)} \propto \sigma_{k}^{2}$ under the variance profile of one
estimator family while the final analysis uses the other family. The
achieved worst case variance then exceeds that of the matched
recruitment--analysis pair by the factor $\rho(F \to G)$ of~\eqref{eq:mismatch} below, which is at least one and equals one only when
the two variance profiles are proportional.
\end{proposition}
\noindent
Allocating a control budget $N = \sum_{k} n_{k}^{(0)}$ to minimize
$\max_{k} \sigma_{k}^{2}/n_{k}^{(0)}$, a stylized version of
sensitivity-driven recruitment, equalizes the variances of the
subpopulations, $n_{k}^{(0)} \propto \sigma_{k}^{2}$. If recruitment
follows the profile $\{\sigma_{k,F}^{2}\}$ of one family $F$ while the
analysis estimator has the profile $\{\sigma_{k,G}^{2}\}$ of the other
family $G$, the achieved worst case 
variance exceeds that of the matched design by
\begin{equation}
\rho(F \!\to\! G)
= \Big[\max_{k} \frac{\sigma_{k,G}^{2}}{\sigma_{k,F}^{2}}\Big]
\cdot
\frac{\sum_{l} \sigma_{l,F}^{2}}{\sum_{l} \sigma_{l,G}^{2}}
\;\ge\; 1,
\label{eq:mismatch}
\end{equation}
with equality if and only if the profiles are proportional. For the
present families the ratio
$\sigma_{k,\mathrm{eff}}^{2}/\sigma_{k,\mathrm{cond}}^{2}
= 1 + E_{k}/\sigma_{k,\mathrm{cond}}^{2}$ varies with the share of outcome
variance explained by the covariates, so the penalty is positive whenever
that share varies between subpopulations. Nothing in the proposition is
specific to this design: any adaptive experiment that allocates by the
minimax-variance rule $n_{k} \propto \sigma_{k}^{2}$ under one variance
functional and is analyzed with an estimator with a nonproportional
variance profile pays the same factor. Designs that allocate to minimize
the variance of the analysis estimator, from the propensity score design
of \citet{hahn2011} to the adaptive Neyman allocations cited in the
Introduction, are design--analysis pairs that are matched in this sense.

\subsection{A Pre-Specified Selector}
\label{sec:selector}
We propose to select one estimator family per trial before recruitment begins; the selection is per trial because the family fixes the final analysis and, in the matched protocol evaluated here, in which the selector arm recruits and analyzes with the family it selects, the recruitment variances as well (a mixture would, for the sample mean analysis, incur the penalty of Proposition~\ref{prop:mismatch}). The mismatch arm of Section~\ref{sec:mismatch-results} shows the doubly robust analysis gaining from the sample mean path in every core scenario, so the protocol recommended in Section~\ref{sec:discussion} recruits with the sample mean functional in every trial and routes the analysis only; steps (ii) and (iii) then choose the analysis, not the recruitment functional. The selector arm as evaluated follows the doubly robust path where it commits, and its discrimination cost in Table~\ref{tab:selector} is that path's cost: the same selector with the recommended pairing, built from the doubly robust analysis of the sample-mean-recruited data in the environments it routes, is within two standard errors of the better family in AUC in every core scenario, above the matched arm by $0.001$ to $0.035$ AUC (beyond two standard errors in six of the eight scenarios), and within $0.09$ of the sample mean family's design-level mean squared error, which it exceeds by $+0.08 \pm 0.02$ and $+0.09 \pm 0.03$ at $K = 8$ under aligned and random confounding (Supplementary Table~S6). Which side of the boundary a trial lies
on can be decided largely at the design stage: the anticipated number of
patients per subpopulation and arm is set by the budget,
$\bar n = H/(2K) + 1$ with the burn-in patient; the variance inflation
curve is estimated before the trial by simulating the pre-specified
nuisance estimators at the sample sizes of the design; the degeneracy status is
arithmetic; design-controlled assignment makes $\alpha = 0$ true by
construction; and the edge $E_{k}$ and the conditional variance
$\sigma_{k,\mathrm{cond}}^{2}$ must be anticipated, as the outcome
variance is in a power calculation. The alignment $\alpha^{\top} w_{T,k}$
is the only coordinate that the protocol can neither fix nor anticipate
from pilot data, and the selector never estimates it in advance: it enters
through the worst case  over the alignment, $|c_{k}|\,\|\alpha\|\,\|w_{T,k}\|$,
which needs the anticipated confounding strength, the outcome-weight norm
and, through $c_{k}$, the anticipated treated fraction $\pi_{k}$. The
theorem is per subpopulation, whereas the selector evaluates it once per
trial at the anticipated mean count and environment means; under adaptive
recruitment the realized counts spread around the mean, so the rule is a
design-level approximation whose adequacy the simulations test.
Pre-specifying the estimator by a rule fixed before the data are seen is
the principle of adaptive pre-specification \citep{balzer2016}, which
chooses among adjustment sets; here the choice is between families and
is made from design stage quantities.

The procedure is as follows. (i) Compute $m = 1 + \Dx + \Tpre$ and report
whether $K \le m$; if so (with the centroids in general position, which
the solver confirms by the realized rank), hard-constrained borrowing is
inert and the choice reduces to the estimators of the individual
subpopulations (the
noiseless count $1 + \Dx + \min(\Tpre, \Dz)$ is reported for
soft-constrained weighting). (ii) If the assignment is randomized by
design, evaluate the crossing condition~\eqref{eq:crossing} with
$\alpha = 0$ at $\bar n$ and select the doubly robust family only when
$E_{k} > \kappa(\bar n)\,\sigma_{k,\mathrm{cond}}^{2}$, with $\kappa$
evaluated at the median of the simulated values (median rule) or at
their upper quartile, about half an interquartile range above the median
(upper-quartile rule). (iii) If the assignment is confounded or cannot be
guaranteed to be randomized, evaluate the crossing
condition~\eqref{eq:crossing} at $\bar n$ with the anticipated worst case 
bias band $|c_{k}|\,\|\alpha\|\,\|w_{T,k}\|$ of Proposition~\ref{prop:bias} on
its left-hand side, $\kappa$ simulated at fixed sample sizes at the
anticipated confounding strength $\|\alpha\|\,\sigma_{x}$ (the
design stage curve), the balanced $s = 2/\bar n$ (under confounded
assignment the realized $s$ exceeds it by the factor
$1/\{4\pi_{k}(1-\pi_{k})\}$, so the balanced value favors the doubly
robust family; Supplementary Section~S5.8 quantifies
this approximation), and the anticipated $E_{k}$ in place of
$\bar E_{k}^{c}$ (an upper bound under the linear-index mechanisms, which
also favors the doubly robust family, slightly; Supplementary
Section~S2), and select the
doubly robust
family when it holds; otherwise retain the sample mean family and report
the band as a pre-specified caveat. Since
$\kappa_{k,\infty} \sigma_{k,\mathrm{cond}}^{2} \ge E_{k}$ for the
design stage curve at the simulated strength, a variance-only version of
step (iii) would have an infinite threshold; the band is what routes a
confounded trial. At the analysis stage the inequality is evaluated again
with plug-in estimates $(\hat\sigma_{x}, \hat w_{T,k}, \hat\alpha,
\hat c_{k})$ and the realized sample sizes; a contradiction with the pre-specified choice is reported as a sensitivity flag, with the other family as a supporting analysis (this audit is proposed here and not evaluated in the simulations). Supplementary Figure~S1 shows
the procedure as a flowchart with the simulated scenarios routed through
it. In the simulations the selector arm uses in step (ii) the
design-embedded variance inflation curve, estimated inside the adaptive
design on held-out environments (Section~\ref{sec:crossing-selector}),
and in step (iii) the design stage curve simulated under aligned
confounding, because the design-embedded confounded curve is dominated by
the sample mean fallback of the held-out runs and would make step (iii)
vacuous (Section~\ref{sec:crossing-selector}); it
is evaluated under exact anticipation, with the edge, the conditional
variance, and the bands computed from the true parameters of each
environment, so that its record measures the rules and not the quality of the anticipation (Supplementary Section~S2). That record is the rules' record with their intended inputs, not a bound: an anticipation error can lower or raise it, since a protocol that under-anticipates the band routes fewer environments to the family that is behind in AUC, while an under-anticipated strength can route more, because the curve falls with it (Supplementary Section~S5.8 reports the routing under each setting, and the selector arm's deficits where they change, with the anticipated band halved or doubled, the anticipated edge at one half and one and a half times its value, and the anticipated strength at $1$ and $3$ against the true $2$, with the band and the curve of that strength). Three elements of the rule, the curve of step (iii), the upper-quartile rule, and the worst case form of the band, were fixed during this study with the routing of the core scenarios in view, so the selector is pre-specified in the sense of being fixed before any future trial, not validated prospectively here; and the design stage confounded curve is simulated from the first ten $K = 3$ evaluation environments, so step (iii) is in-sample for those ten (the same curve estimated on held-out environments changes the routing by at most one environment per scenario under the median rule and two under the upper-quartile rule, Supplementary Section~S5.9).

% ===========================================================================
\section{SIMULATION STUDY}
\label{sec:simstudy}
% ===========================================================================
Two couplings distinguish the design from $K$ parallel two-arm trials, and
the scenario grid is constructed so that each can be switched off:
recruitment is coupled to estimation, since the next subpopulation is
chosen by a sensitivity index computed from the running estimates, so the
estimator family changes which data are collected
(Proposition~\ref{prop:mismatch}), and estimation is coupled across
subpopulations through the constrained weights, a channel that can
substitute for sample size but not for bias correction and that is
switched off entirely when $K \le m$ (Proposition~\ref{prop:degeneracy}).
Uniform recruitment with inert borrowing is $K$ independent trials, the
baseline against which the contrasts along the recruitment and borrowing
factors are measured.

\subsection{Data Generating Process}
\label{sec:dgp}
Each subpopulation $k$ has a covariate centroid $x_{k} \in \R^{\Dx}$, which
the design knows and uses in the matching constraints, and a latent
loading $z_{k} \in \R^{\Dz}$, which it does not. A patient $j$ in
subpopulation $k$ has
\begin{equation}
x_{kj} = x_{k} + \eta_{kj},\ \ \eta_{kj} \sim \mathcal{N}(0, \sigma_{x}^{2} I_{\Dx}),
\qquad
z_{kj} = z_{k} + \zeta_{kj},\ \ \zeta_{kj} \sim \mathcal{N}(0, \sigma_{z}^{2} I_{\Dz}),
\end{equation}
of which the design observes $x_{kj}$ only. The $\Tpre$ pre-treatment
outcomes use weights $w_{t} \in \R^{\Dx}$ and $\mu_{t} \in \R^{\Dz}$ that
are global, that is, shared across subpopulations as in the base design,
\begin{equation}
Y_{kjt} = \delta_{t} + w_{t}^{\top} x_{kj} + \mu_{t}^{\top} z_{kj} + \varepsilon_{kjt},
\qquad \varepsilon_{kjt} \sim \mathcal{N}(0, \sigma^{2}), \quad t = 1,\dots,\Tpre,
\end{equation}
and the single post-treatment outcome, the potential outcome
$Y_{kj}(a)$ of Section~\ref{sec:setup} written $Y_{kjT}^{(a)}$, uses
subpopulation-specific weights $w_{T,k} \in \R^{\Dx}$ and
$\mu_{T,k} \in \R^{\Dz}$,
\begin{equation}
Y_{kjT}^{(a)} = \delta_{T} + w_{T,k}^{\top} x_{kj} + \mu_{T,k}^{\top} z_{kj}
             + r_{k}\, a + \varepsilon_{kjT}^{(a)},
\qquad \varepsilon_{kjT}^{(a)} \sim \mathcal{N}(0, \sigma^{2})
\text{ independently across } a,
\end{equation}
so that $r_{k}$ is the effect in subpopulation $k$. The marginal
post-treatment variance is
$\sigma_{k,\mathrm{eff}}^{2} = \sigma^{2} + \sigma_{x}^{2}\|w_{T,k}\|^{2}
+ \sigma_{z}^{2}\|\mu_{T,k}\|^{2}$, whereas conditioning on $x_{kj}$
leaves $\sigma_{k,\mathrm{cond}}^{2} = \sigma^{2}
+ \sigma_{z}^{2}\|\mu_{T,k}\|^{2}$. The environment-level parameters (centroids, loadings, intercepts,
outcome weights, and effects) are drawn once per environment from
absolutely continuous laws and held fixed (Supplementary
Section~S3); the absolute continuity is what the proof of
Proposition~\ref{prop:degeneracy} uses, and only the patient draws, the
arm assignments, and the outcome noise vary between replications. Under
confounded assignment every backdoor path from $A$ into $Y_{T}$ passes
through the observed $x$, since the latent loading influences the
outcomes but not the assignment, so adjustment for $x$ within
subpopulation identifies the arm means, and adaptive recruitment changes
which subpopulation contributes the next patient but not the outcome law
given subpopulation and arm (Supplementary Figure~S2).

\subsection{Scenarios and Arms}
\label{sec:metrics-reporting}
Table~\ref{tab:design} lists the design factors, each chosen to isolate
one mechanism of Section~\ref{sec:theory}. All scenarios use
$\sigma^{2} = 1$, $\sigma_{x} = \sigma_{z} = 1$ unless varied,
$\Dx = 2$, $\Dz = 2$, and $\Tpre = 4$, so that $m = 7$ and
$\rank(\Gamma^{\circ}) = \min(K, 5)$. The numbers of subpopulations, $K \in \{3, 8\}$, lie on both sides of
the degeneracy threshold, and each $K$ is crossed with four assignment
mechanisms at horizon $H = 150$: randomized assignment, and three confounded mechanisms at
$\|\alpha\| = 2$, namely aligned ($\alpha$ along the mean outcome-weight
direction $\bar w_{T} = K^{-1}\sum_{k} w_{T,k}$ of the environment, so
that the alignment $\alpha^{\top} w_{T,k}$ is positive on average across
the subpopulations but not in every one: in $240$ of $300$ subpopulations at $K = 3$ and $275$ of $400$ at $K = 8$), randomly oriented
($\alpha$ redrawn for each environment, uniformly on the sphere), and
misspecified (the true propensity is
$\expit(\alpha^{\top} x + (\alpha^{\top} x)^{2})$ with $\alpha$ as in the
aligned scenario, while the working model stays logistic in $x$; under
the quadratic index the between-arm shift reverses where
$\alpha^{\top} x_{k} \lesssim -1/2$, so the shift coefficient $c_{k}$ is
negative in $131$ of $300$ subpopulations at $K = 3$ and $171$ of $400$ at $K = 8$ and the bias is positive in $163$ and $211$, close to random orientation in sign). These eight core scenarios are reported in full. Seven further
configurations are probes: the $K = 8$ scenarios at $H = 400$
(randomized and aligned), uniform recruitment at $K = 8$, $\sigma_{x} = 2$
with $\sigma_{z} = 1$, two that replace the oracle variances in the
weighting and the recruitment index by running estimates (randomized and
aligned; Supplementary Section~S5.4), and an anchor
configuration with effects $r = (-10, 0, +10)$, an implementation check.
The mismatch pairs of Proposition~\ref{prop:mismatch} and the same-path
runs behind the variance inflation curve complete the inventory
(Supplementary Table~S2). The eight core scenarios are averaged over $50$ independently drawn environments with $20$ trial replications each ($100$ in the three confounded $K = 3$ scenarios, extended from $50$ after the analysis of the first $50$, as described below) and the probes over ten; every arm runs on common seeds, and the first $50$ environments are shared across the four scenarios at each $K$, so contrasts are paired within scenarios and, by environment, across the scenarios at each $K$ on the common environments.

\begin{table}[!htbp]
\centering
\caption{Design factors of the simulation study.}
\label{tab:design}
\small
\setlength{\tabcolsep}{4pt}
\resizebox{\ifdim\width>\linewidth\linewidth\else\width\fi}{!}{%
\begin{tabular}{@{}>{\raggedright\arraybackslash}p{0.27\textwidth}>{\raggedright\arraybackslash}p{0.33\textwidth}>{\raggedright\arraybackslash}p{0.34\textwidth}@{}}
\toprule
Factor & Levels & Mechanism isolated \\
\midrule
Recruitment policy & adaptive; uniform & recruitment effect on the families \\
Arm assignment & randomized (design-controlled, Section~\ref{sec:arm}); confounded & bias correction \\
Confounding direction & aligned (on average); randomly oriented & directional consistency
(Corollary~\ref{cor:consistency}) \\
Propensity-model form & correctly specified; quadratic index & robustness to propensity misspecification (the outcome model is correct in the core scenarios) \\
Sample size per subpopulation & varied via $K$ at fixed $H$ & efficiency vs.\ finite-sample
penalty \\
Degeneracy regime & $K \le m$; $K > m$ & borrowing collapse
(Proposition~\ref{prop:degeneracy}) \\
\bottomrule
\end{tabular}}
\end{table}

In each scenario we evaluate twelve arms: the six variants of the grid,
the AIPW and cross-fitted comparators at the weights of \textbf{(2b)},
the selector in its median and upper-quartile rules, and the
optimal-transport relaxation of Supplementary Section~S4 with
its oracle version. The nuisance models of family 2 are a linear outcome
model in the covariates and the arm and a logistic propensity model,
truncated to $[0.05, 0.95]$, with a weak ridge penalty; when a
subpopulation holds fewer than six patients, or fewer than two in an arm,
the doubly robust arm means fall back to the sample means, and the share
of estimates affected is reported. Under randomization the
subpopulation-level arms \textbf{(1a, 2a)} and the transport relaxations
run on a subpopulation-level patient pool in which every patient sits at
the centroid of its subpopulation, as in the base design, while the
individual-level arms run on the individual-level pool. Under
confounded assignment every arm draws from one individual-level pool, the
balanced burn-in is forced, and the arm of each recruited patient is drawn
from the propensity at observation time; the subpopulation-level arms
then store the centroid as the covariate, so only the individual-level
doubly robust arms and the comparators remove confounding bias. The
individual-level family-2 arms \textbf{(2b, 2c)} are recruited and
weighted with the conditional variances
multiplied by the design-embedded inflation factor of
Section~\ref{sec:crossing-selector}, estimated on ten separate
environments and then held fixed; the comparators are analysis-only arms
on the patients recruited by \textbf{(2b)} at its weights, and the
selector arm copies, for each environment, the arrays of the family it
selects. The remaining implementation details, the tie-breaking of the
look-ahead, and the recruitment episode of the base design are recorded
in Supplementary Section~S3.

For each arm we report TPR and FPR as functions of $H$, averaged over
environments and replications. Since a location shift moves an estimator
along its own receiver operating characteristic curve rather than onto a
better one, inflating TPR and FPR together under confounding, we also
report the area under that curve (AUC), the probability that a randomly
chosen responsive subpopulation receives a higher estimate than a
nonresponsive one, which is invariant to a common shift. The AUC of an environment pools the $K \times 20$
estimates of its replications at the horizon into one Mann--Whitney
statistic against the sign pattern (Supplementary
Section~S3); environments in which all subpopulations are
responsive or all nonresponsive have no AUC ($39$ of the $50$ environments have one in the randomized $K = 3$ scenario, $80$ of the $100$ in the confounded $K = 3$ scenarios, all $50$ at $K = 8$), and the raw rates
keep them with the convention of the base design that they contribute a
rate of zero. Because the AUC is invariant to the part of a confounding bias that is common to the subpopulations of an environment and responds only to the part that differs between them (in the simulated scenarios the part that differs is as large as the common part at $K = 3$ under aligned confounding and two to three times it at $K = 8$, so the AUC does see the bias; Supplementary Section~S5.4), under confounding we report as co-primary evidence the zero-threshold false positive rate, which a common bias moves, and
the true positive rates of \textbf{(1b)} and \textbf{(2b)} at a common
false positive rate, the zero-threshold rate of \textbf{(2b)}, to which
both arms are re-thresholded by the same rule (``matched FPR''; the rule is stated in Supplementary Section~S3; like the AUC, this comparison is invariant to a common shift and compares the two rankings at one operating point). Both measures were designated after the core runs had been examined, when the blindness of the AUC to a common shift became apparent, and are not among the pre-specified predictions of Supplementary Table~S5. All uncertainty statements are paired differences with standard errors across environments; ``within simulation error'' means within two paired standard errors, a statement about the resolution of $50$ environments per core scenario ($100$ in the three confounded $K = 3$ scenarios, extended from $50$ after the analysis of the first $50$, a data-dependent choice made to improve their precision; ten per probe) and not a demonstration of equivalence. The two-standard-error criterion is descriptive and unadjusted for multiplicity (about two of the $48$ contrasts of Table~\ref{tab:contrasts} would exceed it under the null), so conclusions rest on patterns that repeat across scenarios; at $K = 3$ under confounding the \textbf{(1b)}$-$\textbf{(2b)} contrasts have standard errors of about $0.01$--$0.015$ AUC with $100$ environments, about two to three times those at $K = 8$ (the \textbf{(1a)}$-$\textbf{(2a)} contrasts are about equally precise at both $K$), so ``within error'' there leaves differences of about $\pm 0.03$ open. No precision target was set in advance, and the ten-environment probes are exploratory (Supplementary Section~S3 summarizes the design in the ADEMP structure of \citet{morris2019}: aims, data-generating mechanisms, estimands, methods, performance measures).
The inflation $\kappa_{k}(\bar n)$ is estimated by simulation twice, at
fixed sample sizes before the trial and inside the adaptive design on ten
held-out environments (Section~\ref{sec:crossing-selector}); the selector
arm uses the latter in step (ii) and the former, simulated under aligned
confounding, in step (iii).

% ===========================================================================
\section{RESULTS}
\label{sec:results}
% ===========================================================================
Three protocol decisions depend on the results, which
Tables~\ref{tab:contrasts} and~\ref{tab:selector} summarize; the AUC of
every arm, the operating characteristics, and the trajectories are in
Supplementary Section~S5 (Supplementary
Table~S3), and the optimal transport relaxation,
which borrows in every regime but improves discrimination only under confounding (beyond two standard errors only in the misspecified $K = 8$ scenario), in Supplementary Section~S4.1. First,
the estimator family is a pre-specified choice keyed to the regime. With
about $10$ patients per subpopulation and arm and adaptive recruitment,
the sample mean family is ahead of the doubly robust family in the
$K = 8$ scenarios by $0.013$ to $0.020$ AUC at the individual-level; with
about $26$ under randomization the two families are within simulation
error at $K = 3$ at the individual-level; both sample sizes lie beyond
the median-rule boundary of about $9$ patients per subpopulation and arm
that the variance inflation curve estimated on separate trials places
under
randomization, and below its upper-quartile counterpart of about $29$.
Under
aligned confounding the bias removal of the doubly robust family lowers the false positive rate by $6$ to $13$ points in the core scenarios ($14$ in the $H = 400$ probe), while the comparison at matched false positive rate is within error in the core scenarios and at $2.0$ standard errors in that probe. Second, the gap in
discrimination at small sample
sizes is a property of the recruitment path and not of the estimator: the
doubly robust analysis of a trial recruited by the sample mean family
matches or exceeds the discrimination of the sample mean analysis of the
same patients wherever the pairing was run, and the map of
Section~\ref{sec:map} shows that the boundary predicts the realized
sign of the mean squared error difference wherever the sample size is
held fixed; in mean squared error the doubly robust family is behind in
every adaptive configuration but the anchor configuration, through its fits on the few
patients of starved subpopulations. Third, the selector removes the confounding bias where it commits to the doubly robust family and, evaluated as a matched protocol that also recruits along that family's path, pays that family's deficit in threshold-free discrimination at small counts and its larger mean squared error at both sampled sizes; the recommended pairing removes the first cost and about nine tenths of the second (Table~\ref{tab:selector}; Supplementary Table~S6; Supplementary
Section~S5.8).

\begin{table}[!htbp]
\centering
\caption{Paired contrasts at the end of the trial in the eight core
scenarios, mean $\pm 1$ paired standard error across environments ($39$ of the $50$ environments with a defined AUC in the randomized $K = 3$ scenario, $80$ of the $100$ in the confounded $K = 3$ scenarios, all $50$ at $K = 8$).}
\label{tab:contrasts}
\footnotesize
\setlength{\tabcolsep}{3pt}
\resizebox{\ifdim\width>\linewidth\linewidth\else\width\fi}{!}{%
\begin{tabular}{@{}lcccccc@{}}
\toprule
& \multicolumn{2}{c}{AUC, family contrasts} & \multicolumn{2}{c}{Percentage points, (1b) vs.\ (2b)} & \multicolumn{2}{c}{AUC, comparators} \\
\cmidrule(lr){2-3} \cmidrule(lr){4-5} \cmidrule(lr){6-7}
Scenario & (1b)$-$(2b) & (1a)$-$(2a) & $\Delta$FPR & $\Delta$TPR (matched) & AIPW$-$(2b) & CF$-$(2b) \\
\midrule
\multicolumn{7}{@{}l}{$K = 3$, $\bar n^{(a)} \approx 26$} \\
Randomized & $+0.010 \pm 0.008$ & $+0.028 \pm 0.008$ & $-0.2 \pm 0.7$ & $-10.1 \pm 4.4$ & $+0.002 \pm 0.001$ & $-0.001 \pm 0.001$ \\
Aligned & $+0.026 \pm 0.014$ & $+0.007 \pm 0.005$ & $+13.1 \pm 2.0$ & $-1.8 \pm 2.5$ & $+0.019 \pm 0.003$ & $-0.018 \pm 0.003$ \\
Random & $+0.007 \pm 0.012$ & $+0.005 \pm 0.005$ & $+3.2 \pm 1.9$ & $-9.4 \pm 3.2$ & $+0.017 \pm 0.002$ & $-0.028 \pm 0.004$ \\
Misspecified & $+0.036 \pm 0.009$ & $+0.005 \pm 0.004$ & $-0.2 \pm 1.4$ & $+0.6 \pm 1.8$ & $+0.017 \pm 0.003$ & $-0.010 \pm 0.003$ \\
\addlinespace[4pt]
\multicolumn{7}{@{}l}{$K = 8$, $\bar n^{(a)} \approx 10$} \\
Randomized & $+0.017 \pm 0.004$ & $+0.018 \pm 0.003$ & $-0.7 \pm 0.5$ & $-3.9 \pm 2.0$ & $+0.003 \pm 0.000$ & $-0.002 \pm 0.001$ \\
Aligned & $+0.016 \pm 0.005$ & $+0.010 \pm 0.004$ & $+5.7 \pm 1.1$ & $-2.7 \pm 1.8$ & $+0.013 \pm 0.001$ & $-0.009 \pm 0.002$ \\
Random & $+0.020 \pm 0.007$ & $+0.010 \pm 0.005$ & $+0.4 \pm 1.1$ & $-1.7 \pm 1.6$ & $+0.009 \pm 0.001$ & $-0.011 \pm 0.002$ \\
Misspecified & $+0.013 \pm 0.005$ & $-0.001 \pm 0.004$ & $+0.5 \pm 0.8$ & $-0.9 \pm 1.6$ & $+0.009 \pm 0.001$ & $-0.001 \pm 0.001$ \\
\bottomrule
\end{tabular}}
\par\vspace{4pt}
\begin{minipage}{\linewidth}\footnotesize
NOTE: A positive
AUC difference favors the arm named first; CF is the cross-fitted
comparator. $\Delta$FPR is the difference of the zero-threshold
false positive rates of \textbf{(1b)} and \textbf{(2b)} in percentage
points; the matched column is the true positive rate of
\textbf{(1b)} minus that of \textbf{(2b)}, both re-thresholded by the same
rule to the zero-threshold false positive rate of \textbf{(2b)}, in
percentage points, so a negative value favors the doubly robust arm. The $\Delta$FPR column uses all environments of a scenario, the matched column those with a defined AUC. The (1a)$-$(2a) column compares arms with identical point estimates (on centroid covariates the targeting shift is zero, Supplementary Section~S3), recruited with different variance inputs, so it isolates the recruitment-input channel rather than the estimator. Supplementary Figure~S3 draws the AUC
contrasts together with the transport relaxation.
\end{minipage}
\end{table}

\begin{table}[!htbp]
\centering
\caption{The pre-specified selector in the eight core scenarios, evaluated under exact anticipation as a matched protocol that recruits and analyzes with the family it selects (Section~\ref{sec:selector}; Supplementary Table~S6 evaluates the recommended pairing).}
\label{tab:selector}
\footnotesize
\setlength{\tabcolsep}{4pt}
\resizebox{\ifdim\width>\linewidth\linewidth\else\width\fi}{!}{%
\begin{tabular}{@{}llcccccc@{}}
\toprule
& & \multicolumn{3}{c}{Median rule} & \multicolumn{3}{c}{Upper-quartile rule} \\
\cmidrule(lr){3-5} \cmidrule(lr){6-8}
Scenario & Better family & Routed to DR & AUC $-$ better & $|\Delta|$/SE & Routed to DR & AUC $-$ better & $|\Delta|$/SE \\
\midrule
\multicolumn{8}{@{}l}{$K = 3$, $\bar n^{(a)} \approx 26$} \\
Randomized & sample mean & 35 & $-0.002 \pm 0.006$ & 0.3 & 12 & $-0.003 \pm 0.002$ & 1.3 \\
Aligned & sample mean & 99 & $-0.026 \pm 0.014$ & 1.8 & 99 & $-0.026 \pm 0.014$ & 1.8 \\
Random & sample mean & 99 & $-0.006 \pm 0.012$ & 0.5 & 99 & $-0.006 \pm 0.012$ & 0.5 \\
Misspecified & sample mean & 62 & $-0.020 \pm 0.009$ & 2.3 & 56 & $-0.019 \pm 0.008$ & 2.2 \\
\addlinespace[4pt]
\multicolumn{8}{@{}l}{$K = 8$, $\bar n^{(a)} \approx 10$} \\
Randomized & sample mean & 21 & $-0.004 \pm 0.002$ & 2.0 & 0 & $0$ & -- \\
Aligned & sample mean & 40 & $-0.014 \pm 0.005$ & 2.8 & 30 & $-0.009 \pm 0.004$ & 2.2 \\
Random & sample mean & 40 & $-0.017 \pm 0.007$ & 2.6 & 27 & $-0.010 \pm 0.005$ & 1.8 \\
Misspecified & sample mean & 4 & $-0.002 \pm 0.001$ & 1.6 & 2 & $-0.001 \pm 0.001$ & 1.3 \\
\bottomrule
\end{tabular}}
\par\vspace{4pt}
\begin{minipage}{\linewidth}\footnotesize
NOTE: The columns give the family with the higher AUC at the end of the trial, and for each rule the
number of environments ($50$; $100$ in the confounded $K = 3$ scenarios) routed to the doubly robust family (DR), the
AUC of the selector arm minus that of the better family (mean $\pm 1$
paired standard error), and that deficit in standard errors (the
randomized $K = 8$ value $2.0$ is $1.95$). The routing
under randomization uses the design-embedded variance inflation curve
and under
confounding the anticipated worst case band against the design stage
confounded curve (Section~\ref{sec:selector}). Supplementary
Figure~S14 draws the median-rule record with the verdicts
at matched false positive rate.
\end{minipage}
\end{table}

\subsection{The Degenerate Randomized Configuration}
\label{sec:mainK3}
With $K = 3$, adaptive recruitment, and randomized assignment, a balanced
burn-in of $2K = 6$ patients is followed by $H = 150$ adaptive episodes,
so the number of patients per subpopulation and arm averages
$\bar n^{(a)} \approx 26$ at the horizon (about $57$ per arm in the most
recruited subpopulation and $5$ in the least). Since $K = 3 \le m = 7$
the configuration is degenerate: the constrained solve returns $e_{k}$,
the variants within a family collect identical data, and the operating
characteristics of \textbf{(1b)} and \textbf{(1c)}, and of \textbf{(2b)}
and \textbf{(2c)}, coincide to the last digit (Supplementary
Table~S4). With $\alpha = 0$ the crossing condition reduces
to the explained-variance edge against the inflation term, and at this
sample size
the two families are within simulation error (Table~\ref{tab:contrasts}):
$+0.010 \pm 0.008$ AUC for \textbf{(1b)} against \textbf{(2b)} on the
individual-level pool, with the zero-threshold rates within about a point
of each other and the doubly robust arm ahead by $10.1 \pm 4.4$ points of
true positive rate at the false positive rate of its zero threshold
($2.3$ standard errors, consistent with crossing receiver operating characteristic (ROC) curves), while on the
subpopulation-level pool \textbf{(1a)} keeps
a margin of $+0.028 \pm 0.008$ over \textbf{(2a)}, whose nuisance models
see only the centroid and whose influence curve variances steer its
recruitment differently. About $26$ patients per subpopulation and arm
thus lies beyond the median-rule boundary of about $9$ and near the
upper-quartile one of about $29$, between which the simulated
variance inflation curves place the parity of the two families.
The comparators show that the estimator itself is not at issue at this
count: the targeting step and the one-step AIPW recombination nearly
coincide ($+0.002 \pm 0.001$ AUC), and cross-fitting costs almost nothing
($-0.001 \pm 0.001$ AUC); its cost appears where the halved nuisance sample and the per-fold targeting meet nontrivial propensities ($0.01$--$0.03$ in the confounded
$K = 3$ scenarios and about $0.01$ under aligned and random confounding at
$K = 8$) and where every subpopulation is split ($0.040$ under uniform
recruitment). The selector chose the doubly robust
family in
$35$ of the $50$ environments under the median rule and in $12$ under the
upper-quartile rule and is within error of the better family under both
(Table~\ref{tab:selector}). The optimal transport relaxation, the only
variant whose borrowing is active in this regime, does not improve on
\textbf{(1a)} under randomization (Supplementary
Section~S4.1).

\subsection{Confounded Configurations at \texorpdfstring{$K = 3$}{K = 3}}
\label{sec:confK3}
The three confounded scenarios rerun the environments and seeds of the randomized scenario, and $50$ further environments, with the arm drawn from the propensity; since every arm draws from one
individual-level pool, only the individual-level doubly robust arms and
their comparators remove confounding bias.

\emph{A common bias moves the operating point; the uneven part also moves the ranking.} At the zero threshold the sample mean family seems to lead clearly under aligned
confounding, by $+8.0 \pm 1.3$ points of TPR for \textbf{(1b)} over \textbf{(2b)}, but at $+13.1 \pm 2.0$ points of FPR: aligned confounding
moves the biased estimator up along its own receiver operating
characteristic curve (Supplementary Figure~S7).
At the false positive rate of \textbf{(2b)} the point estimate favors the doubly robust arm in true positive rate by $1.8 \pm 2.5$ points, within error, and the AUC difference is $+0.026 \pm 0.014$, $1.8$ standard errors from zero in favor of the sample mean, an interval that leaves $\pm 0.03$ open. Under random orientation the doubly robust arm is ahead at matched false positive rate by $9.4 \pm 3.2$ points ($3.0$ standard errors) while the AUC contrast is within error ($+0.007 \pm 0.012$): a sign-random bias widens the spread of the sample mean's null estimates, and its removal by the doubly robust arm shows at matched false positive rate; the AUC contrast is a net of two sides, since the bias also differs between the subpopulations of an environment and so lowers the AUC of the sample mean arm (by $0.034 \pm 0.010$ under random orientation and $0.029 \pm 0.010$ under aligned confounding at $K = 3$ when the predicted bias is removed from its estimates), which the variance side of the doubly robust family, its estimator's inflation on identical patients and its recruitment path, offsets (Supplementary Section~S5.4). Across
the three orientations and both levels one of the six family contrasts is
individually significant (Table~\ref{tab:contrasts}): \textbf{(1b)} exceeds \textbf{(2b)} by $+0.036 \pm 0.009$ AUC in the misspecified scenario, whose predicted bias magnitude $|c_{k}\,\alpha^{\top} w_{T,k}|$ is the smallest of the three orientations ($0.39$ in outcome units on average against $0.62$ and $0.55$), so that
the doubly robust family incurs its variance inflation to remove the
smallest bias. The
doubly robust arms reduce the estimation bias: the realized bias of
\textbf{(1b)} in each environment and subpopulation follows the
covariate-shift prediction of Proposition~\ref{prop:bias} given the
realized arm counts (Supplementary Figure~S6), and the aligned
scenarios have biases of one sign on average ($\alpha^{\top} w_{T,k} > 0$
in $240$ of the $300$ subpopulations) while the randomly oriented ones
scatter around zero, which is the mechanism of
Corollary~\ref{cor:consistency} in the data.

\emph{Comparators and selector.} Once the propensities are nontrivial
the one-step recombination and the targeting step separate: AIPW exceeds
targeted maximum likelihood at the same weights by about $0.02$ AUC in
all three scenarios, and cross-fitting costs $0.01$ to $0.03$, where the halved nuisance sample and the per-fold targeting meet nontrivial propensities (Table~\ref{tab:contrasts}). The selector's step (iii) enters the
anticipated worst case band into the crossing condition with the
design stage confounded variance inflation; at $\bar n = 26$ the squared
band (the band is about $0.9$ in outcome units under the linear-index
mechanisms) exceeds the net variance term
$s(\kappa\sigma^{2}_{k,\mathrm{cond}} - E_{k})$ (about $0.2$, so that a
band above about $0.45$ decides) in all but one of the $100$ environments of the aligned and random scenarios, so both rules route $99$ to the doubly robust family, whereas in the misspecified scenario, where the band is the worst case  over the alignment of the exact between-arm shift ($0.1$--$1.4$), the median rule routes $62$ environments and the upper-quartile rule $56$ (Table~\ref{tab:selector}). The selector arm is therefore the doubly robust arm in all but one environment of the aligned and random scenarios, within error of \textbf{(1b)} in AUC ($-0.026 \pm 0.014$ and $-0.006 \pm 0.012$) and, under aligned confounding, at matched false positive rate ($-1.8 \pm 2.5$ points), while under random orientation it is ahead there by $9.4 \pm 3.2$ points; it is a mixture in the misspecified scenario, $-0.020 \pm 0.009$ AUC below \textbf{(1b)}, its one significant miss at this sample size: the anticipated band is the worst case  over the alignment and exceeds the realized shift of the quadratic mechanism (about $0.6$ against $0.4$ on average over the environments), so the rule
commits to the family whose variance inflation the smaller bias does not repay. With the recommended pairing, the sample mean functional recruiting and the routed family analyzing, the selector is within error of the better family in every core scenario (Supplementary Table~S6).

\subsection{The Nondegenerate Scenarios: \texorpdfstring{$K = 8$}{K = 8}}
\label{sec:K8}
With $K = 8 > m = 7$, borrowing is active for every hard-constrained arm
with a single free dimension. The burn-in is $16$ patients, the number of
patients per subpopulation and arm averages $\bar n^{(a)} \approx 10$ at
the horizon (about $43$ per arm in the most recruited subpopulation and
one in the least), and the AUC is defined in all $50$ environments. At
these sample sizes the fallback rule applies to $35$--$48\%$ of the
subpopulation cells of \textbf{(2b)} at the end of the trial, against $16$--$23\%$ at $K = 3$ and none under uniform recruitment, so the family
contrasts compare the sample mean with a mixture of targeted and plain
arm means (Supplementary Section~S5.3).

\emph{Family contrast at small sample sizes.} The sample mean family is
ahead in seven of the eight paired contrasts and significantly so in six,
by $0.013$ to $0.020$ AUC at the individual-level
(Table~\ref{tab:contrasts}); at matched false positive rate the point estimates favor the doubly robust arm by $3.9 \pm 2.0$ points of true positive rate in the randomized scenario ($1.96$ standard errors) and by $2.7 \pm 1.8$ under aligned confounding, both within two standard errors. This is the small-sample regime that the
crossing condition assigns to the sample mean family, but
Section~\ref{sec:mismatch-results} shows that the gap is carried by the
recruitment path: on identical patients the two estimators discriminate
equally well at these sample sizes. AIPW exceeds targeted maximum likelihood by about a hundredth of AUC under confounding at fixed weights, so a family built on the one-step recombination would narrow every contrast by about that much through the estimator alone, as the $\kappa$-indexed reading of the boundary predicts (run as a recruiting arm with its own curve it moves the $K = 8$ contrasts by $0.01$--$0.02$, Supplementary Section~S5.9), and
cross-fitting costs about as much under aligned and random confounding,
almost nothing under randomization and the misspecified mechanism, and
$0.040 \pm 0.006$ under uniform recruitment, where every subpopulation
is split (Table~\ref{tab:contrasts}; Supplementary
Section~S5.3).

\emph{Selector.} At the anticipated $\bar n = 10.4$ the median rule
routes $21$ of the $50$ randomized environments to the doubly robust
family and ends $0.004 \pm 0.002$ AUC below the sample mean arm; the
upper-quartile rule keeps the sample mean family everywhere and has zero
regret. In the confounded scenarios the squared band must exceed a net
variance term of about $0.57$ at this count, that is, the band must exceed
about $0.75$ in outcome units: the median rule routes $40$ of $50$
environments to the doubly robust family under aligned and random
confounding (the upper-quartile rule $30$ and $27$) and $4$ under the
misspecified mechanism, whose exact shift is smaller. The selector arm is
therefore a mixture that pays the doubly robust family's deficit at ten
patients per arm, $0.014 \pm 0.005$ and $0.017 \pm 0.007$ AUC against
\textbf{(1b)} under aligned and random confounding
(Table~\ref{tab:selector}). Across the eight core scenarios the median
rule is within two paired standard errors of the better family in five
and beyond it in three, the misspecified $K = 3$ scenario, where the bias side is within error of zero ($+0.007 \pm 0.007$) and the miss is the variance side, and the aligned
and random $K = 8$ scenarios, where the band selects the family that removes the bias and the AUC contrast nets that family's variance side, its inflation on identical patients and its recruitment path, against the sample mean's uneven bias (Supplementary Section~S5.4); the upper-quartile rule is beyond two
standard errors in the misspecified $K = 3$ and the aligned $K = 8$
scenario.

\subsection{Probes Along the Coordinates of the Regime Map}
\label{sec:probes}
Seven probes complete the design, four of which move one coordinate of
the map at a time; Supplementary
Table~S5 sets each against the direction pre-specified
for it before the runs reported here, and Supplementary
Section~S5.4 gives the full account. With $H = 400$ the
$K = 8$ scenarios reach $\bar n^{(a)} \approx 26$ with borrowing active:
under randomization the family gap narrows to $+0.011 \pm 0.006$ AUC for
\textbf{(1b)} against \textbf{(2b)}, from $+0.031 \pm 0.007$ at $H = 150$
on the same ten environments, while the path remains concentrated (about
$117$ patients per arm in the most recruited subpopulation and two in the
least), leaving a gap of the size the $K = 3$ scenario shows at the same
mean over the $39$ of its $50$ environments with a defined AUC
($+0.010 \pm 0.008$);
under aligned confounding the AUC contrast is within error ($+0.007 \pm 0.013$, an interval of about $\pm 0.026$)
while the false positive rate of the sample mean arm is $14.1 \pm 3.6$
points higher and its true positive rate at matched false positive rate $8.5 \pm 4.2$ points lower ($2.0$ standard errors; the pre-specified prediction of no crossing at this sample size fails in this metric, Supplementary Table~S5). Uniform recruitment at $K = 8$
reverses the family gap at the same $\bar n^{(a)} \approx 10$
($-0.009 \pm 0.005$ AUC, $1.8$ standard errors, with the point estimate favoring the doubly robust arm, against $+0.017 \pm 0.004$ under adaptive recruitment and
$+0.031 \pm 0.007$ on the same ten environments) and raises
the discrimination of every arm: with every subpopulation just beyond
the median-rule threshold the doubly robust family is favored within error, as the crossing condition implies, and the adaptive gap in the other direction is the recruitment-path effect that uniform recruitment switches off; in mean squared error at the design level the doubly robust family is also the better one there ($0.36$ against $0.43$, a paired difference of $-0.065 \pm 0.007$ over the ten environments), the one configuration without adaptive recruitment, whereas at $H = 400$ under adaptive recruitment it stays behind ($+0.68 \pm 0.15$ under randomization and $+1.07 \pm 0.18$ under aligned confounding; Supplementary Section~S5.4 gives the values of the other probes). Doubling $\sigma_{x}$ quadruples the edge and the
design-embedded variance inflation with it, so the gap widens only
slightly
($+0.038 \pm 0.015$ from $+0.031 \pm 0.007$ on the same ten
environments); running variance estimates in place of the oracle inputs
cost both families discrimination and leave the gap where it was ($+0.029 \pm 0.008$ against $+0.031 \pm 0.007$ under randomization, $+0.006 \pm 0.012$ against $+0.015 \pm 0.014$ under aligned confounding), the predicted widening not confirmed (two further pre-specified predictions are contradicted by the data, the ``no crossing'' of the aligned $H = 400$ probe in the co-primary metric, above, and the threshold $n^{*}(0) > 26$ of the $\kappa$ row of Supplementary Table~S5; the table also scores the mismatch prediction as contradicted for the doubly robust analysis, Section~\ref{sec:mismatch-results}); and in the anchor
configuration every arm declares the true positive in every replication
and the zero-effect subpopulation in about half of them.

\subsection{The Regime Map and the variance inflation Curves}
\label{sec:map}
\begin{figure}[!htbp]
\centering
\includegraphics[width=\linewidth]{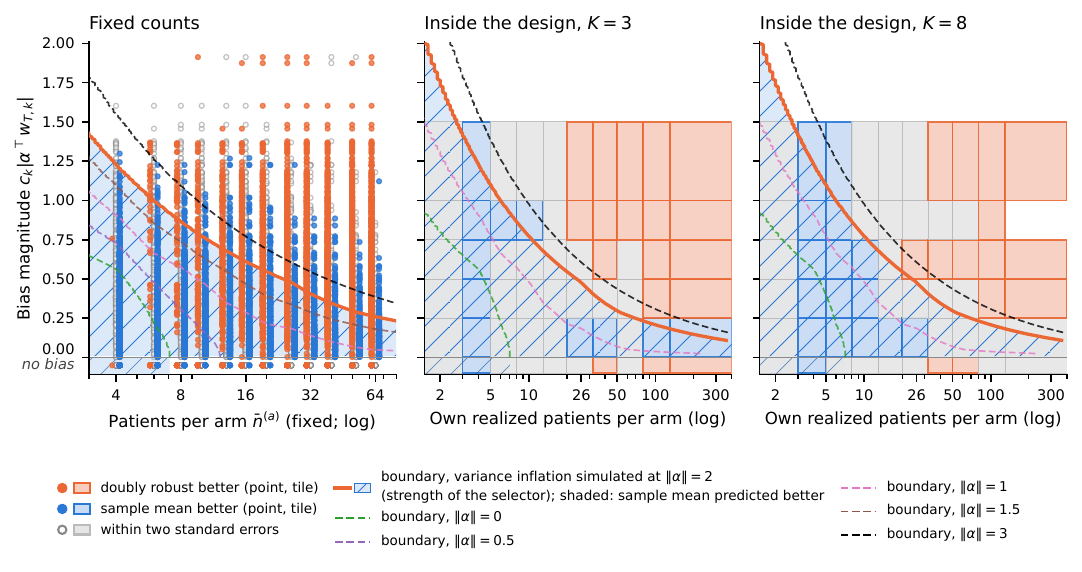}
\caption{The regime map in the coordinates of
Theorem~\ref{thm:surface}: patients per subpopulation and arm (log
scale) against the bias magnitude $c_{k}|\alpha^{\top} w_{T,k}|$ of the
sample mean contrast. Left: fixed counts, one point per evaluation
environment (ten per $K$), subpopulation, per-arm count, confounding
direction, and strength;
filled orange where the doubly robust family has the smaller realized
mean squared error by more than two standard errors, filled blue where
the sample mean has, hollow otherwise; the six strengths share the same
counts, so the doubly robust points are offset to the left of the count
and the sample mean points to the right, and the zero-bias points are
drawn in the band below zero (labeled no bias and separated by a
horizontal line), as the zero-bias tiles are. Lines: the
boundary $n^{*}(u)$
of~\eqref{eq:deltaR} at the balanced $s = 2/\bar n$ and the environment
means, with the variance inflation curve estimated at each strength on
the ten held-out $K = 3$ environments (each point's verdict is compared
with the prediction at its own constants and realized $s$,
Section~\ref{sec:map}); the curve of the anticipated
strength $2$, a replicate on the held-out environments of the
design stage curve that step (iii) of the selector uses, is drawn thick,
with the side on which the sample mean is predicted better tinted and
hatched (the zero-bias row by the strength-$0$ curve, the fixed-count
replicate of the design stage randomized curve), and the other strengths
dashed; a curve is drawn where a crossing exists below $400$ patients per
arm, and where none does the tinted side extends to the right edge.
Middle and right: the same coordinates inside the adaptive design at
$K = 3$ and $K = 8$, each family placed at the per-arm count its own
recruitment gave the subpopulation and the plane tiled; a tile shows the
sign of the difference of the two families' mean squared errors, filled
when it exceeds two cluster-robust standard errors (clusters: environment
and subpopulation, pooled over the panel's configurations), grey otherwise, and
blank with fewer than $30$ observations (no decided tile differs from the
prediction); the randomized configurations form the band below zero bias.}
\label{fig:map}
\end{figure}

Figure~\ref{fig:map} draws the equal-error boundary of
Theorem~\ref{thm:surface} and places realized mean squared error
differences on it. The left panel is the setting of
Proposition~\ref{prop:crossing}: for each of the first ten evaluation
environments per $K$ and each subpopulation, for $\bar n^{(a)}$ from $4$
to $64$ patients per arm, and
for confounding strengths $\|\alpha\|\sigma_{x}$ from $0$ to $3$ along the
environment's mean outcome-weight direction and along a random direction,
the two contrasts at $\beta = e_{k}$ are computed at fixed counts in up to
$200$ replications (replications in which an arm holds fewer than two
patients are discarded, and the $70$ of the $14{,}520$ points with fewer
than ten valid replications are dropped), and the sign of
$\mathrm{MSE}_{2} - \mathrm{MSE}_{1}$ is recorded where it is more than two
standard errors from zero. The prediction evaluates~\eqref{eq:deltaR} at
the subpopulation's own $E_{k}$ and $\sigma^{2}_{k,\mathrm{cond}}$ and at
its mean realized $s_{k}$, against which the variance inflation curve is
estimated at each strength on the ten held-out $K = 3$ environments, so
it uses nothing from the evaluation environments beyond their design
constants and realized counts; the drawn lines are the roots
of~\eqref{eq:deltaR} at the balanced $s = 2/\bar n$ and the means of $E_{k}$ and $\sigma^{2}_{k,\mathrm{cond}}$ over all records. Of the $14{,}450$ points, $71\%$ are decided; the strength-specific
prediction has the realized sign in $99.6\%$ of them ($99.3$--$99.8\%$ at each strength; $97.0\%$ at the balanced $s = 2/\bar n$ with each point's own constants, and $94.6\%$ for the drawn lines themselves, which use those means), and the
$42$ disagreements lie close to the boundary (a median factor of $1.3$ in the sample size) and are marginal decisions (median $2.2$ standard errors). Of the decided points $48\%$ favor the doubly robust family, so the agreement is not carried by an unbalanced outcome, and the two cruder predictors scored on the adaptive panels below agree in $78\%$ (a constant $\kappa = 2$) and $75\%$ (the sample mean when the count is below $13$ or the bias below $0.25$) of the same points, failing mostly at the weak strengths: the constant inflation agrees in $15\%$ of the decided points at strength $0$ and $41\%$ at $0.5$.
The bias mechanism is confirmed point by point: the realized bias of the
sample mean is $1.00$ times the predicted $c_{k}\alpha^{\top}w_{T,k}$ and
that of the doubly robust estimator $0.00$ times it. The inflation is
strength-specific (median $\kappa$ at ten patients per arm $0.2$, $0.4$,
$1.1$, $1.7$, $2.4$, and $3.3$ at strengths $0$ to $3$), so the single
curve estimated at the anticipated strength $2$, a replicate on the
held-out environments of the design stage curve the selector uses,
predicts the sign in $78\%$ of the decided points overall and in
only $15\%$, $41\%$, and $77\%$ at strengths $0$, $0.5$, and $1$, where it
overstates the variance inflation of the doubly robust family
(Supplementary
Section~S5.7 records a departure from
Assumption~4 at strength $3$). The other two panels are the
same map inside the adaptive design, for trials with $K = 3$ and $K = 8$,
horizons $H \in \{50, 150, 300, 600\}$, and strengths $0$, $1$, $2$, and
$3$ (ten environments and twenty replications each; the doubly robust
family recruits with the design-embedded curve of the aligned scenario at
every positive strength and with the randomized one at strength $0$). At
the design
level the doubly robust family has the larger mean squared error in all $32$ grid configurations and in all eight core scenarios ($4.9$ to $11$ standard errors over environments; six aligned $K = 3$ grid configurations at $H = 300$ and $600$ are within two standard errors), because its targeted fits on the six to nine patients of starved subpopulations, just above the fallback threshold, produce occasional large errors (raising the threshold to ten or fourteen patients removes most of the deficit: on the first $50$ environments of the aligned $K = 3$ scenario the gap falls from $+1.78 \pm 0.35$ to $+0.12 \pm 0.10$, and in the random $K = 8$ scenario from $+1.37 \pm 0.16$ to $+0.16 \pm 0.04$ and $+0.05 \pm 0.04$, with the false positive rate advantage of the doubly robust family at $K = 3$ kept; Supplementary Section~S5.9): observations with an arm of at most
three patients carry about nine tenths of its squared error, and the
targeted cells among them carry about the whole gap ($91$--$122\%$),
whereas at own
counts of $13$ or more per arm it has the smaller error in every core
scenario and in $28$ of the $32$ grid configurations. Placed at the per-arm count its
own path gave the subpopulation (the two paths differ by a median factor
of $1.8$), each family's error can be set against the prediction at that
count, a comparison that conditions on outcome-dependent counts (the
remark after Assumption~3): the prediction has the sign of
all $60$ decided tiles of $119$, $30$ at each $K$,
with the doubly robust family the better one beyond the boundary and the
sample mean below it, although cruder predictors agree nearly as well and
the $K = 8$ panel, where the families borrow with different weights, is
an empirical extension of the theorem. Starvation is what the adaptive design adds (under uniform recruitment at $K = 8$ the doubly robust family has the smaller design-level error, $-0.065 \pm 0.007$, Section~\ref{sec:probes}), and it is symmetric in frequency but not in cost: on the
randomized band at $8$ to $80$ patients per arm the doubly robust path
gives a subpopulation less than half the patients of the sample mean
path about as often as the reverse, but where it does its squared error
exceeds the sample mean's by $10.6$ and $4.6$ at $K = 3$ and $K = 8$,
against $2.3$ and $2.5$ the other way; the deficit is the
estimator's fragility at these counts, exposed by the path
(Supplementary Section~S5.7). On the randomized axis
the picture in discrimination is thus the one the crossing condition
draws once the recruitment path is accounted for: the sample mean family
is significantly ahead at $\bar n^{(a)} \approx 10$ under adaptive
recruitment, the doubly robust family is favored within error at the same sample size under uniform recruitment, and the two families are within error at
$\bar n^{(a)} \approx 26$ at $K = 3$ at the individual-level, which does
not separate the median-rule threshold of about $9$ from the
upper-quartile one of about $29$. On the confounded side the
boundary depends on the variance inflation of the doubly robust
estimator under confounding, $2.1$--$2.3$ at the sampled sizes on the
design stage curve:
the root of~\eqref{eq:deltaR} at the balanced $s = 2/\bar n$, $n^{*}(u) \approx
2\,[\kappa\,\sigma^{2}_{k,\mathrm{cond}} - \bar E_{k}^{c}]/(c_{k}u)^{2}$,
lies at about $15$ ($K = 3$) to $18$ ($K = 8$) patients per arm for the
aligned scenarios (bias magnitude $|c_{k} u| \approx 0.62$ and $0.56$) and
about $19$ for the randomly oriented ones ($\approx 0.55$), so
the aligned $K = 3$ and $H = 400$ scenarios, at $\bar n^{(a)} \approx 26$,
lie beyond the boundary, where the doubly robust family removes the bias
and the AUC contrast is within error, and the $K = 8$ scenarios at
$H = 150$ on the sample mean side, in agreement with
Table~\ref{tab:contrasts}; under the misspecified mechanism the
inflation is lower ($\kappa_{k,\infty} \approx 0.1$--$0.2$), the variance terms alone
favor the doubly robust family in about half of the subpopulations, and
the boundary lies at the smallest sample sizes, although in
discrimination the sample mean is ahead there by $0.036 \pm 0.009$ at $K = 3$ (Table~\ref{tab:contrasts}), the one significant confounded $K = 3$ family contrast. The operating characteristics reflect the crossing more slowly than the mean squared error, because the AUC is invariant to the part of the bias that is common to an environment's subpopulations and the matched-FPR comparison to a common shift, and the pooled zero-threshold rates of the random orientation scenarios show no directional shift, the first-order statement of Corollary~\ref{cor:consistency} (the sample mean's false positive rate is $+0.4 \pm 1.1$ points from the doubly robust arm's at $K = 8$ and $+3.2 \pm 1.9$ at $K = 3$, against $+5.7 \pm 1.1$ and $+13.1 \pm 2.0$ under aligned confounding); these pooled contrasts are nets as well, since the spread of a sign-random bias raises the sample mean's false positive rate and lowers its true positive rate at second order, by $2.9$ and $6.1$ points of false positive rate at $K = 8$ and $K = 3$ when the predicted bias is removed from its estimates, which the doubly robust family's inflation offsets. At $K = 3$, beyond the boundary, the doubly robust arm is ahead at matched false positive rate by $9.4 \pm 3.2$ points with the AUC contrast within error, a net of the variance side ($+0.041 \pm 0.007$ with the predicted bias removed from the sample mean estimates) and the uneven bias, which costs the sample mean arm $0.034 \pm 0.010$ AUC (Supplementary Section~S5.4), whereas the aligned scenarios show a directional deficit for the sample mean family in the false positive rate.

\label{sec:crossing-selector}
\emph{Two variance inflation curves.} The inflation $\kappa_{k}(\bar n)$
was estimated by simulation in two ways (Supplementary
Figure~S13 and Supplementary Section~S5.8):
at fixed sample sizes before the trial (design stage), where the
inflation of the targeted estimator under randomization falls from
$0.77$ at $\bar n^{(a)} = 4$ to zero at $64$, its inflation term crossing
the edge at $n^{*}(0) \approx 7$, while under aligned confounding
it stays at $2.1$--$2.5$ up to $\bar n^{(a)} = 26$ and a variance-only
threshold is infinite; and inside the adaptive trial on ten held-out
environments (design-embedded), as the variance ratio of the two
analyses of identical patients, which is about $1.03$ below $9$ patients
per arm, $0.98$ up to $23$, and $0.75$--$0.86$ beyond $27$, and the median
and upper-quartile curves give $n^{*}(0) \approx 9$ and $29$. The embedded curves measure the fallback
rule as much as the nuisance fits; the family-2 recruitment and step (ii)
use them, and step (iii) takes the design stage confounded curve.

\emph{Crossing condition and selector.} The sampled sizes, about $10$
and $26$ patients per arm, lie between the median-rule threshold of about
$9$ and the upper-quartile one of about $29$, and on a fixed recruitment
path the two families are within error of each other in discrimination
there (Sections~\ref{sec:probes} and~\ref{sec:mismatch-results}).
Table~\ref{tab:selector} gives the selector's record on the core
scenarios with the curves the selector uses (the design-embedded curve
under randomization, the design stage confounded curve otherwise); among
the probes, the median rule
routes the randomized $H = 400$ scenario to the doubly robust family in
eight of ten environments (within $0.008 \pm 0.006$ of the better family)
and the uniform-recruitment probe to the sample mean family in seven,
its one miss there ($-0.010 \pm 0.004$, $2.5$ standard errors), where the
sample size sits on the threshold. The record is in discrimination; in
mean squared error the doubly robust family is behind in every adaptive
configuration at the sampled counts, the anchor configuration apart
(Supplementary Section~S5.7), so the matched selector arm pays in mean squared error at both sampled sizes, and in discrimination beyond two standard errors in the aligned and random $K = 8$ scenarios and the misspecified $K = 3$ one (Table~\ref{tab:selector}, over all environments;
over the routed environments alone, Supplementary
Section~S5.8; median rule; under
the upper-quartile rule the aligned $K = 8$ and misspecified $K = 3$
deficits remain beyond two standard errors); the arm with the recommended pairing pays no discrimination cost and about a tenth of the error cost (Supplementary Table~S6). The confounded step is
decided by the squared anticipated band against the net variance term of
a variance inflation of order one (Supplementary
Section~S5.8).

\subsection{Recruitment--Analysis Consistency}
\label{sec:mismatch-results}
The two mismatched pairs at $K = 8$ under randomization, run on the seeds
of Section~\ref{sec:K8} with the same variance inflation as the matched
doubly robust arm, give an asymmetric answer (Supplementary
Section~S5.5). Recruiting with the doubly robust family
and analyzing with the sample mean is $-0.012 \pm 0.004$ AUC below the
matched sample mean arm, the direction that Proposition~\ref{prop:mismatch}
anticipates for a penalty factor of $1.28$ on the ten mismatch
environments ($1.26$ for the reverse pair; Supplementary Figure~S9). Recruiting with the sample mean family and
analyzing with the doubly robust family, however, beats the matched
doubly robust arm by $+0.034 \pm 0.009$ AUC, and its discrimination,
$0.837$, equals that of the matched sample mean arm on the identical
patients ($0.834$ on those ten environments; $+0.003 \pm 0.005$), a
pattern the same-path runs on the held-out environments repeat in all
five configurations and the mismatch arm run on every core environment repeats in every core scenario ($+0.019 \pm 0.004$ on all $50$ randomized $K = 8$ environments; $+0.011$ to $+0.036$ over the eight scenarios, Supplementary Section~S5.5).
The family gap of the adaptive scenarios in discrimination is therefore a
recruitment-path effect: on identical patients the doubly robust
estimator discriminates within error of the sample mean in every core scenario, and the deficit of the doubly
robust family is
what its own recruitment does with the noisier early estimates. Proposition~\ref{prop:mismatch} describes matched recruitment under minimax allocation, from which the realized doubly robust path deviates at the sampled sizes: the principle is supported for the sample mean analysis, whose mismatched pair loses in the predicted direction, and not for the doubly robust analysis in these runs, whose matched path is the worse one; inflating the recruitment variance by the design-embedded factor does not repair this, because that inflation is nearly constant over the sample sizes at which the allocation is decided (the steeper design stage curves of the AIPW and regression-adjusted recruiting arms do move the allocation, Supplementary Section~S5.9).

% ===========================================================================
\section{DISCUSSION}
\label{sec:discussion}
% ===========================================================================
\subsection{Practical Recommendations}
\label{sec:recommendations}
The regime map translates into pre-specification guidance in six
steps. First, check degeneracy: compute $m = 1 + \Dx + \Tpre$ and report
whether $K \le m$, with the noiseless count $1 + \Dx + \min(\Tpre, \Dz)$
for soft-constrained weighting; if $K \le m$ (with the centroids in
general position, which the solver confirms by the realized rank),
borrowing is inert and the
comparison reduces to the estimators of the individual subpopulations.
Second, under design-controlled assignment with adaptive recruitment
across many subpopulations, so that most subpopulations are held at a few
patients per arm, pre-specify the sample mean family: in this regime the
recruitment path of the doubly robust family costs it more discrimination
than its explained-variance edge returns. With uniform recruitment, or with few subpopulations so that every subpopulation passes the threshold, pre-specify the doubly robust family, which under uniform recruitment at $K = 8$ has the smaller design-level mean squared error ($-0.065 \pm 0.007$, Section~\ref{sec:probes}) and whose same-path variance ratio is
one up to about $9$ patients per subpopulation and arm, a few percent
below one up to $23$, and $0.75$--$0.86$ beyond $27$,
provided adaptive recruitment does not leave starved subpopulations
(about five patients per arm in the least recruited at $K = 3$), where its error, with the six-patient fallback rule, is larger in every simulated adaptive configuration but the anchor configuration (within error of the sample mean's in the aligned $K = 3$ scenario once the threshold is raised to ten or fourteen, Supplementary Section~S5.9); the adaptive
$K = 8$ trial at about $26$ leaned to the sample mean through its path,
within error. Third, when the assignment is confounded or randomization cannot be guaranteed, prefer the doubly robust family if the aim is sign identification with a controlled false positive rate: its bias removal lowers the false positive rate by six to thirteen points under aligned confounding, and the point estimates favor it at matched false positive rate at about ten and at about twenty-six patients per subpopulation and arm (within two standard errors in the aligned core scenarios; $3.0$ standard errors in its favor under random orientation at $K = 3$, and $+0.6 \pm 1.8$ for the sample mean in the misspecified $K = 3$ scenario); if the aim is estimation accuracy at the trial level, note that in mean squared error the confounded step chose the family with the larger error in the five confounded core scenarios in which it routes most environments to the doubly robust family (in the misspecified $K = 8$ scenario it routes $4$ of $50$ and its excess is $+0.04 \pm 0.03$), through that family's starved subpopulations, a deficit that the fallback threshold controls (Supplementary Section~S5.9) and of which about nine tenths disappears when the selector recruits with the sample mean functional and routes the analysis only; $+0.08 \pm 0.02$ and $+0.09 \pm 0.03$ remain at $K = 8$ under aligned and random confounding, the doubly robust analysis's own error on the starved subpopulations of the sample mean path (Supplementary Table~S6). The confounded step of the selector
enters the anticipated worst case band $|c_{k}|\,\|\alpha\|\,\|w_{T,k}\|$ into
the crossing condition against the design stage confounded variance
inflation, about $2$ at these sample sizes; at the simulated strength the band decides for
the linear-index mechanisms, while the smaller shifts of the misspecified
mechanism leave the sample mean family in place in $38$ of $100$ environments at $K = 3$ and $46$ of $50$ at $K = 8$, with the band as a reported caveat. The cost of this rule in the matched protocol is visible in threshold-free discrimination, up to $0.02$ AUC at ten patients per arm and
$0.020 \pm 0.009$ under the misspecified mechanism at $26$ for the selector arm over all environments ($0.032 \pm 0.013$ in the environments it routes), where
the anticipated band overstates the shift (Table~\ref{tab:selector}), and
in
mean squared error at both sampled sizes, since the doubly robust family
is behind in every adaptive configuration but the anchor configuration; with the recommended pairing of the fourth recommendation the discrimination cost disappears ($-0.005$ to $+0.012$ AUC against the better family) while the bias removal is kept, so a protocol need not trade the ranking of subpopulations against the removal of a common bias, and one that must recruit and analyze with a single family and weighs the ranking above the bias removal may pre-specify the
sample mean family there and report the band. Fourth, pre-specify the recruitment variance functional together with the final estimator; in the $K = 8$ runs the sample mean functional is the one to pre-specify for either analysis, since its path serves both analyses whereas the doubly robust path is worse for both, and the doubly robust analysis gains from the sample mean path in every core scenario (beyond two standard errors in seven of the eight), so the penalty of Proposition~\ref{prop:mismatch} is realized only when the sample mean analysis follows the doubly robust path, and the runs do not identify sample sizes at which the matched doubly robust pair would win. The selector of Section~\ref{sec:selector} was evaluated as a matched protocol, recruiting along the doubly robust path where it commits; run with this recommendation, routing the analysis only, it is within error of the better family in discrimination in every core scenario and sheds about nine tenths of the matched arm's mean squared error excess (Supplementary Table~S6). Fifth, since pooled operating characteristics hide a directionally inconsistent bias at first order (Corollary~\ref{cor:consistency}) and the AUC and the matched-FPR comparison hide the part of a bias that is common to the subpopulations of an environment, simulation-based validation should report per-environment error measures and zero-threshold error rates, which reveal a common bias (pooled over environments, only when its sign is consistent), together with the pooled TPR, FPR, and AUC and a comparison at matched false positive
rate. Sixth, capture patient-level records wherever governance permits: the doubly robust family and the analysis-stage audit require them, and aggregates can be computed from records but never the reverse (Supplementary Section~S1).

\subsection{Limitations}
\label{sec:limitations}
The evidence beyond the propositions is simulation-based, under a linear
factor model with Gaussian noise, homogeneous covariate dispersion within
subpopulations, differences between subpopulations in location only, and a
single post-treatment measurement. The inflation $\kappa_{k}(\bar n)$ is
estimated within this process and carries over to other nuisance
estimators only qualitatively: the location of the equal-error boundary
moves with $\kappa_{k}$, while its existence, single-crossing structure,
and comparative statics do not; its two simulation-based estimates agree
on the threshold under randomization ($7$ against $9$ patients per subpopulation and arm) and differ in level. With more covariates the inflation grows (by the small-sample form after Proposition~\ref{prop:crossing}, from about $0.13$ at ten patients per arm with two covariates to about $0.55$ with six), effect heterogeneity within a subpopulation shrinks the edge, and flexible nuisance learners raise the inflation, so the second and third recommendations are scoped to low-dimensional, correctly specified parametric nuisance models and homogeneous effects within subpopulation. The inflation is also specific to the confounding strength, so step (iii) depends on the inputs anticipated for it: the band halved or doubled at the true strength curve moves the routing from none to nearly all ($144$ of $150$) of the $K = 8$ confounded environments, whereas a consistent error in the anticipated strength, which changes the band and the curve together, moves it less under the linear-index mechanisms ($41$ and $33$ of $50$ under aligned confounding and $41$ and $29$ under random orientation at anticipated strengths $1$ and $3$, against $40$ at the true strength $2$) but not under the quadratic one, where an under-anticipated strength routes $29$ instead of $4$ of $50$ environments at $K = 8$ and $92$ instead of $57$ of $100$ at $K = 3$ (Supplementary Section~S5.8). Every family-2 result is for one implementation, the linear-fluctuation targeted estimator with truncation to $[0.05, 0.95]$ and the six-patient fallback; the AIPW comparator at the same weights has the lower inflation and the higher AUC, and its curve would place the confounded threshold at smaller bands; run as recruiting arms with their own design stage curves, AIPW and the regression-adjusted estimator close most of the targeted arm's discrimination gap at $K = 8$, under randomization mainly, for the AIPW arm, because its steeper curve changes the recruitment (the regression-adjusted arm was not decomposed), and remain behind in mean squared error through the same starved cells (Supplementary Section~S5.9). The mean-squared-error--optimal variants
\textbf{(1c, 2c)} and the oracle transport relaxation require the true
outcome weights and latent loadings and are benchmarks. Proposition~\ref{prop:mismatch}
stylizes sensitivity-driven recruitment as minimax variance allocation,
from which the realized paths deviate, most visibly for the doubly robust
family, whose own path is worse for its analysis than the sample mean
family's. The crossing condition compares the estimators on a fixed path
and says nothing about the path itself, which dominates the family
contrast in the adaptive $K = 8$ scenarios; its comparative statics are consistent with the probes
within simulation error, except that estimated variances left the margin
unchanged rather than widening it, and with ten environments per probe
those margins range from one half to about four standard errors, and three of the pre-specified predictions are contradicted and one is not confirmed (Supplementary Table~S5). The
fallback threshold
of six patients, below which the doubly robust arm means revert to sample
means, is the one design lever the mean squared error deficit points to,
and it was varied in a sensitivity run (Supplementary Section~S5.9): a threshold of ten or fourteen patients gives up the targeted fits in the cells with six to nine or six to thirteen patients and with them most of the deficit, which falls within error on the first $50$ environments of the aligned $K = 3$ scenario and by about nine tenths in the random $K = 8$ scenario ($+0.16 \pm 0.04$ remains at ten, $4.0$ standard errors), while the false positive rate advantage of the doubly robust family at $K = 3$ is kept (these runs recruit with the embedded curve estimated under the six-patient rule). The statement that the doubly robust family is behind in mean squared error in every adaptive configuration is therefore largely a statement about the six-patient rule, and the fallback mixture is large: the rule applies to $35$--$48\%$ of the subpopulation cells of that family at $K = 8$, and the fitted propensity is truncated for $20$--$25\%$ of the patients under the linear-index mechanisms (the true propensity lies outside $[0.05, 0.95]$ for $20$--$48\%$; Supplementary Section~S5.8). ``Correctly specified'' propensity means the parametric logistic mechanism of Section~\ref{sec:arm}; misspecification is examined only within a quadratic-index alternative, and the outcome model is correctly specified in every core scenario, so bias removal by the doubly robust family is shown mainly where its parametric outcome model is exactly right; in one probe with a misspecified outcome model the false positive rate advantage persists because the linear part of the bias is removed by the linear adjustment whatever the propensity, and with the truncation binding that probe does not test double robustness (Supplementary Section~S5.9). The main arms recruit with oracle variance inputs and the selector is evaluated under exact anticipation; the estimated-variance probe on ten environments is the only evidence on the practical version (Supplementary Section~S5.4). The fixed-count statements are applied inside an adaptive design whose counts depend on earlier outcomes, a distortion that the remark after Assumption~3 describes and does not bound. Finally, the confounded mechanism is a stylization: a per-patient propensity acts on both arms of every subpopulation, including the target's own treated arm, whereas in hybrid-control settings the trial's treated arm is randomized and only the external controls are selected, a different mechanism whose bias constant we have not derived.

\subsection{Future Work and Conclusion}
\label{sec:conclusion}
Three directions follow: the extension of the transport relaxation to
subpopulations that differ in distributional shape, where the penalty is
computed between empirical distributions, exactly by linear programming
or by an entropic approximation \citep{cuturi2013,peyre2019}; a
semi-synthetic illustration fitted to a
completed enrichment trial; and a transportability formalization of the
directional consistency result \citep{pearl2014,bareinboim2016}. In
adaptive subpopulation identifying trials the choice of estimator is a
design decision and not a default: the crossing condition trades a
squared bias that only the doubly robust family removes against a
finite-sample variance inflation that only the sample mean family avoids,
the resulting equal-error boundary is defined by the theory and located
by the simulations, and the pre-specified selector collects
these results into a committed and auditable choice suitable for a
statistical analysis plan.

% ===========================================================================
% Declarations (the journal's checklist: funding, disclosure of interest,
% generative AI, data availability). None of them identifies the authors, so
% the anonymous version (\anon = 0) carries the same statements; the
% corresponding author's details are on the title page of the version with
% author details only. There is no acknowledgments section.
% ===========================================================================
\section*{Funding}
The authors declare that no funding was obtained for the reported work.

\section*{Disclosure of Interest}
The authors report no conflict of interest.

\section*{Declaration of Generative AI Use}
Generative AI tools were used for copyediting and language refinement of the
manuscript and for software coding assistance. The
authors take full responsibility for the content of the article.

\section*{Data Availability Statement}
This article reports simulation results only, and no empirical data were used.
The simulation code, the scripts that produce every table and figure, the
figures, and a README that describes their organization and the run order
are provided as supplementary material (the folder \texttt{simulation
files}); the stored simulation outputs about 7~GB are not included, because
of their size and are regenerated by the scripts.

\bigskip
\begin{center}
{\large\bf SUPPLEMENTARY MATERIAL}
\end{center}
\begin{description}
\item[Supplementary document:] the data resolution (Section~S1), the
  selector as a decision procedure (Section~S2), the causal structure, run
  inventory, and implementation details of the simulation study
  (Section~S3), the optimal transport relaxation of the
  borrowing weights (Section~S4), the operating characteristics,
  trajectories, probes, variance inflation curves, and sensitivity runs
  behind the results (Section~S5), and the assumptions and proofs
  (Section~S6). (PDF, separate file)
\item[Simulation files:] the folder \texttt{simulation files}: \texttt{scr/}
  with every script (the simulation, the estimators and the selector, the
  variance inflation curves, the regime map, the summaries, tables, and
  figures, the checks, the shell drivers, and the test suite),
  \texttt{fig/} with the figures of the article and the supplement, and a
  README with the organization and the run order. The stored simulation
  outputs about 7~GB are not included; the scripts regenerate them. (zip
  file)
\end{description}

% The supplement cites these works in author-year form and has no reference
% list of its own; they are kept in this list.
\nocite{athey2016,nie2021,wager2018}
\bibliography{refs}   % the style, asa-sbr, is set in the preamble (the template's line)

\end{document}

\def\spacingset#1{\renewcommand{\baselinestretch}%
{#1}\small\normalsize} \spacingset{1}

\bigskip
\bigskip
\bigskip
\begin{center}
{\LARGE\bf Supplementary Material for ``Characterizing Estimator-Regime Transitions in Adaptive Enrichment Trials''}
\end{center}
\medskip

\bigskip
\spacingset{1.8} % DON'T change the spacing!

\appendix
\setcounter{section}{0}
\renewcommand{\thesection}{S\arabic{section}}
\setcounter{figure}{0}
\renewcommand{\thefigure}{S\arabic{figure}}
\setcounter{table}{0}
\renewcommand{\thetable}{S\arabic{table}}
\setcounter{equation}{0}
\renewcommand{\theequation}{S\arabic{equation}}
\renewcommand{\theHsection}{S\arabic{section}}
\renewcommand{\theHfigure}{S\arabic{figure}}
\renewcommand{\theHtable}{S\arabic{table}}
\renewcommand{\theHequation}{S\arabic{equation}}
\setcounter{page}{1}
\renewcommand{\thepage}{S\arabic{page}}
\medskip
\noindent
This supplement contains the discussion of the data resolution
(Section~\ref{sec:levels}), the selector flowchart
(Section~\ref{sec:supp-flow}), the implementation details of the simulation
study (Section~\ref{sec:supp-sim}), the optimal transport relaxation of the borrowing weights and its results (Section~\ref{sec:ot}), the operating characteristics,
trajectories, and full accounts of the scenarios that the main text
summarizes (Section~\ref{sec:supp-results}), and the assumptions and proofs
(Section~\ref{app:proofs}). Section and equation numbers without the prefix
S refer to the main text, whose reference list contains the works cited here.

\section{SUMMARY-LEVEL VERSUS INDIVIDUAL-LEVEL DEVELOPMENT}
\label{sec:levels}
The first step of the variant axis, from the summary-level variant (a) to
the individual-level variant (b), is a data-resolution step, and we
describe it separately. The base design is a summary-level method: every
quantity it uses is an aggregate over a subpopulation, namely the
covariate moments of the subpopulations in the constraint matrix $\Gamma$, the
running pre-treatment means $\Ybar_{k,\neg T}$, and the running arm means
$\Ybar_{k,a}$. Patients enter the analysis only through the
subpopulation-level statistics to which they contribute. Developing the
methodology for individual-level data instead, as variants
\textbf{(2b, 2c)} and the comparators of Section~3.2 do
(variants \textbf{(1b, 1c)} are evaluated on individual-level data but use
them only through subpopulation aggregates and oracle variance inputs),
and as the shape-aware form of the transport penalty must
(Section~\ref{sec:ot}), means that the estimates of the subpopulations,
the borrowing objective, and the recruitment variances become functionals
of the patient-level records, that is, of the covariate $x_{kj}$, the arm
$A_{kj}$, and the outcome trajectory $\{Y_{kjt}\}$. This choice is often
treated as an implementation detail. In this design it is a commitment
with consequences for identification, efficiency, borrowing,
auditability, operations, and data governance, and since
Section~4 quantifies several of these consequences, they
can be stated with some precision. We collect
them here with the summary-level base variant \textbf{(1a)} as the
reference point; Table~\ref{tab:levels} gives a summary.

\begin{table}[!htbp]
\centering
\caption{Consequences of the data resolution, with the summary-level base
variant \textbf{(1a)} as the reference point.}
\label{tab:levels}
\small
\resizebox{\ifdim\width>\linewidth\linewidth\else\width\fi}{!}{%
\begin{tabular}{@{}>{\raggedright\arraybackslash}p{0.24\linewidth}>{\raggedright\arraybackslash}p{0.34\linewidth}>{\raggedright\arraybackslash}p{0.34\linewidth}@{}}
\toprule
Dimension & summary-level (base) & individual-level \\
\midrule
Data consumed & subpopulation moments and running means $\Ybar_{k,\neg T}$,
$\Ybar_{k,a}$ & patient-level records $(x_{kj}, A_{kj}, \{Y_{kjt}\})$ \\
Confounding bias $b_{k}$ & retained, $\mathcal{O}(1)$ in $n$; pre-specified bias band
only (Proposition~3) & removable by nuisance fitting when the outcome
model is correct, or the propensity is correct and untruncated
(Proposition~3) \\
Efficiency & marginal variance $\sigma_{k,\mathrm{eff}}^{2}$
(Proposition~2) & conditional variance
$\sigma_{k,\mathrm{cond}}^{2}$; edge $E_{k}$ (Proposition~2) \\
Finite-sample variance inflation & none, $\kappa \equiv 0$ & $\kappa_{k}(\bar n)$ of~(10); estimated by simulation at $0.36$ inside the design at
$\bar n^{(a)} = 10$ and $0.28$ at $26$ (Section~\ref{sec:supp-inflation}) \\
Borrowing resolution & moment matching; centroid transport form~\eqref{eq:wasserstein} & shape-aware transport penalty, exact or entropic
(Section~\ref{sec:ot}) \\
Diagnostics and audit & rank check of Proposition~1 &
comparators and selector plug-in audit (Sections~3.2
and~4.6) \\
Availability and governance & aggregate statistics suffice; shareable
across sponsors and sources & row-level capture, cleaning, and sharing
agreements required \\
\bottomrule
\end{tabular}}
\par\vspace{4pt}
\begin{minipage}{\linewidth}\footnotesize
NOTE: Entries refer to the result
that states or quantifies them where one exists.
\end{minipage}
\end{table}

Three advantages keep the summary-level competitive. The first is
availability. Aggregate statistics are exactly what remains available when
patient-level records are not: historical and external control arms
reported as summaries, registry and multi-sponsor data under sharing
restrictions, and published trials. This is the data regime the
synthetic control literature was developed for
(Abadie and Gardeazabal 2003; Abadie et~al. 2010; Abadie 2021) and that the base design inherits
(H{\"u}y{\"u}k et~al. 2024). Nothing in the weighting, in the estimates of the
subpopulations, or in the sensitivity index requires a row-level record.
The second is parsimony. Subpopulation moments average the patient-level
dispersion before it reaches the estimator, no nuisance function is
fitted, and the finite-sample variance inflation of the crossing
condition is identically zero, $\kappa \equiv 0$ in~(10).
The inflation that the small-sample scenarios of
Section~6 record is incurred only by the nuisance fitting
that individual-level data make possible. The
third is pre-specifiability. The estimator is a deterministic function of
a short list of pre-registered aggregates and of the design constants
$\sigma^{2}$ and $\lambda_{k}$ of the factor model, with no fitted model
class, tuning parameter, or simulated inflation curve to write into the
protocol.
This form of
auditability is rewarded by the estimand and adaptive-design guidances
(International Council for Harmonisation 2019; U.S. Food and Drug Administration 2019), and the one design stage computation the summary
level requires, the rank check of Proposition~1, is
arithmetic in $(K, \Dx, \Tpre, \Dz)$ for centroids in general position.

What the summary-level cannot do is remove confounding. The mechanism of
Proposition~3 works through the covariate of the individual
patient, $A_{kj} \sim \mathrm{Bernoulli}(g(x_{kj}))$, and the backdoor path
it opens is blocked by conditioning on $x_{kj}$ (Figure~\ref{fig:dag}),
which is an individual-level operation. The doubly robust construction fits
nuisance functions of patient-level covariates and cannot be defined on
subpopulation means. A summary-level analysis averages exactly the biased
arm means, borrowing redistributes the bias and, under a homogeneous
confounding geometry, cannot remove it, by~(6), and
the bias term is $\mathcal{O}(1)$ in $n$ while every variance term
decreases.
Under confounding, individual-level development is therefore a requirement
for consistency, and beyond the boundary of Theorem~1 it
also gives the smaller mean squared error. Below the
threshold, the summary-level treatment that is available is the
pre-specified worst case  bias band $|c_{k}|\,\|\alpha\|\,\|w_{T,k}\|$ carried
by the selector at the design stage, reported as the realized
$|c_{k}\,\alpha^{\top} w_{T,k}|$ at the analysis stage
(Section~4.6).

Enlarging the summary interface does not remove this boundary, although
it moves it, and it is useful to see how. Arm-conditional covariate means,
which are summaries themselves, reveal the between-arm covariate shift
$c_{k}\alpha$ behind Proposition~3. Converting the shift into an outcome
correction requires the outcome-weight direction $w_{T,k}$ of the
subpopulation, which first moments do not identify, because the
post-treatment weights are subpopulation-specific and a regression of
subpopulation means across subpopulations estimates a single pooled slope,
which is exactly what $w_{T,k}$ is not. Adding second cross moments within
subpopulation and arm does identify the linear outcome model of
Section~5.1, and with it a regression correction, but only
because that model is linear. The summaries that suffice are the
sufficient statistics of an assumed outcome model, so the summarization
carries exactly the specification error that double robustness is meant to
protect against, and the propensity side of that protection is out of
reach, since the logistic fit and the overlap functionals it feeds depend
on the full empirical covariate distribution and not on any fixed set of
moments. The boundary between the levels is therefore not which
statistics are stored but which functionals can be estimated from them: a
correction under a maintained model can be moved to a rich summary
interface, whereas a model-robust correction cannot.

On the efficiency side the consequences are quantified by
Propositions~2 and~4. The
explained-variance edge $E_{k} = \sigma_{x}^{2}\|w_{T,k}\|^{2}$ exists only
by conditioning on the individual covariate. Summary-level arm means have
the marginal variance $\sigma_{k,\mathrm{eff}}^{2}$, conditioning attains
$\sigma_{k,\mathrm{cond}}^{2}$, and when the data are captured at the
summary-level there is nothing to condition on: the subpopulation is
represented by a single covariate point, the clever covariate is constant
within each arm, and the targeted arm means are the sample means. The edge is then lost by construction rather than by choice. The
edge is bought at the variance inflation $\kappa_{k}(\bar n_{k})$ of the
nuisance fits, and the simulations quantify the exchange: on the held-out
environments the variance of the linear-fluctuation targeted estimator
matches that of the sample mean up to about nine patients per
subpopulation and arm, lies a few percent below it up to about $23$, and
$14$--$25\%$ below beyond $27$ (Section~6.5),
while on the adaptive path of the
$K = 8$ scenarios the recruitment effect dominates. In the small-sample
region, individual-level development is thus a capability that is
available but not yet realized as a gain, and at moderate sample sizes it
begins to pay. The two points go together: the same records that permit
bias removal supply the conditioning that generates the edge, and the same
small subpopulation samples that adaptive recruitment produces make both
expensive.

For the borrowing weights the resolution question is one of sufficiency.
Under the linear outcome model the bias of a weighted control mean depends
on the donors only through their centroids, whatever the shapes of their
covariate distributions, so the centroid is a sufficient description for
the matching problem, and under the location-only heterogeneity of
Section~5.1 the pairwise transport cost between two
subpopulations is the squared centroid distance. This is why the
summary-level objective is adequate here, and why the evaluated transport
penalty~\eqref{eq:wasserstein}, which is itself a summary-level (centroid)
form, is within simulation error of the best hard-constrained arm under
confounding and behind the summary-level sample mean arm under
randomization (Section~\ref{sec:supp-ot-results}). The sufficiency is a property of
the data generating process, not of the method. When the subpopulations
differ in distributional shape, the moments are no longer sufficient and
the penalty has to be computed from patient-level empirical
distributions, exactly or by entropic approximation
(Cuturi 2013; Peyr{\'e} and Cuturi 2019), so the borrowing component itself moves to
the individual-level (Section~\ref{sec:ot}); the simulations do not cover
this case. In the confounded configurations, where a single patient pool
serves every arm, the summary-level arms store the centroid as the
covariate of each patient; for the sample mean family the two levels then
differ only through the recruitment path at $K = 3$ (at $K = 8$ the
level-specific inputs also change the weights), whereas for the doubly robust
family they differ in the point estimation as well, since the targeting
step of \textbf{(2a)} is a zero shift on centroid covariates while
\textbf{(2b)} adjusts for $x_{kj}$ (Section~\ref{sec:supp-degdiag}). Under
randomization the two levels are
instantiated as two patient pools, where the subpopulation-level pool is
the data model of the base design itself, so contrasts across levels there
compare data regimes as much as estimators. This is why the results are
paired within pool (Section~6.1).

Three further consequences are operational. The diagnostic layer of this
article is individual-level throughout: the comparators that attribute the
finite-sample penalty, the overlap factor discussed after
Proposition~4, and the analysis-stage audit of the
selector, whose plug-ins $(\hat\sigma_{x}, \hat w_{T,k}, \hat\alpha,
\hat c_{k})$ are within-subpopulation dispersion, regression, and
propensity estimates. A trial captured at the summary-level therefore runs
the base design but loses the audit that Section~4.6
attaches to it. The level also enters the recruitment rule in the same way
as the family does. The sensitivity index uses level-specific variance
formulas (Section~\ref{sec:supp-degdiag}), so the consistency principle of
Proposition~5 applies with level-specific profiles in
place of family profiles in~(12), and the level, like the
family, should be pre-specified jointly for recruitment and analysis.
Finally, the adaptive loop changes the operational comparison. A
conventional summary-level analysis uses the aggregates once, after the
trial, whereas the present design uses them episode by episode. The
summary-level here therefore means a thin real-time interface of running
means, and the individual-level means row-level capture, cleaning, and
monitoring inside the recruitment loop, which is a considerably heavier
apparatus to validate and to pre-specify.

The consequences are asymmetric. Little in the sampled region rewards individual-level machinery under randomization at small sample sizes: the summary-level base variant, on its own pool, posts the highest AUC of the hard-constrained arms there (a cross-pool comparison under randomization, Section~\ref{sec:supp-sim}),
and the summary-level transport relaxation is behind it under
randomization and within simulation error of it under confounding except
in the misspecified $K = 8$ scenario (Section~\ref{sec:supp-ot-results}). What individual-level
development buys is capability that may be needed: consistency under
confounding, whose bias removal shows in the false positive rate, with the point estimates favoring the doubly robust arm at matched false positive rate (Section~6.4);
the explained-variance edge at moderate sample sizes; shape-aware
borrowing; and the audit layer. Each of these has a cost, and the costs
are the quantities that Section~4 derives or the
simulations estimate. Unlike the analysis, the capability cannot be
recovered afterward: aggregates can be computed from records, but never
the reverse. The recommendation below follows from this asymmetry.

\paragraph{Recommendation.}
The level at which the trial data are captured fixes the regime map
before any question about the regime within it is reached
(Table~\ref{tab:levels}). With summary-level capture the doubly robust
family, its comparators, shape-aware borrowing, and the analysis-stage
audit of the selector are unavailable, and under confounding the only
treatment of the $\mathcal{O}(1)$ bias term is the pre-specified band of
Proposition~3. Since aggregates can be computed from
patient-level records and never the reverse, the recommendation is
asymmetric: capture and retain patient-level records wherever governance
permits, pose the analysis at the level the regime warrants, and
pre-specify the level jointly with the family for both recruitment and
analysis, since the sensitivity index is level-specific and the mismatch
penalty~(12) applies to level profiles as it does to
family profiles.

% ===========================================================================
\section{THE SELECTOR AS A DECISION PROCEDURE}
\label{sec:supp-flow}
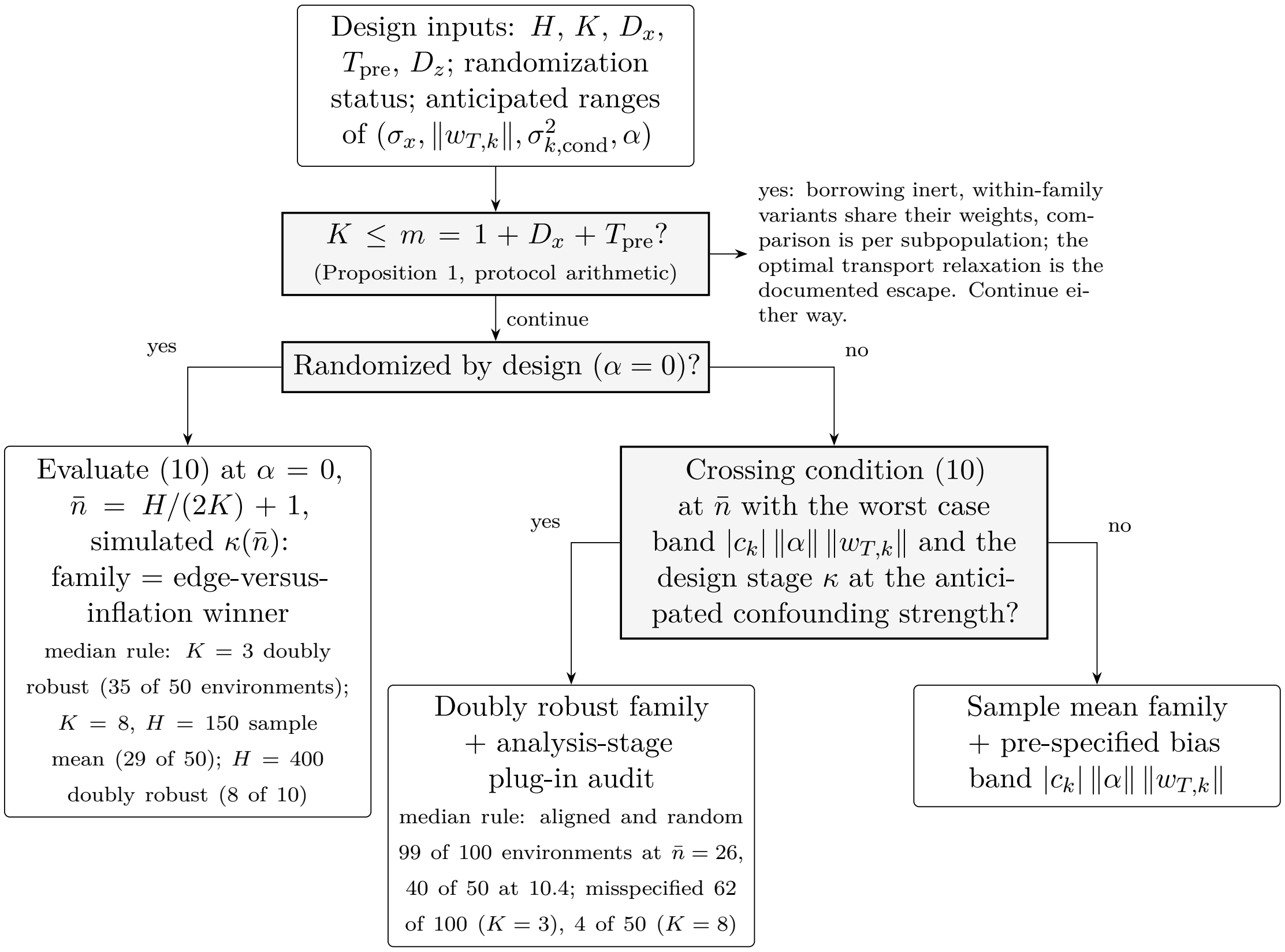
\begin{figure}[!htbp]
\centering
\resizebox{0.9\textwidth}{!}{%
\begin{tikzpicture}[
  >=Stealth, font=\small,
  box/.style={rectangle, draw, rounded corners=2pt, align=center,
              inner sep=4pt, text width=5.0cm},
  dec/.style={rectangle, draw, thick, align=center, inner sep=4pt,
              text width=5.4cm, fill=black!4},
  term/.style={rectangle, draw, rounded corners=2pt, align=center,
               inner sep=4pt, text width=4.6cm},
  note/.style={font=\scriptsize, align=left, text width=4.6cm}
]
\node[box] (in) {Design inputs: $H$, $K$, $\Dx$, $\Tpre$, $\Dz$;
  randomization status; anticipated ranges of
  $(\sigma_{x}, \|w_{T,k}\|, \sigma_{k,\mathrm{cond}}^{2}, \alpha)$};
\node[dec, below=6mm of in] (d1) {$K \le m = 1 + \Dx + \Tpre$?\\
  \scriptsize (Proposition~1, protocol arithmetic)};
\node[note, right=5mm of d1] (n1) {yes: borrowing inert, within-family
  variants share their weights, comparison is per subpopulation; the
  optimal transport relaxation is the documented escape. Continue either way.};
\node[dec, below=6mm of d1] (d2) {Randomized by design ($\alpha = 0$)?};
\node[term, below left=7mm and -12mm of d2] (t1) {Evaluate~(10)
  at $\alpha = 0$, $\bar n = H/(2K) + 1$, simulated $\kappa(\bar n)$:\\
  family $=$ edge-versus-inflation winner\\
  \scriptsize median rule: $K = 3$ doubly robust (35 of 50 environments); $K = 8$, $H = 150$ sample mean (29 of 50); $H = 400$ doubly robust (8 of 10)};
\node[dec, below right=7mm and -12mm of d2] (d3) {Crossing condition~(10) at $\bar n$ with the
  worst case band $|c_{k}|\,\|\alpha\|\,\|w_{T,k}\|$ and the design stage $\kappa$ at the anticipated confounding strength?};
\node[term, below left=6mm and -18mm of d3] (t2) {Doubly robust family\\
  $+$ analysis-stage plug-in audit\\
  \scriptsize median rule: aligned and random 99 of 100 environments at $\bar n = 26$, 40 of 50 at $10.4$; misspecified 62 of 100 ($K = 3$), 4 of 50 ($K = 8$)};
\node[term, below right=6mm and -18mm of d3] (t3) {Sample mean family\\
  $+$ pre-specified bias band $|c_{k}|\,\|\alpha\|\,\|w_{T,k}\|$};
\draw[->] (in) -- (d1);
\draw[->] (d1) -- (n1);
\draw[->] (d1) -- node[right, font=\scriptsize] {continue} (d2);
\draw[->] (d2) -| node[above left, font=\scriptsize] {yes} (t1);
\draw[->] (d2) -| node[above right, font=\scriptsize] {no} (d3);
\draw[->] (d3) -| node[above left, font=\scriptsize] {yes} (t2);
\draw[->] (d3) -| node[above right, font=\scriptsize] {no} (t3);
\end{tikzpicture}}
\caption{The pre-specified selector as a decision procedure, with the
results that its decisions rest on and the routing of the simulated
scenarios under the variance inflation curves the selector uses, the
design-embedded curve in the evaluation that follows the second decision
(step (ii)) and the design stage confounded curve in the third; the decisions choose the analysis, which under the recommended pairing (Recommendation 4) follows recruitment with the sample mean functional in every trial, and in the matched protocol of Table~3 the recruitment functional as well.}
\label{fig:flow}
\end{figure}

% ===========================================================================
Figure~\ref{fig:flow} shows the pre-specified selector of
Section~4.6 as a flowchart, with the results its decisions
rest on and the routing of the simulated scenarios under the
variance inflation curves the selector arm uses. The third
decision applies the crossing condition with the anticipated worst case
bias band and the design stage variance inflation curve simulated at the
anticipated confounding strength; at the simulated
strength the band routes the aligned and randomly oriented scenarios to the doubly robust family in $99$ of $100$ and $40$ of $50$ environments and the misspecified scenarios, whose exact shift is smaller, in $62$ of $100$ environments at $K = 3$ and $4$ of $50$ at $K = 8$ (Table~3).

\emph{Details of the evaluation.} Step (iii) uses $E_{k}$ in place of the
count-weighted $\bar E_{k}^{c}$ of Proposition~4:
$E_{k}$ is anticipated at the design stage without the mechanism and is
an upper bound for $\bar E_{k}^{c}$ under the linear-index mechanisms, so
the substitution favors the doubly robust family slightly; with
$\bar E_{k}^{c}$ five decisions of the simulated record change, one under
the median rule in the misspecified $K = 3$ scenario, two in the
misspecified $K = 8$ scenario, and one each under the upper-quartile rule
in the random and misspecified $K = 8$ scenarios (on the first $50$ environments of each scenario). A variance-only version of step (iii), which
compares $\bar n$ with the threshold $n_{k}^{*}(0)$ of the confounded
variance inflation curve, has an infinite threshold whenever
$\kappa_{k,\infty} \sigma_{k,\mathrm{cond}}^{2} \ge E_{k}$, which is the
case for the design stage curve at the simulated strength
(Section~\ref{sec:supp-inflation}); the band is what routes a confounded
trial. In the simulations the selector is evaluated under exact
anticipation: the edge $E_{k}$, the conditional variance
$\sigma_{k,\mathrm{cond}}^{2}$, and the bands are computed from the true
parameters of each environment (means over its subpopulations), the
information a well-anticipated protocol would have, so its record
measures the rules and not the quality of the anticipation. The realized
band, the between-arm covariate shift times the outcome weights
($c_{k}|\alpha^{\top} w_{T,k}|$ for the linear-index mechanisms) averaged
over the subpopulations, is reported alongside the worst case band that
enters the decision (Section~\ref{sec:supp-inflation}).

% ===========================================================================
\section{SIMULATION STUDY: CAUSAL STRUCTURE, RUN INVENTORY, AND IMPLEMENTATION DETAILS}
\label{sec:supp-sim}
% ===========================================================================
\paragraph{Design summary in the ADEMP structure (aims, data-generating mechanisms, estimands, methods, performance measures; Morris et~al. 2019).} \emph{Aims}: to locate the equal-error boundary between the two estimator families in the plane of patients per subpopulation and arm and confounding bias, and to evaluate the pre-specified selector. \emph{Data generating mechanisms}: the linear Gaussian factor model of Section~5.1 with $K \in \{3, 8\}$ subpopulations, two observed covariates, two latent loadings, and four pre-treatment outcomes; design-controlled assignment and three confounded mechanisms at strength $2$; adaptive and uniform recruitment; horizons $H = 150$ (core) and $400$; eight core scenarios with $50$ environments and $20$ replications each ($100$ environments in the three confounded $K = 3$ scenarios, extended from $50$ after the analysis of the first $50$), seven probes with ten environments each, a fixed-count map with up to $200$ replications per point, an adaptive map with ten environments per configuration, the two mismatch pairs of Section~\ref{sec:supp-mismatch} on ten randomized $K = 8$ environments, the held-out same-path runs behind the design-embedded curves (ten environments per configuration), the sensitivity runs of Section~\ref{sec:supp-sens} on the first $50$ environments of four core scenarios, and the mismatch arm recruit-1b/analyze-2b on every environment of the eight core scenarios (the selector with the recommended pairing, Section~\ref{sec:supp-inflation}). \emph{Estimands}: the subpopulation treatment effects $r_{k}$ and their signs. \emph{Methods}: the six variants of the estimator grid of Section~5, the AIPW and cross-fitted comparators at the weights of \textbf{(2b)}, the selector under its two rules (as a matched protocol and with the recommended pairing), and the transport relaxations. \emph{Performance measures}: true and false positive rates at the zero threshold, the AUC, the true positive rate at matched false positive rate, the mean squared error per subpopulation, and the sign of the mean squared error difference against the prediction, all as paired differences with standard errors across environments. No Monte Carlo precision target was set in advance: the core cells were raised from ten to $50$ environments after the first runs had left the margins at one to two standard errors, the three confounded $K = 3$ cells from $50$ to $100$ after the analysis of the first $50$ for the same reason, and the probes and the sensitivity runs are exploratory.

\subsection{Causal Structure of One Recruitment Episode}
Figure~\ref{fig:dag} shows the causal structure of one recruitment episode
and the two facts on which the analysis rests. The first is
identification. Under confounded assignment the backdoor paths from $A$
into $Y_{T}$ are $A \leftarrow x \rightarrow Y_{T}$,
$A \leftarrow x \leftarrow k \rightarrow Y_{T}$, and
$A \leftarrow x \leftarrow k \rightarrow z \rightarrow Y_{T}$, because the
latent loading $z$ influences the outcomes but not the assignment; all of
them run through the observed $x$. Conditioning on $x$ within the subpopulation therefore
blocks every noncausal path, there is no unmeasured confounding by
construction, and the propensity $g(x) = \expit(\alpha^{\top} x)$ is a
function of observables. This is what
makes the doubly robust correction of Proposition~3 feasible
with $x$ alone. The second is selection. Adaptive recruitment and
design-controlled assignment act through the accrued record
$\mathcal{D}_{h-1}$ before episode $h$, which decides which subpopulation, and under randomization which arm, contributes the next patient. Conditional on the
subpopulation and the arm, the outcome law is unchanged. Recruitment
therefore moves the information terms of the boundary between the two
regimes, namely the sample sizes entering $s_{k}$, while the assignment
mechanism moves its
bias term $b_{k}$. The two panels of the figure differ exactly in the
single arrow into $A$, which comes from the record under randomization and
from the covariate of the patient under confounding.

\begin{figure}[!htbp]
\centering
\begin{tikzpicture}[
  >=Stealth,
  obs/.style={circle, draw, minimum size=9mm, inner sep=1pt},
  lat/.style={circle, draw, dashed, minimum size=9mm, inner sep=1pt},
  rec/.style={rectangle, draw, minimum height=8mm, minimum width=13mm,
              inner sep=2pt}
]
\node[rec] (D)  at (0.0,  0.0) {$\mathcal{D}_{h-1}$};
\node[obs] (k)  at (1.9,  1.1) {$k$};
\node[obs] (x)  at (3.6,  1.7) {$x$};
\node[lat] (z)  at (3.6,  0.3) {$z$};
\node[obs] (A)  at (2.6, -1.4) {$A$};
\node[obs] (Yp) at (5.6,  1.7) {$Y_{\neg T}$};
\node[obs] (Yt) at (5.6,  0.0) {$Y_{T}$};
\draw[->] (D) -- (k);
\draw[->, very thick, red!70!black] (D) -- (A);
\draw[->] (k) -- (x);
\draw[->] (k) -- (z);
\draw[->] (k) to[bend right=34] (Yt);
\draw[->] (x) -- (Yp);
\draw[->] (x) -- (Yt);
\draw[->] (z) -- (Yp);
\draw[->] (z) -- (Yt);
\draw[->] (A) -- (Yt);
\draw[->, densely dashed, line width=0.8pt, black!75]
  (Yt.south) .. controls (5.6,-2.8) and (0,-2.8) .. (D.south)
    node[pos=0.08, right, font=\scriptsize, black!75] {enters $\mathcal{D}_{h}$};
\draw[->, densely dashed, line width=0.8pt, black!75]
  (Yp.north) .. controls (5.6,3.3) and (0,3.3) .. (D.north);
\node[font=\small] at (2.8, -3.3) {(a) Randomized assignment};
\end{tikzpicture}
\hspace{9mm}
\begin{tikzpicture}[
  >=Stealth,
  obs/.style={circle, draw, minimum size=9mm, inner sep=1pt},
  lat/.style={circle, draw, dashed, minimum size=9mm, inner sep=1pt},
  rec/.style={rectangle, draw, minimum height=8mm, minimum width=13mm,
              inner sep=2pt}
]
\node[rec] (D)  at (0.0,  0.0) {$\mathcal{D}_{h-1}$};
\node[obs] (k)  at (1.9,  1.1) {$k$};
\node[obs] (x)  at (3.6,  1.7) {$x$};
\node[lat] (z)  at (3.6,  0.3) {$z$};
\node[obs] (A)  at (2.6, -1.4) {$A$};
\node[obs] (Yp) at (5.6,  1.7) {$Y_{\neg T}$};
\node[obs] (Yt) at (5.6,  0.0) {$Y_{T}$};
\draw[->] (D) -- (k);
\draw[->, very thick, red!70!black] (x) to[bend right=38] (A);
\draw[->] (k) -- (x);
\draw[->] (k) -- (z);
\draw[->] (k) to[bend right=34] (Yt);
\draw[->] (x) -- (Yp);
\draw[->] (x) -- (Yt);
\draw[->] (z) -- (Yp);
\draw[->] (z) -- (Yt);
\draw[->] (A) -- (Yt);
\draw[->, densely dashed, line width=0.8pt, black!75]
  (Yt.south) .. controls (5.6,-2.8) and (0,-2.8) .. (D.south)
    node[pos=0.08, right, font=\scriptsize, black!75] {enters $\mathcal{D}_{h}$};
\draw[->, densely dashed, line width=0.8pt, black!75]
  (Yp.north) .. controls (5.6,3.3) and (0,3.3) .. (D.north);
\node[font=\small] at (2.8, -3.3) {(b) Confounded assignment};
\end{tikzpicture}
\caption{Causal structure of one recruitment episode. The one arrow that
differs between the panels, the arrow into $A$, is drawn heavy; the arrow
$k \to Y_{T}$ carries the subpopulation-specific outcome weights
$(w_{T,k}, \mu_{T,k})$ and effect $r_{k}$. Solid
circles are observed variables, the dashed circle is the latent loading,
the rectangle is the accrued trial record, and the dashed arrows close the
adaptive loop: the data of the episode enter the record that drives the
later recruitment. (a)~Randomized (design-controlled) assignment: the
design selects the subpopulation $k$ and the arm $A$ from the record
$\mathcal{D}_{h-1}$, so $A \perp (x, z)$ given the record and the
subpopulation, and every noncausal path into $Y_{T}$ runs through the
record and the subpopulation ($A \leftarrow \mathcal{D}_{h-1} \rightarrow
k \rightarrow Y_{T}$, and through $k$ to $x$ and $z$), so all are blocked
by the analysis within subpopulation. (b)~Confounded assignment:
the arm is drawn from the propensity $g(x) = \expit(\alpha^{\top} x)$,
which opens the backdoor paths $A \leftarrow x \rightarrow Y_{T}$,
$A \leftarrow x \leftarrow k \rightarrow Y_{T}$, and
$A \leftarrow x \leftarrow k \rightarrow z \rightarrow Y_{T}$; all are
blocked by conditioning on the observed $x$ within subpopulation, and $z$ influences
the outcomes but not the assignment, so adjustment for $x$ alone
identifies the arm means. In both panels
recruitment acts by selection: it changes the sample sizes of the
subpopulations but not the outcome law given subpopulation and arm. This
is why recruitment moves the variance terms of the boundary between the
two regimes while the assignment moves its bias term.}
\label{fig:dag}
\end{figure}

\subsection{Run Inventory}
Table~\ref{tab:status} lists the realized configurations against the
factor grid: eight core scenarios that are reported in full, the probes
(four that move one coordinate of the map at a time, the
estimated-variance runs, and the anchor), the mismatch pairs, the
same-path runs on ten further environments from which the
design-embedded variance inflation curve is estimated, the grids and curves of the regime map, the sensitivity runs of Section~\ref{sec:supp-sens}, and the mismatch arm behind the selector with the recommended pairing.
\begin{table}[!htbp]
\centering
\caption{Run inventory (first part; the second part continues the table with the same columns).}
\label{tab:status}
\footnotesize\setlength{\tabcolsep}{3pt}
\resizebox{\ifdim\width>\linewidth\linewidth\else\width\fi}{!}{%
\begin{tabular}{@{}>{\raggedright\arraybackslash}p{0.20\linewidth}>{\raggedright\arraybackslash}p{0.32\linewidth}>{\raggedright\arraybackslash}p{0.11\linewidth}cc>{\raggedright\arraybackslash}p{0.15\linewidth}@{}}
\toprule
Configuration & Assignment & Recruitment & $H$ & $\bar n^{(a)}$ & Role \\
\midrule
$K = 3$ & randomized; aligned; random; misspecified & adaptive & 150 & $\approx 26$ & core scenarios, borrowing inert \\
$K = 8$ & randomized; aligned; random; misspecified & adaptive & 150 & $\approx 10$ & core scenarios, borrowing active \\
$K = 8$ & randomized; aligned & adaptive & 400 & $\approx 26$ & boundary probe \\
$K = 8$ & randomized & uniform & 150 & $\approx 10$ & recruitment axis \\
$K = 8$, $\sigma_{x} = 2$, $\sigma_{z} = 1$ & randomized & adaptive & 150 & $\approx 10$ & coordinate probe (covariate signal) \\
$K = 8$, estimated variances & randomized; aligned & adaptive & 150 & $\approx 10$ & one-factor variant \\
Mismatch pairs, $K = 8$ & randomized & adaptive & 150 & $\approx 10$ & Proposition~5 magnitudes \\
Anchor $r = (-10, 0, +10)$, $K = 3$ & randomized & adaptive & 150 & $\approx 26$ & implementation check \\
\bottomrule
\end{tabular}}
\end{table}

\begin{table}[!htbp]
\centering
\ContinuedFloat
\caption{Run inventory (second part, continued; the columns are those of the first part).}
\footnotesize\setlength{\tabcolsep}{3pt}
\resizebox{\ifdim\width>\linewidth\linewidth\else\width\fi}{!}{%
\begin{tabular}{@{}>{\raggedright\arraybackslash}p{0.20\linewidth}>{\raggedright\arraybackslash}p{0.32\linewidth}>{\raggedright\arraybackslash}p{0.11\linewidth}cc>{\raggedright\arraybackslash}p{0.15\linewidth}@{}}
\toprule
Configuration & Assignment & Recruitment & $H$ & $\bar n^{(a)}$ & Role \\
\midrule
Map, fixed counts & randomized; aligned and random directions at $\|\alpha\| = 0.5$--$3$ & fixed sample sizes & --- & 4--64 & regime map (Figure~1, left) \\
Map, adaptive grid, $K = 3, 8$ & randomized; aligned at $\|\alpha\| = 1, 2, 3$ & adaptive & 50--600 & varies & regime map (Figure~1, middle and right) \\
$\kappa_{k}(\bar n)$, design stage & randomized; aligned & fixed sample sizes (first ten $K = 3$ evaluation environments) & --- & 4--64 & variance inflation curve \\
$\kappa_{k}(\bar n)$, design-embedded & randomized; aligned; $\sigma_{x} = 2$ & adaptive, path of \textbf{(1b)}, held-out environments & 150 & $\approx 10$, $26$ & variance inflation curve, $n^{*}$ \\
Sensitivity runs (Section~\ref{sec:supp-sens}) & aligned $K = 3$ and random $K = 8$ (fallback at $10$ and $14$); aligned $K = 3$, $8$ (misspecified outcome; AIPW and regression-adjusted arms); randomized $K = 8$ (AIPW and regression-adjusted arms); aligned and randomized $K = 8$ (targeted arm with the AIPW curve) & adaptive & 150 & $\approx 26$, $10$ & first $50$ environments of the core scenarios \\
Selector with the recommended pairing (Section~\ref{sec:supp-inflation}) & the eight core scenarios: the doubly robust analysis of the data recruited by \textbf{(1b)} (recruit-1b/analyze-2b) & adaptive, path of \textbf{(1b)} & 150 & $\approx 26$, $10$ & all environments of the core scenarios \\
$\kappa_{k}(\bar n)$, sensitivity & randomized (bounded fluctuation, $60$ replications; AIPW; regression adjustment); aligned (AIPW; regression adjustment; held-out environments) & fixed sample sizes & --- & 4--64 & curves of Section~\ref{sec:supp-sens} \\
\bottomrule
\end{tabular}}
\par\vspace{4pt}
\begin{minipage}{\linewidth}\footnotesize
NOTE: Every trial configuration is averaged over ten
environments with $20$ replications each ($50$ environments for the eight core scenarios, extended to $100$ for the three confounded $K = 3$ scenarios after the analysis of the first $50$; the sensitivity runs over the first $50$ environments of their scenarios), and the design stage simulation uses $200$ replications per environment, subpopulation, and sample size ($60$ for the bounded-outcome fluctuation); $\bar n^{(a)}$ is the mean number of patients per
subpopulation and arm at the horizon. The core
scenarios are reported in Sections~6.1--6.3, the
probes and the anchor in Section~6.4, the mismatch pairs in
Section~6.6, and the variance inflation curves in
Section~6.5.
\end{minipage}
\end{table}

\subsection{Implementation Details of the Arms}
The implementation details of the arms are recorded here, since
they enter the variance functionals of the arms, their level labels, or
the reading of the contrasts. \textbf{(1a)} and \textbf{(2a)} use the rates
$\lambda_{k} = \|M_{\neg T}^{+}\mu_{T,k}\|^{2}$ of the base design for each
subpopulation, which are exact on the subpopulation-level pool; the
individual-level arms use $\tilde\lambda_{k}$ and the cross-covariance term~\eqref{eq:crossterm}. The optimal transport arms run on the
subpopulation-level pool with the inputs of that pool ($\sigma^{2}$ and
$\lambda_{k}$) under randomization and, under confounded assignment, on the
shared individual-level pool with the individual-level inputs
$\sigma_{k,\mathrm{eff}}^{2}$, $\tilde\lambda_{k}$, and the cross-covariance
term. The nuisance models of family 2 are a linear outcome model in the
covariates and the arm and a logistic propensity model, truncated to
$[0.05, 0.95]$, with a weak ridge penalty (inverse regularization strength
$C = 100$ in scikit-learn). The influence curve variances of \textbf{(2a)}
and of the estimated-variance variant are divided by the sample size
$n_{k}$ of the subpopulation, and the covariance of the two arm means
enters the variance of the contrast. At small counts these estimates
understate the realized variance: in balanced randomized cells the
realized variance of the targeted contrast is $3.2$, $1.5$, $1.2$, and
$1.04$ times $\sigma_{k,\mathrm{cond}}^{2} s_{k}$ at $3$, $5$, $10$, and
$26$ patients per arm, while the mean influence curve estimate is $0.31$,
$0.62$, $0.87$, and $0.97$ times it ($4{,}000$ replications per size). Under confounding these estimates
converge to $\tilde\omega\,\sigma^{2}_{k,\mathrm{cond}} s_{k}$, which omits
the outcome model term of the influence curve as implemented and lies
about a third below the realized factor $\omega_{k}^{\dagger}$
($\tilde\omega/\omega^{\dagger}$ of $0.62$--$0.69$ under the linear-index mechanisms, $0.78$ and $0.63$ under the quadratic one), which bears on the confounded
estimated-variance probe of Section~\ref{sec:supp-probes}. When a
subpopulation holds fewer than
six patients, or fewer than two in an arm, the doubly robust arm means
fall back to the sample means, and the fallback returns for each arm its
sample variance divided by its count ($\sigma^{2}$ for a one-patient
arm); the
estimated-variance variant of \textbf{(1b)} uses $\sigma^{2} = 1$ while a
subpopulation holds fewer than four patients. The shares of subpopulation
cells on these fallbacks are recorded in the summary file of the
accompanying code (for example,
$35\%$ of the subpopulation cells of \textbf{(2b)} at the horizon of the
randomized $K = 8$ scenario are on the sample mean fallback). The cross-fitted
comparator splits each subpopulation into two folds stratified by arm and
uses the split only when every training fold holds at least two patients
of each arm and $2(\Dx + 2)$ patients in total, two observations per
coefficient of the linear outcome model, that is, when
$\lfloor n_{k}^{(1)}/2 \rfloor + \lfloor n_{k}^{(0)}/2 \rfloor \ge 8$ with
at least four patients in the smaller arm;
otherwise it uses the full-sample targeted fit, so that it measures the
cost of sample splitting and not that of saturated outcome model fits or
of training folds with a single patient of an arm; a fold with two
patients of an arm can still separate the logistic fit, which the ridge
penalty keeps finite and the truncation bounds. Under
confounded assignment every arm draws from one patient-level
potential outcomes pool; the balanced burn-in of one patient per
subpopulation and arm is forced (drawn without regard to the propensity,
so that, for subpopulation sizes fixed in advance and given the arm
assignments (Assumption~\ref{ass:counts}), the expected bias of the
no-borrowing contrast is $c_{k}\alpha^{\top} w_{T,k}$, with $c_{k}$ the
shift coefficient of Proposition~3, diluted by the factor
$(1 - 1/n_{k}^{(1)})(1 - \pi_{k}) + (1 - 1/n_{k}^{(0)})\pi_{k}$; under
the adaptive recruitment run here the realized bias departs from it,
Figure~\ref{fig:bias}), and thereafter the arm of each recruited patient
is drawn from the propensity at observation time, so common seeds give
common potential outcomes but not common assignments across the arms. The summary-level arms \textbf{(1a, 2a)} and the transport
relaxations store the centroid of the subpopulation as the covariate of
each patient, so the nuisance models of \textbf{(2a)} condition on a
constant within the subpopulation, its targeting shift is zero, and its
arm means are the sample means. This is the summary-level reading of
Section~\ref{sec:levels} taken literally: only the individual-level
doubly robust arms and the comparators remove confounding bias in these
scenarios. The hard-constrained solvers return the no-borrowing weight
whenever the realized constraint matrix has rank $K$, which on running
means is the case, almost surely and for centroids in general position,
whenever $K \le m$. Exact ties in the look-ahead, which
arise between the two arms of the target subpopulation when their counts
are equal, are broken deterministically (fewest patients, then the lowest
subpopulation and arm index, which sends such a tie to the control arm),
so variants that share an objective also share the recruitment path.
Finally, the family-2 arms \textbf{(2b, 2c)} are recruited and weighted
with the oracle conditional variances inflated by the design-embedded
factor $1 + \kappa(\bar n_{k})$ of Section~6.5,
which is estimated on a separate set of ten simulated environments and
then held fixed, and with the covariate sampling terms discussed after
Proposition~2 in the weighting; the comparators
\textbf{(3a, 3b)} are analysis-only arms evaluated on the patients
recruited by \textbf{(2b)} at its weights and inflated in the same way;
\textbf{(2a)} and the estimated-variance variant use influence curve
estimates and no inflation factor. The confounded scenarios use the design-embedded curve simulated under aligned confounding and the $\sigma_{x} = 2$ probe its own curve. The selector arm is not recruited separately: for each
environment it copies the arrays of the family it selects, which share
the environment's seeds.

The list of arms is short by design. The two comparators of
Section~3.2 vary one axis at a time
from variant \textbf{(2b)}: the recombination axis (targeting versus the
one-step correction) and the fitting axis (own sample versus cross-fit;
the cross-fitted comparator also fits its fluctuation within each
held-out fold and averages the per-fold targeted means, so at $8$--$10$
patients per arm part of its cost is the per-fold targeting rather than
the halved nuisance sample).
The fourth combination in that factorial, cross-fitted AIPW, differs from
\textbf{(2b)} on both axes at once and is informative only about the
interaction of the two axes, whose main effects the pair already isolates;
we omit it to keep the number of arms fixed. Super-learner nuisance fits
(van~der Laan and Rose 2011) protect against misspecification of the nuisance
form. Here the outcome model is linear and, except in the
misspecified scenario, the propensity is logistic by construction, so the
parametric working models are correctly specified in the other scenarios,
and an ensemble would add the finite-sample cost of its internal
cross-validation without a specification benefit; that cost differs from
the halved-sample fitting that the cross-fitted arm measures, since a
super learner refits on the full sample after selecting its weights. Causal trees and forests and the R-learner
(Athey and Imbens 2016; Wager and Athey 2018; Nie and Wager 2021) estimate the conditional effect
surface and thereby discover subgroup structure. The estimand here
is the effect vector over pre-specified subpopulations whose membership is
known at recruitment, which is a different question, and the sample
splitting those methods use (honest splitting for trees and forests,
cross-fitting for the R-learner) would add to the small-sample problem
that already binds. Bayesian hierarchical borrowing
shrinks each subpopulation toward a precision-weighted common mean, a
borrowing geometry different from the constrained projection of the base
design, since it fades continuously with the estimated heterogeneity rather
than collapsing at a rank threshold (the contrast is made precise after
Proposition~1). Replacing the borrowing mechanism while
also replacing the estimator family would confound the two changes, so we
keep the borrowing of the base design fixed and vary the family.

For reference, one recruitment episode of the design proceeds as follows;
this is Algorithm~1 of the base design (H{\"u}y{\"u}k et~al. 2024), stated with the
family-specific variance. Given the accrued record, (1) compute for every
subpopulation the effect estimate $\hat r_{k}$ and its variance $V_{k}$
from the chosen family and weighting, and the sensitivity index
$S_{k} = |\hat r_{k}| / \sqrt{V_{k}}$; (2) select the least certain
subpopulation $k^{*} = \arg\min_{k} S_{k}$; (3) choose the subpopulation
and arm $(l, a)$ whose incremented sample size reduces the arm's working
objective at $k^{*}$ the most ($V_{k^{*}}$; the full or penalized objective
for \textbf{(1c, 2c)} and the transport arms; under confounded assignment
the arm is instead drawn from the propensity); (4) observe the patient, update the running means, and return
to (1). Uniform recruitment replaces (2)--(3) by a fixed balanced schedule.
The index is computed from the same family-specific $V_{k}(\beta)$ of~(3) that the final analysis uses, which is what
Proposition~5 requires of a matched design; the
MSE-optimal variants \textbf{(1c, 2c)} and the oracle relaxation divide by
the square root of their full objective including the squared bias, and
the transport arms by the square root of the penalized objective they
minimize, the working functional plus the transport and trajectory
penalties. Since the index is a ratio, a common
inflation of all variance terms of all subpopulations leaves the choice of
subpopulation unchanged; an inflation that differs across subpopulations
moves it, and the simulated variance inflation, which multiplies only
the conditional-variance terms, also changes the weights and the index
when
borrowing is active (on identical states at $K = 8$ a constant
$\kappa = 0.36$ changed the recruited subpopulation in about $2\%$ of the
decisions of a six-environment check).

\paragraph{Estimators and metrics as computed.}
The targeting step of family 2 updates the fitted outcome model as
$Q^{*}(a,x) = Q(a,x) + \varpi_{a} \mathsf{H}_{a}(a,x)$ along the clever
covariates $\mathsf{H}_{1}(A,x) = \ind\{A=1\}/g(x)$ and
$\mathsf{H}_{0}(A,x) = \ind\{A=0\}/(1-g(x))$, with $\varpi_{a}$ the
least-squares coefficient of the residual on $\mathsf{H}_{a}$ in arm $a$.
The environment-level parameters of Section~5.1 are drawn once
per environment: $x_{k}, z_{k} \sim \mathcal{N}(0, I)$,
$\delta_{t} \sim \mathcal{N}(0, 1)$, $w_{t}$ and $\mu_{t}$ uniform on the
unit disk scaled by $2 - 10^{t-5}$ for $t = 1, \dots, 5$ (about $2$ in
the four pre-treatment periods and $1$ in the post-treatment period),
$w_{T,k}$ and $\mu_{T,k}$ uniform on the unit disk, and
$r_{k} \sim \mathcal{N}(0, 1)$.
The AUC of an environment pools the $K \times 20$ estimates of its
replications at the horizon into one Mann--Whitney statistic against the
sign pattern, and the matched-FPR comparison re-thresholds both arms
over the same pooled estimates: the threshold of an arm is the $(1 - f)$
quantile of its pooled negative estimates, $f$ the zero-threshold false
positive rate of \textbf{(2b)} in that environment, applied to
\textbf{(2b)} as well, because the quantile threshold sits inside the
vertical segment of the pooled ROC and re-reads \textbf{(2b)} $0.8$--$2.5$ points above its zero-threshold true positive rate in the core scenarios (reading \textbf{(2b)} at its zero threshold instead would favor the sample mean arm by that much); averaging per trial instead
gives slightly different values ($0.897$ against $0.902$ for
\textbf{(1b)} in the randomized $K = 3$ scenario). The two patient pools of the randomized scenarios share seeds, so contrasts are paired within a pool.

% ===========================================================================
\section{AN OPTIMAL TRANSPORT RELAXATION OF THE BORROWING WEIGHTS}
% ===========================================================================
\label{sec:ot}
The hard covariate constraint $x_{k} = X\beta$ can be replaced by a soft
penalty derived from the squared 2-Wasserstein distance of optimal
transport (OT) between the mixture of donor centroids
$\mathbb{P}_{\beta} = \sum_{l}\beta_{l}\,\boldsymbol{\delta}_{x_{l}}$ and
the target centroid $\mathbb{P}_{k} = \boldsymbol{\delta}_{x_{k}}$, with
$\boldsymbol{\delta}_{x}$ the point mass at $x$. Under the linear outcome model the bias of a weighted control
mean depends on the donors only through their centroids, whatever the
shapes of their covariate distributions, so the centroid is a sufficient
description for the matching problem; for these point masses the distance
is linear in $\beta$,
\begin{equation}
W_{2}^{2}(\mathbb{P}_{\beta}, \mathbb{P}_{k})
= \sum_{l} \beta_{l}\,\|x_{l} - x_{k}\|^{2},
\label{eq:wasserstein}
\end{equation}
which is a transport cost only when $\beta$ is a probability vector; off
the simplex its minimization rewards negative weight on distant donors.
(Location-only heterogeneity, the case of Section~5.1, is what
makes the pairwise transport cost between two subpopulations' full
covariate distributions equal $\|x_{l} - x_{k}\|^{2}$; for the mixture
against the target the right-hand side is the cost of translating each
component and bounds the transport distance from above.) The weights are
therefore
constrained to the simplex, which preserves the interpretation of
$\hat r_{k}(\beta)$ as an average effect and keeps the synthetic control
inside the convex hull of the donors. On the simplex,~\eqref{eq:wasserstein} is the pairwise discrepancy penalty of penalized
synthetic control (Abadie and L'Hour 2021). Here the penalty replaces
the hard covariate constraint, and the trajectory constraint is relaxed
to the quadratic penalty
$\varsigma_{y}\,\|\Ybar_{k,\neg T} - \widehat{Y}_{\neg T}\beta\|^{2}$ on
the running pre-treatment means, so that only the simplex constraints
remain hard. The covariate penalty enters the objective as
$\varsigma_{x}\,h_{k}^{\top}\beta$ with $(h_{k})_{l} = \|x_{l} - x_{k}\|^{2}$,
and the two penalty weights are set from the variance scale of the
functional, $\varsigma_{x} = \bar\sigma^{2}/(\bar n^{(0)}\,\bar h_{k})$ and
$\varsigma_{y} = \bar\sigma^{2}/(\bar n^{(0)}\,\|\widehat{Y}_{\neg T}\|_{F}^{2})$,
with $\bar\sigma^{2}$ and $\bar n^{(0)}$ the mean outcome variance and
control count over the subpopulations and $\bar h_{k}$ the mean of the
$K$ squared centroid distances to the target, the zero self-distance
included. The arm evaluated in the simulations minimizes the working
functional~(3) of the sample mean family plus the two
penalties. It uses the inputs of the sample mean variant run on the same
pool, \textbf{(1a)} on the subpopulation-level and \textbf{(1b)} on the
individual-level pool, and is implementable in the same sense as those
variants: its objective has no oracle bias term, the bias evaluated at
the true outcome weights and the latent loadings $z_{l}$ that
\textbf{(1c)} carries, while its
variance inputs ($\sigma^{2}$ and $\lambda_{k}$, or
$\sigma_{l,\mathrm{eff}}^{2}$, $\tilde\lambda_{k}$, and $C_{kl}$) are the
oracle inputs that the family-1 arms and the oracle family-2 arms receive
(\textbf{(2a)} and the estimated-variance probe estimate theirs). The
corresponding relaxation of the oracle variant \textbf{(1c)} adds the
squared bias of the weighted control mean at the true outcome weights and
latent loadings, $(\tilde d_{k}^{\top}\beta - \tilde d_{kk})^{2}$ with
$\tilde d_{kl} = w_{T,l}^{\top} x_{l} + \mu_{T,l}^{\top} z_{l}$, which
reduces to the representation bias of~(4) when both the
covariate match $X\beta = x_{k}$, which the relaxation drops, and the
latent-loading match $\sum_{l} \beta_{l} z_{l} = z_{k}$ hold; it is run alongside as
a benchmark. With the sum-to-one condition
as the only equality the problem is a closed-form $(K+1)\times(K+1)$
Karush--Kuhn--Tucker system, solved by an active-set rule that binds the
most negative weight and releases bound weights with negative multipliers
(capped at $3K + 10$ solves; in the test trials of the accompanying code it converged within eight solves in every call).
Since inequality constraints do not reduce the affine dimension of the
feasible set, the relaxation removes the overdetermination that causes the
borrowing collapse of Proposition~1 at small $K$. When
subpopulations differ in distributional shape the penalty has to be
computed between the empirical distributions, exactly by linear
programming or by an entropic (Sinkhorn) approximation
(Cuturi 2013; Peyr{\'e} and Cuturi 2019), which we leave as future work.

\subsection{Results of the Relaxation}
\label{sec:supp-ot-results}
The relaxation and its oracle version run alongside the other arms of
every scenario (Table~\ref{tab:auc}; right panel of
Figure~\ref{fig:forest}). Under randomization the implementable relaxation
does not improve on the subpopulation-level sample mean arm
\textbf{(1a)}: it is $-0.024 \pm 0.017$ AUC below it at $K = 3$, where
its borrowing is the only active one, and $-0.026 \pm 0.011$ below it at
$K = 8$; at $K = 3$ its borrowing moves the treated share to $0.522$
(standard deviation $0.017$ across environments) against $0.496$ for the
constrained arms. Under confounded assignment it has the highest AUC of the nine nonoracle arms in the aligned scenario at both $K$ and in the misspecified scenario at $K = 8$ (at $K = 3$ the AIPW comparator is above it by $0.003$ under random orientation and \textbf{(1b)} by $0.001$ in the misspecified scenario), exceeding \textbf{(1a)} by $+0.006 \pm 0.013$, $+0.011 \pm 0.016$, and $+0.007 \pm 0.013$ at $K = 3$ and by $+0.007 \pm 0.010$,
$-0.003 \pm 0.010$, and $+0.022 \pm 0.008$ at $K = 8$ (aligned, random,
and misspecified); there it also uses the individual-level variance
inputs (Section~\ref{sec:supp-sim}), so these margins combine the
relaxation with that change of inputs, and only the misspecified $K = 8$
margin is beyond two standard errors. The oracle relaxation exceeds \textbf{(1a)} in every confounded scenario ($+0.022$ to $+0.045$ AUC, all beyond two standard errors) and the implementable arm by $+0.012$ to $+0.032$; under randomization it matches \textbf{(1a)} at $K = 3$
($+0.002 \pm 0.003$) and exceeds it at $K = 8$ ($+0.025 \pm 0.004$).
Knowing the true post-treatment weights and latent loadings is thus worth up to five
hundredths of AUC at $H = 150$ ($0.012$--$0.032$ under confounding and
$0.051$ in the randomized $K = 8$ scenario, Table~\ref{tab:auc}), the
information that the oracle variants
\textbf{(1c, 2c)} carry and no implementable arm has. The relaxation
borrows at every $K$ (Figure~\ref{fig:degeneracy}) and is the documented
escape from the degenerate regime; its discrimination advantage over the
hard-constrained arms is confined to confounded assignment and is within
simulation error in all but one scenario.

% ===========================================================================
\section{ADDITIONAL RESULTS}
\label{sec:supp-results}
\begin{figure}[!htbp]
\centering
\includegraphics[width=\linewidth]{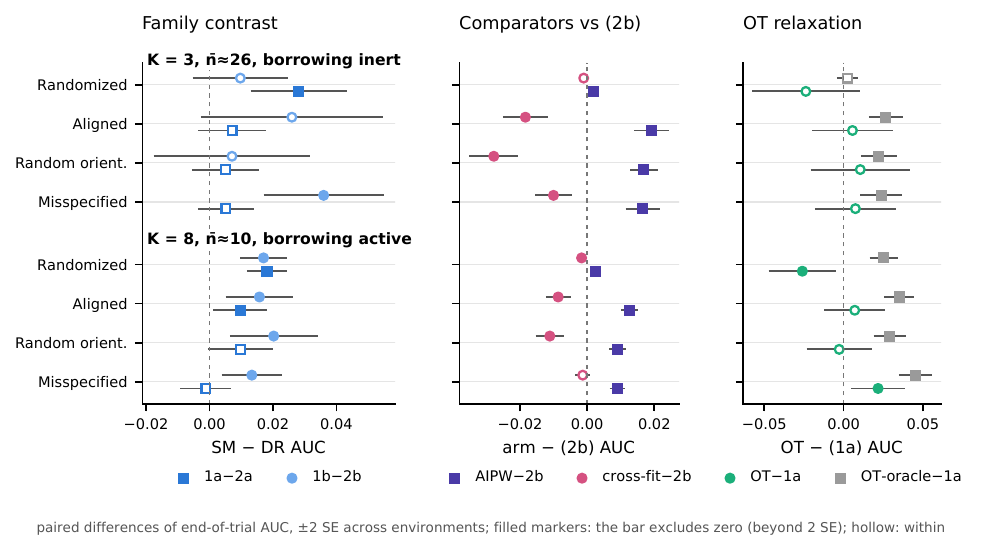}
\caption{Paired contrasts across the eight scenarios, AUC at the end of
the trial, mean $\pm 2$ paired standard errors across environments. Left:
family contrasts (sample mean minus doubly robust) at both levels; a
positive value favors the sample mean family. Middle: comparators against
targeted maximum likelihood at the same weights (AIPW minus \textbf{(2b)},
cross-fitted minus \textbf{(2b)}). Right: the implementable
optimal-transport relaxation and its oracle version against the
subpopulation-level sample mean arm \textbf{(1a)}, the hard-constrained
arm with the highest discrimination under randomization. Filled markers
are individually significant, that is, more than two paired standard
errors from zero.}
\label{fig:forest}
\end{figure}

% ===========================================================================
Table~\ref{tab:auc} gives the AUC of every arm in the eight core
scenarios, and Figure~\ref{fig:forest} draws the paired contrasts of
Table~2 together with the transport relaxation.

\begin{table}[!htbp]
\centering
\caption{AUC at the end of the trial (probability that a truly responsive
subpopulation is ranked above a nonresponsive one) in the eight core
scenarios, averaged over the environments in which the AUC is defined
($39$ of the $50$ in the randomized $K = 3$ scenario and $80$ of the $100$ in the confounded $K = 3$ scenarios, which share their first $50$ environments with it; all $50$ at $K = 8$).}
\label{tab:auc}
\footnotesize
\setlength{\tabcolsep}{4pt}
\resizebox{\ifdim\width>\linewidth\linewidth\else\width\fi}{!}{%
\begin{tabular}{@{}lcccccccc@{}}
\toprule
& \multicolumn{4}{c}{$K = 3$ ($\bar n^{(a)} \approx 26$)}
& \multicolumn{4}{c}{$K = 8$ ($\bar n^{(a)} \approx 10$)} \\
\cmidrule(lr){2-5} \cmidrule(lr){6-9}
Estimator arm & Rand. & Aligned & Random & Missp. & Rand. & Aligned & Random & Missp. \\
\midrule
Sample mean, subpopulation (1a) & 0.962 & 0.866 & 0.857 & 0.871 & 0.894 & 0.799 & 0.815 & 0.801 \\
Sample mean, individual (1b) & 0.902 & 0.868 & 0.860 & 0.880 & 0.837 & 0.804 & 0.818 & 0.807 \\
Sample mean, MSE-optimal oracle (1c) & 0.902 & 0.868 & 0.860 & 0.880 & 0.832 & 0.804 & 0.818 & 0.808 \\
Doubly robust, subpopulation (2a) & 0.933 & 0.859 & 0.852 & 0.866 & 0.876 & 0.789 & 0.806 & 0.802 \\
Doubly robust, individual (2b) & 0.892 & 0.842 & 0.853 & 0.844 & 0.819 & 0.789 & 0.797 & 0.794 \\
Doubly robust, MSE-optimal oracle (2c) & 0.892 & 0.842 & 0.853 & 0.844 & 0.820 & 0.788 & 0.798 & 0.794 \\
AIPW at (2b) weights & 0.894 & 0.861 & 0.870 & 0.861 & 0.822 & 0.801 & 0.807 & 0.803 \\
Cross-fitted TMLE at (2b) weights & 0.891 & 0.824 & 0.826 & 0.834 & 0.818 & 0.780 & 0.786 & 0.793 \\
Selector, median rule & 0.900 & 0.842 & 0.854 & 0.860 & 0.832 & 0.790 & 0.800 & 0.805 \\
Selector, upper-quartile rule & 0.899 & 0.842 & 0.854 & 0.862 & 0.837 & 0.795 & 0.808 & 0.806 \\
OT relaxation (Section~\ref{sec:ot}) & 0.938 & 0.871 & 0.868 & 0.879 & 0.868 & 0.806 & 0.813 & 0.822 \\
OT relaxation of the oracle (1c) & 0.964 & 0.892 & 0.879 & 0.895 & 0.919 & 0.834 & 0.844 & 0.846 \\
\bottomrule
\end{tabular}}
\par\vspace{4pt}
\begin{minipage}{\linewidth}\footnotesize
NOTE: The uncertainty of every comparison quoted in the text is the paired standard error across environments. Within-family pairs are identical at $K = 3$ because borrowing is inert. Under randomization the summary-level arms \textbf{(1a, 2a)} and the transport relaxations run on the subpopulation-level pool (every patient at the centroid of its subpopulation) and the individual-level arms on the individual-level pool, so cross-level comparisons in the two randomized columns compare data regimes as well as estimators (Section~\ref{sec:levels}).
\end{minipage}
\end{table}

\subsection{The Degenerate Randomized Configuration}
\label{sec:supp-mainK3}
\begin{figure}[!htbp]
\centering
\includegraphics[width=\linewidth]{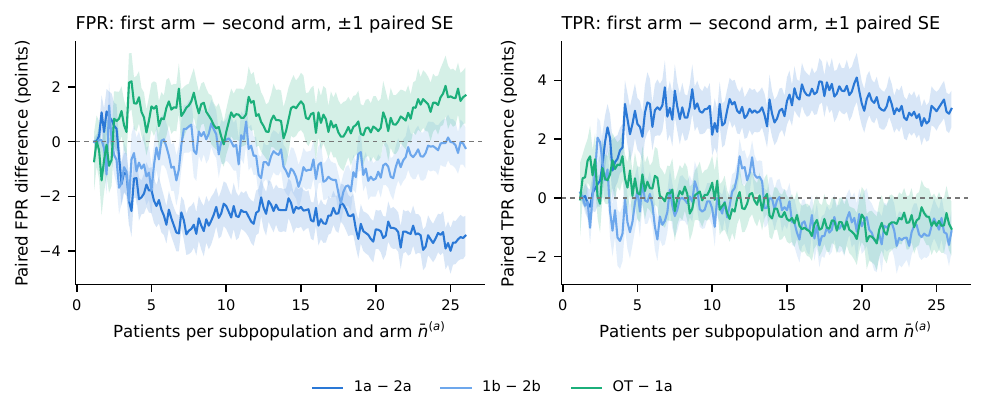}\\[2pt]
\includegraphics[width=0.48\linewidth]{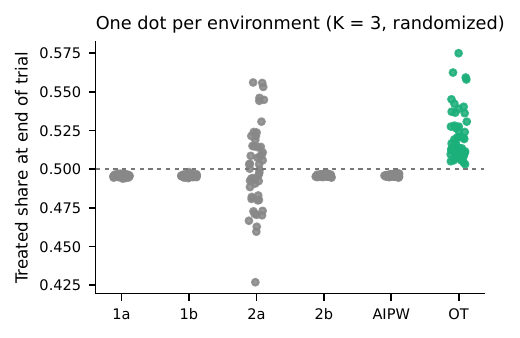}
\caption{Paired-difference trajectories in the degenerate randomized
configuration ($K = 3$, adaptive recruitment, randomized assignment).
Top: the paired differences of the false positive rate (left) and the
true positive rate (right) between the arms of each main contrast, in
percentage points, as functions of the number of patients per
subpopulation and arm, with $\pm 1$ paired standard error across the $50$
environments; the zero line is the crossing. Bottom: treated share at the
end of the trial, one dot per environment. The individual-level
constrained arms hold one half ($0.496$), the summary-level doubly robust
arm
\textbf{(2a)} scatters around it ($0.502$, standard deviation $0.027$
across environments, its look-ahead using influence curve variances), and
the optimal transport arm lies above it in every environment ($0.522$,
standard deviation $0.017$, minimum $0.503$). The absolute
trajectories are shown in Figure~\ref{fig:mainK3abs}; their spread across
environments (Table~\ref{tab:mainK3}) is an order of magnitude wider than
these paired errors.}
\label{fig:mainK3}
\end{figure}

\begin{figure}[!htbp]
\centering
\includegraphics[width=0.49\linewidth]{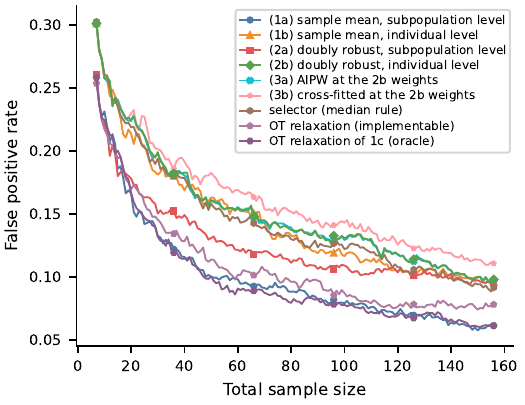}\hfill
\includegraphics[width=0.49\linewidth]{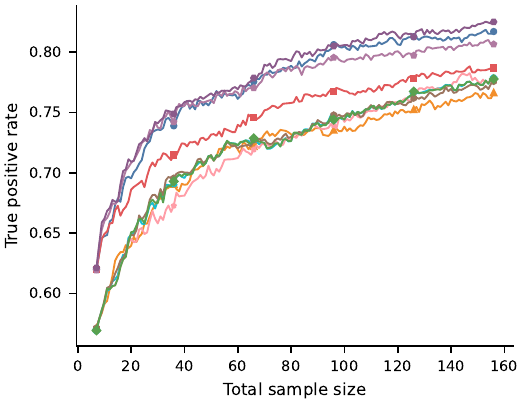}\\[2pt]
\includegraphics[width=0.49\linewidth]{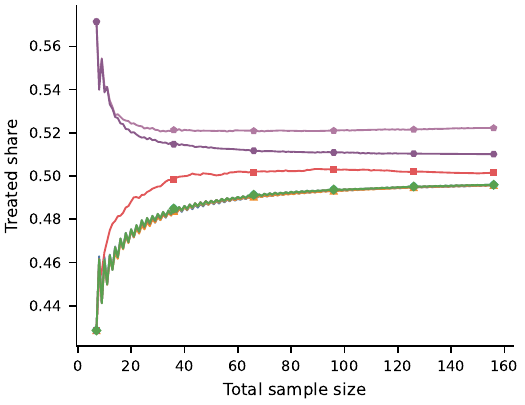}
\caption{Absolute operating characteristics in the degenerate randomized
configuration (companion to Figure~\ref{fig:mainK3}). False positive rate
(top left), true positive rate (top right), and proportion of patients
allocated to treatment (bottom) as functions of the total sample size
($K = 3$, adaptive recruitment, randomized assignment); the legend in the
first panel applies to all three. Lines are means across the $50$
environments, each averaged over $20$ replications, with the one-sided
environments included under the zero-rate convention of
Table~\ref{tab:mainK3}; the spread across environments is in that table
and in Figure~\ref{fig:scatter}. The curves start at $2K + 1 = 7$, after
the balanced burn-in of $2K = 6$ samples; \textbf{(1c)} and \textbf{(2c)}
coincide with \textbf{(1b)} and \textbf{(2b)} in this configuration and
are not drawn, nor is the upper-quartile selector. The constrained
individual-level arms hold the treated
share at one half ($0.496$); the optimal transport arm shifts it to
$0.522$ at similar error rates.}
\label{fig:mainK3abs}
\end{figure}

Table~\ref{tab:mainK3} and Figures~\ref{fig:mainK3} and~\ref{fig:mainK3abs}
give the operating characteristics, the paired-difference trajectories,
and the absolute trajectories of the configuration of
Section~6.1 ($K = 3$, adaptive recruitment, randomized
assignment).

\emph{Attribution.} With the linear fluctuation the targeting step and the
one-step AIPW recombination nearly coincide under randomization
($+0.002 \pm 0.001$ AUC). Cross-fitting, applied wherever both training
folds can support the nuisance fits (Section~\ref{sec:supp-sim}), costs
$-0.001 \pm 0.001$ AUC, $+1.3 \pm 0.4$ points of FPR, and $+0.1 \pm 0.3$
of TPR against \textbf{(2b)}: at about $26$ patients per subpopulation and
arm neither the recombination, own-sample fitting, nor the halved
nuisance sample moves the operating characteristics by more than about a
point.
Within the sample mean family the oracle mean-squared-error weighting
cannot act, since the weights are $e_{k}$ for every variant.

\begin{table}[!htbp]
\centering
\caption{Operating characteristics in the degenerate randomized
configuration ($K = 3$, adaptive recruitment, randomized assignment).}
\label{tab:mainK3}
\footnotesize
\setlength{\tabcolsep}{4pt}
\resizebox{\ifdim\width>\linewidth\linewidth\else\width\fi}{!}{%
\begin{tabular}{@{}lcccc@{}}
\toprule
& \multicolumn{2}{c}{Mid-trial ($n = 81$)}
& \multicolumn{2}{c}{End of trial ($n = 156$)} \\
\cmidrule(lr){2-3} \cmidrule(lr){4-5}
Estimator arm & FPR & TPR & FPR & TPR \\
\midrule
Sample mean, subpopulation (1a) & 8.9 (13.8) & 78.7 (29.0) & 6.1 (12.5) & 81.7 (29.5) \\
Sample mean, individual (1b) & 13.2 (14.8) & 73.0 (28.3) & 9.6 (12.9) & 76.6 (29.2) \\
Sample mean, MSE-optimal oracle (1c) & 13.2 (14.8) & 73.0 (28.3) & 9.6 (12.9) & 76.6 (29.2) \\
Doubly robust, subpopulation (2a) & 11.3 (15.5) & 75.8 (28.7) & 9.6 (14.4) & 78.7 (28.7) \\
Doubly robust, individual (2b) & 13.7 (16.0) & 73.3 (28.3) & 9.8 (13.8) & 77.8 (29.2) \\
Doubly robust, MSE-optimal oracle (2c) & 13.7 (16.0) & 73.3 (28.3) & 9.8 (13.8) & 77.8 (29.2) \\
AIPW at (2b) weights & 13.6 (16.0) & 73.2 (28.2) & 9.8 (13.7) & 77.8 (29.2) \\
Cross-fitted TMLE at (2b) weights & 15.0 (16.8) & 73.0 (27.9) & 11.1 (15.0) & 77.8 (28.9) \\
Pre-specified selector, median rule & 12.9 (14.8) & 73.2 (28.0) & 9.2 (12.6) & 77.6 (29.1) \\
Pre-specified selector, upper-quartile rule & 13.5 (14.6) & 73.6 (28.3) & 9.5 (12.5) & 77.1 (29.0) \\
OT relaxation (Section~\ref{sec:ot}) & 9.6 (15.1) & 78.7 (29.3) & 7.8 (14.6) & 80.7 (29.8) \\
OT relaxation of the oracle (1c) & 8.3 (13.4) & 79.7 (29.2) & 6.2 (12.4) & 82.5 (29.6) \\
\bottomrule
\end{tabular}}
\par\vspace{4pt}
\begin{minipage}{\linewidth}\footnotesize
NOTE: FPR
and TPR in percent, mean (standard deviation) across the $50$ environments,
each averaged over $20$ replications, at total sample sizes $81$
(mid-trial) and $156$ (end of trial). The rates are averaged over all $50$
environments with the convention that an environment without a positive
(negative) subpopulation contributes a true positive (false positive) rate
of zero. Five environments have no positive and six have no negative
subpopulation, which is the source of the large standard deviations across
environments; over the $45$ environments with a positive subpopulation
the end-of-trial TPR of \textbf{(1a)} is $90.8\%$. The standard deviations
measure the diversity of the environments and this convention, not the
uncertainty of a comparison. Every contrast quoted in the text is a paired
difference on common seeds, with standard errors one to two orders of
magnitude smaller (Figure~\ref{fig:forest}). The two rows of each within-family pair are identical because borrowing is inert; the summary-level arms and the transport relaxations run on the subpopulation-level pool and the individual-level arms on the individual-level pool (Section~\ref{sec:levels}), so comparisons across levels in this table compare data regimes as well as estimators.
\end{minipage}
\end{table}

\FloatBarrier
\subsection{Confounded Configurations at \texorpdfstring{$K = 3$}{K = 3}}
\label{sec:supp-confK3}
\begin{figure}[!htbp]
\centering
\includegraphics[width=\linewidth]{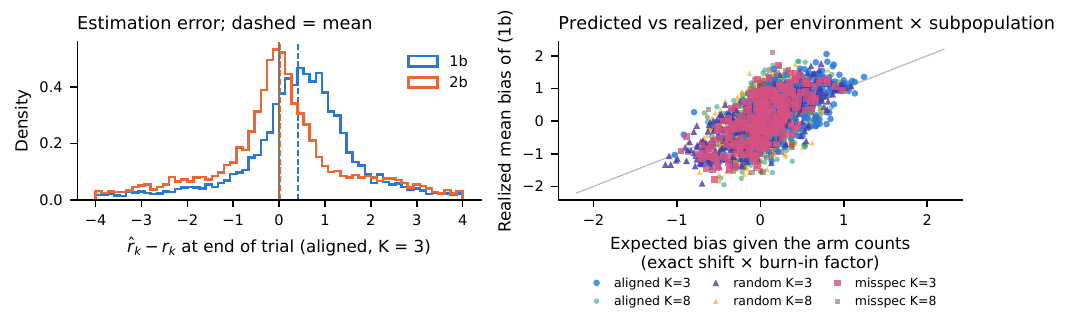}
\caption{Confounding bias. Left: distribution of the estimation error
$\hat r_{k} - r_{k}$ at the end of the trial over environments,
replications, and subpopulations in the aligned $K = 3$ scenario, for the
sample mean arm \textbf{(1b)} and the doubly robust arm \textbf{(2b)};
the dashed lines mark the means (densities over the range $\pm 4$).
Right: realized mean bias of
\textbf{(1b)} for each environment and subpopulation against the expected
bias of the no-borrowing contrast given the realized arm counts, the
between-arm covariate shift of Proposition~3 (with $c_{k}$
the shift coefficient) times the burn-in dilution factor
$(1 - 1/n_{k}^{(1)})(1 - \pi_{k}) + (1 - 1/n_{k}^{(0)})\pi_{k}$ (the
forced burn-in patient of each arm is drawn without regard to the
propensity), computed from the stored environment parameters, the
assignment mechanism, and the realized counts, for the aligned, randomly
oriented, and misspecified scenarios at both $K$; the identity line is
drawn. At $K = 3$ the arm does not borrow; at $K = 8$ it does, so its
realized bias also contains the representation bias and the borrowed
confounding of~(6), which the horizontal axis omits.
Over all replications the correlation is $0.67$--$0.76$ at $K = 3$ and $0.40$--$0.54$ at $K = 8$; restricted to the replications with at least five patients in each arm (cells with at least ten such replications: $212$, $214$, and $189$ of $300$ at $K = 3$; $80$, $94$, and $41$ of $400$ at $K = 8$) it is $0.94$--$0.97$ and $0.92$--$0.97$, with slopes $1.01$--$1.06$ and $0.94$--$1.04$ through the origin. The departures come from the
subpopulations that the sensitivity-based recruitment, which depends on
the outcomes, stops recruiting once their estimate has moved away from
zero, so that the estimate is frozen at few patients (the
adaptive-sampling distortion of the remark after
Assumption~\ref{ass:counts}). The aligned scenarios have biases of one sign on average (in $240$ of $300$ subpopulations at $K = 3$ and $275$ of $400$ at $K = 8$ the alignment is positive) and the randomly oriented scenarios scatter around
zero, which is the cancellation-versus-accumulation mechanism of
Corollary~2 in the data.}
\label{fig:bias}
\end{figure}

Figure~\ref{fig:bias} shows the confounding bias itself in the scenarios of
Section~6.2; the two paragraphs that follow give the full
comparator and selector results of those scenarios.

\begin{figure}[!htbp]
\centering
\includegraphics[width=0.46\linewidth]{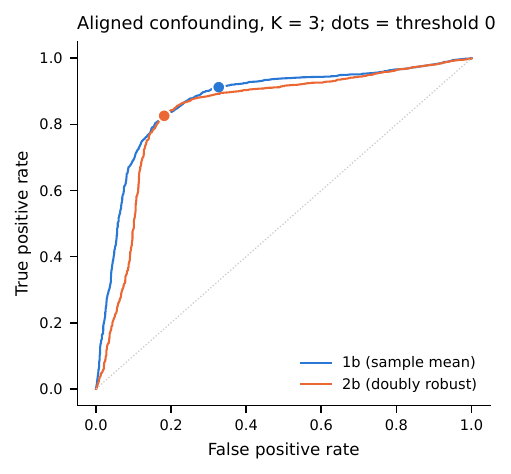}
\caption{Receiver operating characteristic curves of \textbf{(1b)} and
\textbf{(2b)} in the aligned $K = 3$ scenario (rates at common
thresholds averaged over the $80$ environments in which both classes occur, so that the zero-threshold operating points lie on the curves), with the zero-threshold operating point of each arm on the same environments marked. Aligned confounding moves the operating point of the sample mean arm up along its own curve (to false positive rate $0.33$ and true positive rate $0.91$, against $0.18$ and $0.83$ for the doubly robust arm, on the same $80$ environments) rather than onto a better curve; the part of the bias that differs between the subpopulations of an environment also lowers the arm's AUC, by $0.029 \pm 0.010$ per environment on average in this scenario (Section~\ref{sec:supp-probes}). Raw rates at unequal
false positive rates therefore do not compare the families.}
\label{fig:roc}
\end{figure}

\emph{Attribution under confounding.} Once the propensities are
nontrivial, the one-step recombination and the targeting step separate.
AIPW exceeds targeted maximum likelihood at the same weights by $+0.019 \pm 0.003$ AUC in the aligned scenario, with $-0.9 \pm 0.3$ points of FPR, and by $+0.017 \pm 0.002$ and $+0.017 \pm 0.003$ in the random and misspecified scenarios. Cross-fitting, applied in $50$--$60\%$ of the
subpopulation cells at the horizon, with the estimator of \textbf{(2b)}
elsewhere (the full-sample fit, or the sample means in the fallback
cells), costs discrimination in all three scenarios
($-0.018 \pm 0.003$, $-0.028 \pm 0.004$, and $-0.010 \pm 0.003$ AUC; $-2.8 \pm 0.6$, $-3.6 \pm 0.5$, and $-2.1 \pm 0.5$ points of TPR against \textbf{(2b)}).

\emph{Selector.} Step (iii) evaluates the crossing condition with the
squared anticipated worst case band $|c_{k}|\,\|\alpha\|\,\|w_{T,k}\|$ (the
band is about $0.9$ in outcome units at $\|\alpha\| = 2$, against a
realized band of $0.62$ in the aligned scenario) against the net variance
term $s(\kappa\sigma^{2}_{k,\mathrm{cond}} - E_{k})$ of the design stage
confounded curve at $\bar n = 26$, about $0.2$, so that a band above
about $0.45$ decides, and both rules route $99$ of the $100$ environments of the aligned and random scenarios to the doubly robust family. In the misspecified scenario the band is the worst case  over the alignment of the exact between-arm shift of the quadratic mechanism ($0.1$--$1.4$), and the median rule retains the sample mean family in $38$ environments (all but one of them among the $38$ smallest worst case bands), the upper-quartile rule in $44$. The median-rule arm is therefore \textbf{(2b)} in all but one environment of the aligned and random scenarios, $-0.026 \pm 0.014$ AUC below \textbf{(1b)} in the aligned scenario ($1.8$ standard errors) and $-0.006 \pm 0.012$ below in the random one, within error of it at matched false positive rate under aligned confounding ($-1.8 \pm 2.5$ points over the $80$ environments with a defined AUC) and ahead of it there by $9.4 \pm 3.2$ points under random orientation; in the misspecified scenario it is a mixture, $-0.020 \pm 0.009$ below \textbf{(1b)}, and the upper-quartile arm is $-0.019 \pm 0.008$ below it. A variance-only step (iii) with the
design stage curve has an infinite threshold in every environment; with
the design-embedded confounded curve it would have found a finite
threshold in six of the first ten environments, where the
explained-variance edge exceeds that curve's inflation term.

\FloatBarrier
\subsection{The Nondegenerate Scenarios: \texorpdfstring{$K = 8$}{K = 8}}
\label{sec:supp-K8}
\begin{figure}[!htbp]
\centering
\includegraphics[width=\linewidth]{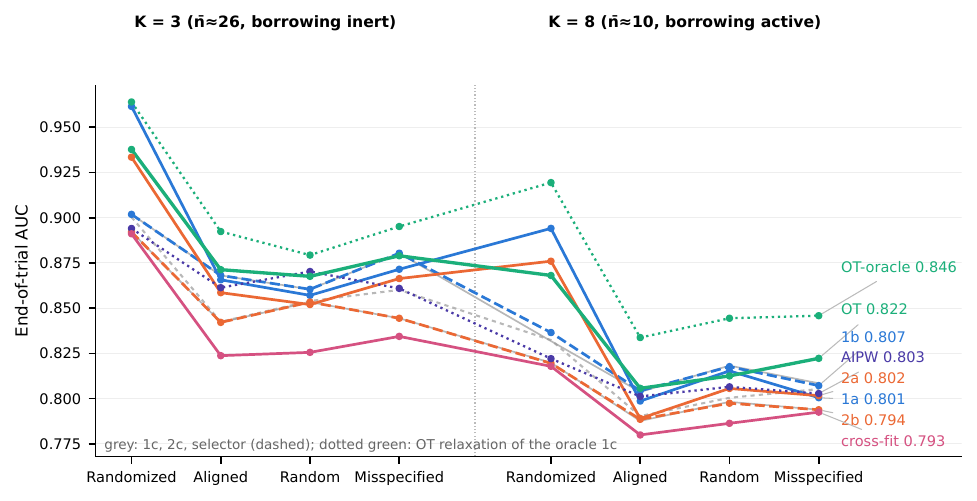}
\caption{AUC at the end of the trial for all arms but the upper-quartile
selector across the eight scenarios (Table~\ref{tab:auc} drawn as a
slopegraph). The implementable optimal transport relaxation is the top nonoracle point in the aligned columns at both $K$ and in the misspecified $K = 8$ column (in the random and misspecified $K = 3$ columns the AIPW comparator and \textbf{(1b)} are above it by $0.003$ and $0.001$), and below \textbf{(1a)} in the two randomized columns (a cross-pool comparison there, Section~\ref{sec:levels}); the oracle relaxation (dotted) is the top point in
every column; the cross-fitted comparator lies at or below
\textbf{(2b)} in every column, by at most $0.028$. The selector (dashed grey, median rule) coincides with \textbf{(2b)} in all but one environment of the aligned and random $K = 3$ columns and is a mixture of the two families in the other six columns. Gaps of $0.01$--$0.03$ within a column are significant in some
scenarios and not in others; the paired contrasts with their standard
errors are given in Figure~\ref{fig:forest} and Table~2.}
\label{fig:slope}
\end{figure}

Figure~\ref{fig:slope} draws the AUC of every arm but the upper-quartile
selector in the eight core scenarios (Table~\ref{tab:auc}) as a
slopegraph. At the horizon of the
$K = 8$ scenarios of Section~6.3 the fallback rule applies to
$35\%$ of the subpopulation cells of \textbf{(2b)} in the randomized
scenario and to $39\%$, $37\%$, and $48\%$ in the aligned, random, and
misspecified scenarios. The cross-fitted comparator applies the split in
$33\%$ of the cells under randomization and the estimator of
\textbf{(2b)} elsewhere (the full-sample fit in $32\%$ of the cells and
the sample means in the $35\%$ on the fallback), and costs
$-0.002 \pm 0.001$ AUC there, $-0.009 \pm 0.002$
under aligned and $-0.011 \pm 0.002$ under random confounding,
$-0.001 \pm 0.001$ under the misspecified mechanism, and
$-0.040 \pm 0.006$ under uniform recruitment, where every subpopulation is
split; AIPW exceeds the targeted step by $+0.003 \pm 0.000$ AUC under
randomization and by $+0.009$ to $+0.013$ ($\pm 0.001$) under
confounding. Within the sample mean family the oracle mean-squared-error
weighting does not improve discrimination with borrowing active:
\textbf{(1b)} exceeds \textbf{(1c)} by $+0.005 \pm 0.002$ AUC under
randomization. The selector arm is $+0.013 \pm 0.003$ AUC above the
doubly robust arm under randomization and $+0.002 \pm 0.002$,
$+0.003 \pm 0.002$, and $+0.011 \pm 0.005$ above it in the aligned,
random, and misspecified scenarios; against \textbf{(1b)} the
upper-quartile rule is $-0.009 \pm 0.004$, $-0.010 \pm 0.005$, and
$-0.001 \pm 0.001$ there.

\FloatBarrier
\subsection{Probes Along the Coordinates of the Regime Map}
\label{sec:supp-probes}

\begin{table}[!htbp]
\centering
\caption{Pre-specified predictions for the probes, the basis of each, and
their outcomes (first part; the second part continues the table with the same columns).}
\label{tab:predictions}
\footnotesize
\setlength{\tabcolsep}{3pt}
\resizebox{\ifdim\width>\linewidth\linewidth\else\width\fi}{!}{%
\begin{tabular}{@{}>{\raggedright\arraybackslash}p{0.10\linewidth}>{\raggedright\arraybackslash}p{0.18\linewidth}>{\raggedright\arraybackslash}p{0.13\linewidth}>{\raggedright\arraybackslash}p{0.53\linewidth}@{}}
\toprule
Probe & Predicted direction & Basis & Outcome \\
\midrule
$K = 8$, $H = 400$, randomized & no crossing at $\bar n^{(a)} \approx 26$ with borrowing active; gap narrower than at $\bar n^{(a)} \approx 10$ & $n^{*}(0) > 26$ from the $K = 3$ cells; Theorem~1(i) & gap narrows from $+0.031 \pm 0.007$ ($H = 150$, same ten environments) to $+0.011 \pm 0.006$ (narrower by $0.020 \pm 0.009$, $2.3$ standard errors), no AUC crossing on the adaptive path; matched FPR $-3.6 \pm 5.5$: as predicted under the pre-specified basis; basis superseded, since the re-estimated $n^{*}(0) \approx 7$--$9$ places $\bar n^{(a)} \approx 26$ beyond the boundary at fixed counts, and the absence of a crossing is attributed to the recruitment path (Section~6.4) \\
$K = 8$, $H = 400$, aligned & no crossing at $\bar n^{(a)} \approx 26$ with borrowing active; gap narrower than at $\bar n^{(a)} \approx 10$ & as above & AUC gap $+0.007 \pm 0.013$ against $+0.015 \pm 0.014$ at $H = 150$ on the same ten environments (narrower by $0.008 \pm 0.012$, $0.65$ standard errors: the predicted direction, within error); no AUC crossing; FPR of \textbf{(1b)} $14.1 \pm 3.6$ points higher; matched FPR $-8.5 \pm 4.2$ ($2.0$ standard errors): the predicted ``no crossing'' fails in the co-primary metric, where the doubly robust arm is ahead (the confounded crossing of the text below) \\
Uniform recruitment, $K = 8$ & gap narrows as the decisive subpopulations grow; ordering reverses only if $\bar n$ exceeds $n^{*}$ & recruitment moves $s(n)$ only (Figure~\ref{fig:dag}) & the predicted narrowing holds ($-0.040 \pm 0.010$ against the same ten adaptive environments, $3.8$ standard errors); the gap reverses at $\bar n^{(a)} \approx 10$ ($-0.009 \pm 0.005$, $1.8$ standard errors): with the pre-specified $n^{*}(0) > 26$ no reversal was predicted at $10$, so the basis of the second clause is superseded and that clause is undecided at two standard errors; against the re-estimated $n^{*}(0) \approx 9$ the reversal is in the predicted direction \\
$\sigma_{x} = 2$ ($\sigma_{z} = 1$) & sample mean margin widens & $\kappa$ and $\tilde\lambda$ both inflate; Theorem~1(iv) & $+0.038 \pm 0.015$ against $+0.031 \pm 0.007$ on the same ten environments; matched FPR $+0.4 \pm 3.5$: direction as predicted, the widening within error \\
\bottomrule
\end{tabular}}
\end{table}

\begin{table}[!htbp]
\centering
\ContinuedFloat
\caption{Pre-specified predictions for the probes, the basis of each, and
their outcomes (second part, continued; the columns are those of the first part).}
\footnotesize
\setlength{\tabcolsep}{3pt}
\resizebox{\ifdim\width>\linewidth\linewidth\else\width\fi}{!}{%
\begin{tabular}{@{}>{\raggedright\arraybackslash}p{0.10\linewidth}>{\raggedright\arraybackslash}p{0.18\linewidth}>{\raggedright\arraybackslash}p{0.13\linewidth}>{\raggedright\arraybackslash}p{0.53\linewidth}@{}}
\toprule
Probe & Predicted direction & Basis & Outcome \\
\midrule
Estimated variances in the weighting & sample mean margin widens & oracle variances are the best case for every arm & $+0.029 \pm 0.008$ (randomized; $+0.031$ with oracle variances on the same ten environments) and $+0.006 \pm 0.012$ (aligned; $+0.015$): not confirmed \\
Mismatch pairs & matched pairs beat mismatched pairs by a margin increasing in the heterogeneity of $\sigma^{2}_{k,\mathrm{eff}}/\sigma^{2}_{k,\mathrm{cond}}$ (Figure~\ref{fig:rho}) & Proposition~5 & recruit-2b/analyze-1b $-0.012 \pm 0.004$ below matched \textbf{(1b)}; recruit-1b/analyze-2b $+0.034 \pm 0.009$ above matched \textbf{(2b)} on the ten environments ($+0.019 \pm 0.004$ on all $50$ of the randomized $K = 8$ scenario, and above matched \textbf{(2b)} in every core scenario, Section~\ref{sec:supp-mismatch}): confirmed for the sample mean analysis, contradicted for the doubly robust analysis ($4.0$ standard errors); the dependence on the heterogeneity was not tested (two pairs on ten environments) \\
Anchor $r = (-10, 0, +10)$ & all variants at $\mathrm{TPR} = 1$, $\mathrm{FPR} \approx 25\%$ & symmetry at $r_{k} = 0$ & every arm at $\mathrm{TPR} = 100\%$, FPR $21.5$--$25.8\%$: as predicted \\
$\kappa_{k}(\bar n)$ curves & single sign change of $B(\bar n) = \kappa_{k}(\bar n)\sigma_{k,\mathrm{cond}}^{2} - E_{k}$; $n^{*}(0) > 26$ for the evaluated targeted estimator & Assumption~\ref{ass:inflation} & single sign change of the isotonic curves, which the isotonic fit imposes (the raw curves are not monotone at strengths $1.5$--$3$, Section~\ref{sec:supp-map}, and the raw medians of the design-embedded confounded and $\sigma_{x} = 2$ curves and the raw upper quartile of the randomized embedded curve each cross the edge four times; with the raw medians the predicted difference changes sign at most once in every record group at strengths $0$--$2$, Section~\ref{sec:supp-map}); $n^{*}(0) > 26$ contradicted: for the linear fluctuation that replaced the bounded-outcome one, $n^{*}(0) = 7$ at fixed sample sizes and $9$ within the design under the median rule (the upper-quartile threshold within the design, $29$, is the one estimate above $26$), and the bounded form's own design stage curve gives $n^{*}(0) \approx 15$ (Section~\ref{sec:supp-sens}) \\
\bottomrule
\end{tabular}}
\par\vspace{4pt}
\begin{minipage}{\linewidth}\footnotesize
NOTE: Family contrasts are \textbf{(1b)} minus \textbf{(2b)} in AUC at
the end of the trial with the paired standard error; ``matched FPR'' is
the true positive rate of \textbf{(1b)} minus that of \textbf{(2b)}, both
re-thresholded to the zero-threshold false positive rate of
\textbf{(2b)}, in percentage points.
\end{minipage}
\end{table}
The predicted directions and their bases were specified before the runs reported here, on the basis of preliminary adaptive runs that used the bounded-outcome fluctuation and, since no variance inflation curve had yet been estimated, a selector with the placeholder $\kappa \equiv 0.5$; the basis $n^{*}(0) > 26$ was inferred from those runs. The two $H = 400$ probes were specified jointly with two predictions, which the table repeats in each of their rows. Each outcome is scored against the pre-specified basis; the rows whose basis the re-estimated curve superseded, the three predictions the data contradict (the aligned $H = 400$ ``no crossing'' in the co-primary metric, the threshold $n^{*}(0) > 26$ of the $\kappa$ row, and the mismatch prediction for the doubly robust analysis), and the one they do not confirm (the widening under estimated variances) say so; two rows, the randomized $H = 400$ and the uniform-recruitment probes, read differently against the pre-specified basis and against the re-estimated threshold. The paragraphs that follow give
the full account of the probes summarized in Section~6.4 and
of the random orientation scenarios of Section~6.5.

\emph{Longer horizon.} With $H = 400$ the $K = 8$ scenarios reach
$\bar n^{(a)} \approx 26$ with borrowing active, which is the point with
active borrowing at moderate sample sizes that the $K = 3$ scenarios could
not provide. Under randomization the family gap narrows from $+0.031 \pm 0.007$ at
$H = 150$ on the same ten environments ($+0.017 \pm 0.004$ over all $50$)
to $+0.011 \pm 0.006$ AUC for \textbf{(1b)} against \textbf{(2b)} and
grows from $+0.023 \pm 0.007$ ($+0.018 \pm 0.003$ over all $50$) to
$+0.041 \pm 0.010$ for \textbf{(1a)} against \textbf{(2a)}; at matched false positive rate the doubly robust arm is favored by $3.6 \pm 5.5$ points, within error. The adaptive path remains concentrated,
with the most recruited subpopulation holding about $117$ patients per
arm and the least recruited about two, so the mean of $26$ is reached by
few subpopulations, and the gap is the one the $K = 3$ scenario has at
the same mean over the $39$ of its $50$ environments with a defined AUC
($+0.010 \pm 0.008$). Under aligned confounding the two individual-level families are within error in AUC ($+0.007 \pm 0.013$), while the
false positive rate of the sample mean arm is $14.1 \pm 3.6$ points higher
($30.8\%$ against $16.7\%$). At the false positive rate of
the doubly robust arm, its true positive rate is $8.5 \pm 4.2$ points
lower ($2.0$ standard errors). This is the confounded crossing of
Theorem~1 in the co-primary metric at this horizon: the
$\mathcal{O}(1)$ bias term has overtaken the variance terms, and the bias
removal of the doubly robust family shows in the operating
characteristics, as the partition of the theorem predicts. The
AIPW comparator is the individual-level arm with the best discrimination
in that scenario ($0.837$ against $0.829$ for \textbf{(1b)}), and the
oracle relaxation leads all arms ($0.872$).

\emph{Uniform recruitment.} Replacing the adaptive rule by a balanced
schedule at $K = 8$ reverses the family gap ($-0.009 \pm 0.005$ AUC for
\textbf{(1b)} against \textbf{(2b)}, $1.8$ standard errors, with the point estimate favoring the doubly robust arm; $+0.001 \pm 0.001$ at the subpopulation-level) at the same
$\bar n^{(a)} \approx 10$ at which the adaptive scenario shows
$+0.017 \pm 0.004$ ($+0.031 \pm 0.007$ on the same ten environments), and
it raises the discrimination of every arm (\textbf{(1a)} $0.934$ against
$0.896$, \textbf{(1b)} $0.887$ against $0.834$, \textbf{(2b)} $0.896$
against $0.803$, the implementable relaxation $0.924$ against $0.888$ on
those environments). The zero-threshold rates of \textbf{(1a)} move the other way (false positive rate $11.8\%$ against $9.9\%$, true positive rate $85.4\%$ against $87.2\%$), while both rates of \textbf{(2a)} improve.
Adaptive recruitment spends the budget on the subpopulations whose signs
are least certain, which improves the decisions in those subpopulations
and worsens the ranking across subpopulations. With every subpopulation
at $\bar n^{(a)} \approx 10$, just beyond the median-rule threshold, the
doubly robust family is favored within simulation error, as the crossing condition implies for a sample size just beyond the threshold, and the
adaptive gap in the other direction, $3.6$ times as large on the same
ten environments, is the recruitment-path effect of
Section~6.6, which
uniform recruitment switches off by construction.

\emph{Covariate noise and estimated variances.} Doubling $\sigma_{x}$ with
$\sigma_{z}$ held at one quadruples the explained-variance edge ($E_{k}$
averages $1.95$ against $0.49$ on the evaluation environments) and raises
the design-embedded variance inflation with
it ($\kappa \approx 1.44$ against $0.36$ at $\bar n^{(a)} = 10$;
Section~\ref{sec:supp-inflation}), so the crossing condition is nearly
unchanged. In the data both individual-level families lose discrimination
(\textbf{(1b)} $0.795$, \textbf{(2b)} $0.757$) and the gap widens slightly,
to $+0.038 \pm 0.015$ from $+0.031 \pm 0.007$ on the same ten
environments, with the two arms at matched false positive rate
within error of each other ($+0.4 \pm 3.5$ points), while the
summary-level arms and the transport relaxations, which do not see the
individual covariates, are unchanged by construction. Replacing the oracle
variances supplied to the weighting and the recruitment index by running
estimates (within-arm sample variances for the sample mean family,
influence curve variances and covariances for the doubly robust family;
the rates $\tilde\lambda_{k}$ stay oracle and the cross-covariance term is
dropped) costs both families discrimination (\textbf{(1b)} $0.813$, \textbf{(2b)}
$0.785$ under randomization) and leaves the gap where it was on the same
ten environments ($+0.029 \pm 0.008$ against $+0.031 \pm 0.007$). Under
aligned confounding the gap is $+0.006 \pm 0.012$ against
$+0.015 \pm 0.014$, with the point estimate at matched false positive rate favoring the doubly robust arm within error ($4.1 \pm 4.5$ points), and the
mean-squared-error--optimal doubly robust variant \textbf{(2c)} is the
better doubly robust arm ($0.796$ against $0.779$). The oracle variances
are the best case for every arm except \textbf{(2c)} under aligned
confounding, and the estimated variances change neither the sign nor the size of the family contrast. At the design level the doubly robust family has the larger mean squared error in these probes as well ($+0.93 \pm 0.19$ with $\sigma_{x} = 2$; $+0.65 \pm 0.20$ and $+1.19 \pm 0.57$ with estimated variances under randomization and aligned confounding), as in the two $H = 400$ probes ($+0.68 \pm 0.15$ and $+1.07 \pm 0.18$) and in contrast to uniform recruitment ($-0.065 \pm 0.007$, Section~6.4).

\emph{Anchor.} In the well-separated configuration with
$r = (-10, 0, +10)$, every arm declares the true positive in every
replication ($\mathrm{TPR} = 100\%$) and the zero-effect subpopulation in
about half of them, which gives false positive rates of $21.5$--$25.8\%$
across the twelve arms against the $25\%$ that symmetry at $r_{k} = 0$
implies, with paired family differences within one to two points and no
difference in discrimination. The anchor configuration thus behaves as
expected.

\emph{The two sides of the confounded AUC contrast.} The AUC of an environment is invariant to a shift common to its subpopulations and responds only to the part of the bias that differs between them; in the simulated scenarios that part is as large as the common part or larger (over the six confounded core scenarios the within-environment standard deviation of the predicted bias $c_{k}\alpha^{\top} w_{T,k}$ of \textbf{(1b)} is $0.38$--$0.59$ against an environment-mean magnitude of $0.14$--$0.47$, ratios of $0.9$ to $3.2$, the larger at $K = 8$). Removing the predicted bias from the stored estimates of \textbf{(1b)} (a script of the accompanying code) raises its AUC by $0.029 \pm 0.010$ and $0.034 \pm 0.010$ in the aligned and random $K = 3$ scenarios, $0.022 \pm 0.005$ and $0.014 \pm 0.006$ at $K = 8$, and $0.023 \pm 0.011$ in the aligned $H = 400$ probe, and leaves it within $0.007 \pm 0.007$ and $0.001 \pm 0.005$ of zero in the misspecified scenarios (whose within-environment bias spread, $0.38$ and $0.46$, is only slightly below the aligned scenarios' $0.44$ and $0.57$, so the size of the bias does not explain the difference); removing the environment mean of the bias alone changes no AUC. The stored family contrasts are therefore nets of a variance side and a bias side: with \textbf{(1b)} de-biased the contrast would be $+0.055 \pm 0.009$, $+0.041 \pm 0.007$, $+0.038 \pm 0.005$, and $+0.034 \pm 0.005$ (aligned $K = 3$, random $K = 3$, aligned $K = 8$, random $K = 8$) against the stored $+0.026$, $+0.007$, $+0.016$, and $+0.020$ (the split is order-dependent: the bias added to the estimates of \textbf{(2b)} instead would cost $0.017$, $0.017$, $0.015$, and $0.006$ AUC, about half), and at matched false positive rate the aligned $K = 3$ verdict would turn from $-1.8 \pm 2.5$ to $+4.6 \pm 1.5$ points. With the mismatch arm recruit-1b/analyze-2b of Section~\ref{sec:supp-mismatch}, run on every core environment, the variance side itself splits into the estimator on identical patients and the recruitment path: $+0.019$ and $+0.036$ (aligned $K = 3$), $+0.022$ and $+0.019$ (random $K = 3$), $+0.022$ and $+0.016$ (aligned $K = 8$), and $+0.019$ and $+0.015$ (random $K = 8$), so the path is $43$--$65\%$ of it. The matched-FPR comparison is itself invariant to a common shift, so neither measure exposes a bias common to the subpopulations of an environment; the zero-threshold rates and the per-environment errors do.

\emph{Random orientation.} Corollary~2 predicts that randomly oriented confounding contributes no directional family gap in the aggregate at first order even though every environment has a bias. The pooled zero-threshold rates of the random orientation scenarios show no directional shift: \textbf{(1b)} minus \textbf{(2b)} is $+3.2 \pm 1.9$ and $+0.4 \pm 1.1$ points of false positive rate at $K = 3$ and $K = 8$, against $+13.1 \pm 2.0$ and $+5.7 \pm 1.1$ under aligned confounding, and $-5.3 \pm 2.1$ and $-0.3 \pm 1.2$ points of true positive rate. These pooled contrasts are nets as well, like the AUC contrasts above: for a nonresponsive subpopulation the probability of a positive estimate is convex in the bias, so a sign-random bias raises the sample mean's false positive rate and lowers its true positive rate at second order, and with the predicted bias removed from its estimates the contrasts would be $-2.9 \pm 0.8$ and $-2.5 \pm 0.7$ points of false positive rate and $+2.2 \pm 0.8$ and $+2.4 \pm 0.7$ of true positive rate, in favor of the sample mean; the sign-random bias therefore costs it $6.1$ and $2.9$ points of false positive rate and $7.6$ and $2.7$ of true positive rate, which the doubly robust family's inflation offsets. The AUC gaps tell the same story: at $K = 8$ the stored gap under random
orientation ($+0.020 \pm 0.007$) is close to the randomized gap
($+0.017 \pm 0.004$), but with \textbf{(1b)} de-biased it would exceed it by $0.017 \pm 0.006$, and at $K = 3$ both are within error of zero ($+0.007 \pm 0.012$ and $+0.010 \pm 0.008$); the comparison at matched false positive rate is within error of zero at $K = 8$ ($-1.7 \pm 1.6$ points; $+3.2 \pm 1.0$ with \textbf{(1b)} de-biased) and favors the doubly robust arm at $K = 3$ ($-9.4 \pm 3.2$ points, $3.0$ standard errors, Section~6.2), whereas the aligned
scenarios show a directional deficit for the sample mean family in the
false positive rate.

\FloatBarrier
\subsection{Recruitment--Analysis Consistency}
\label{sec:supp-mismatch}
\begin{figure}[!htbp]
\centering
\includegraphics[width=0.5\linewidth]{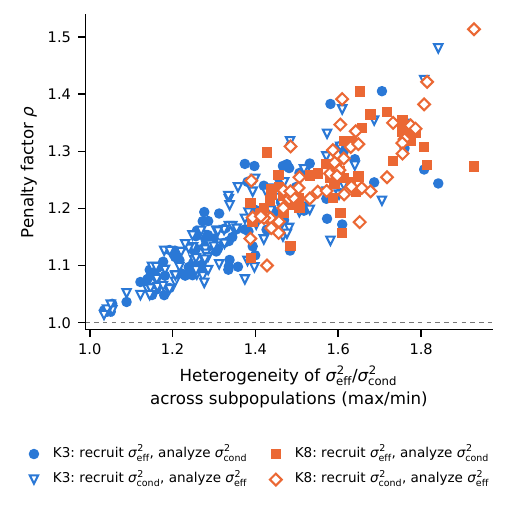}
\caption{The closed-form mismatch penalty $\rho(F \to G)$ of~(12), computed from the stored variance profiles of the core environments at each $K$ ($100$ at $K = 3$, $50$ at $K = 8$), against the heterogeneity of the ratio
$\sigma^{2}_{k,\mathrm{eff}} / \sigma^{2}_{k,\mathrm{cond}}$ across
subpopulations (largest over smallest). The penalty is one only for
homogeneous ratios and averages $1.16$--$1.26$ in these environments
($1.26$ and $1.28$ on the first ten $K = 8$ environments, those of the
mismatch pairs), whose ratios differ between subpopulations by a factor
of up to $1.8$ ($K = 3$) and $1.9$ ($K = 8$).}
\label{fig:rho}
\end{figure}

Crossing the recruitment variance functional and the analysis family
means recruiting with the marginal profile $\{\sigma_{k,\mathrm{eff}}^{2}\}$
but analyzing with the doubly robust family, or the reverse. The two
mismatched pairs at $K = 8$ under randomization, run on the seeds of
Section~6.3 and with the same variance inflation as the matched
doubly robust arm, give an asymmetric answer. Recruiting with the doubly robust family
and analyzing with the sample mean is $-0.012 \pm 0.004$ AUC below the
matched sample mean arm ($+2.4 \pm 0.6$ points of false positive rate,
$-1.3 \pm 1.1$ of true positive rate), which is the direction that
Proposition~5 anticipates for a penalty factor of
$1.28$ on the ten mismatch environments ($1.26$ for the reverse pair;
Figure~\ref{fig:rho}). Recruiting with the sample mean
family and analyzing with the doubly robust family, however, beats the matched doubly robust arm by $+0.034 \pm 0.009$ AUC ($-2.2 \pm 1.2$
points of false positive rate, $+0.7 \pm 0.8$ of true positive rate), and
its discrimination, $0.837$, equals that of the matched sample mean arm on
the identical patients ($0.834$ on those ten environments; $+0.003 \pm 0.005$). The same-path
runs of Section~6.5 repeat the pattern on the ten
held-out environments: the doubly robust analysis of the
data of the sample mean family reaches $0.935$, $0.856$, $0.874$, $0.854$,
and $0.815$ in the randomized $K = 3$ and $K = 8$, aligned $K = 3$ and
$K = 8$, and $\sigma_{x} = 2$ configurations, against $0.932$, $0.857$,
$0.848$, $0.856$, and $0.802$ for the sample mean analysis of the same
data, with false positive rates $26$ and $8$ points lower under aligned
confounding at $K = 3$ and $K = 8$. The family gap of the adaptive
scenarios in discrimination is therefore a recruitment-path effect, and the mismatch arm recruit-1b/analyze-2b, run on every environment of the eight core scenarios for the selector of Section~\ref{sec:supp-inflation}, repeats it in every core scenario: it is above the matched doubly robust arm by $+0.011 \pm 0.008$, $+0.036 \pm 0.008$, $+0.019 \pm 0.007$, and $+0.030 \pm 0.007$ AUC in the randomized, aligned, random, and misspecified $K = 3$ scenarios and by $+0.019 \pm 0.004$, $+0.016 \pm 0.004$, $+0.015 \pm 0.005$, and $+0.011 \pm 0.004$ at $K = 8$ (beyond two standard errors in seven of the eight), within error of \textbf{(1b)} in all eight ($-0.006$ to $+0.012$), and below the matched doubly robust arm in design-level mean squared error everywhere ($-0.50$ to $-1.77$); on all $50$ randomized $K = 8$ environments the contrast is $+0.019 \pm 0.004$ against $+0.034 \pm 0.009$ on the first ten. On
identical patients the doubly robust estimator discriminates within error of the sample mean (above it in five scenarios and below it in three, none beyond $1.3$ standard errors), and the deficit of the doubly robust family is what its own
recruitment does with the noisier
small-sample estimates it produces early in the trial, when the
sensitivity index follows those estimates into subpopulations that the
sample mean rule would not have chosen. Proposition~5
stylizes recruitment as an allocation that equalizes the variances of the
recruiting family, and its premise, that the own path of each family is
the better one for its own analysis, holds for the sample mean family and
fails for the doubly robust family at these sample sizes. Two consequences
follow. The consistency principle holds for the sample mean analysis, whose mismatched pair loses in the predicted direction, and fails for the doubly robust analysis at these sample sizes, whose matched path is worse than the sample mean path (the asymmetric reading of Section~6.6). And the loss that the doubly robust
family incurs at small sample sizes is not primarily estimator variance,
which the same-path pairing shows to be at most a few percent of the
sample mean's (Section~6.5), but the allocation
that its early estimates induce; in mean squared error, too, the two
analyses of the identical patients of the same-path runs differ by only
$+0.15 \pm 0.24$, $-0.02 \pm 0.04$, $-0.14 \pm 0.10$, and
$+0.09 \pm 0.08$ in the randomized $K = 3$, randomized $K = 8$, aligned
$K = 3$, and aligned $K = 8$ configurations (ten held-out environments
each; on the core environments the mismatch arm's error exceeds \textbf{(1b)}'s by $+0.12 \pm 0.03$, $+0.11 \pm 0.03$, and $+0.13 \pm 0.05$ in the aligned, random, and misspecified $K = 8$ scenarios, $+0.03 \pm 0.02$ under randomization, and is within $0.07$ of it at $K = 3$, the doubly robust analysis's own error on the starved subpopulations of the sample mean path), against $+0.90$, $+0.76$, $+1.72$, and $+1.02$ on their own paths over the core environments ($50$; $100$ in the aligned $K = 3$ scenario; Section~\ref{sec:supp-map}), so the
path produces the starved cells in
which the targeted fits fail. Inflating the recruitment variance by the simulated factor does not repair this, because the design-embedded inflation is nearly constant over the sample sizes at which the allocation is decided and a constant factor on the conditional variance moves the index little (the steeper design stage curves of the AIPW and regression-adjusted recruiting arms do move the allocation, Section~\ref{sec:supp-sens}).

\FloatBarrier
\subsection{Degeneracy Diagnostic}
\label{sec:supp-degdiag}
\begin{figure}[!htbp]
\centering
\includegraphics[width=0.5\linewidth]{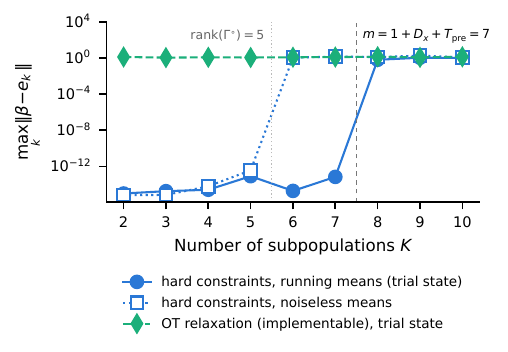}
\caption{Departure of the borrowing weights from no borrowing,
$\max_{k}\|\beta - e_{k}\|$, against the number of subpopulations, for the
plain hard-constrained solve shared by the variants \textbf{(1a, 1b, 2a,
2b)}, run without the rank shortcut of the production solvers, evaluated
on the running means at the end of a simulated \textbf{(1a)} trial of
$2K + 50K$ patients ($26$ per subpopulation and arm on average; the
adaptive state is unbalanced; environment $0$, seed $0$) and on
the noiseless means with the same sample sizes, and for the
implementable optimal transport relaxation on the same trial state. On
realized data the constrained weights are $e_{k}$ to numerical precision
(below $10^{-12}$) for $K \le m = 7$ and depart at $K = 8$; on the
noiseless means they depart at $K = 6 = \rank(\Gamma^{\circ}) + 1$; the
relaxation is never pinned for at least one target.}
\label{fig:degeneracy}
\end{figure}

\begin{figure}[!htbp]
\centering
\includegraphics[width=\linewidth]{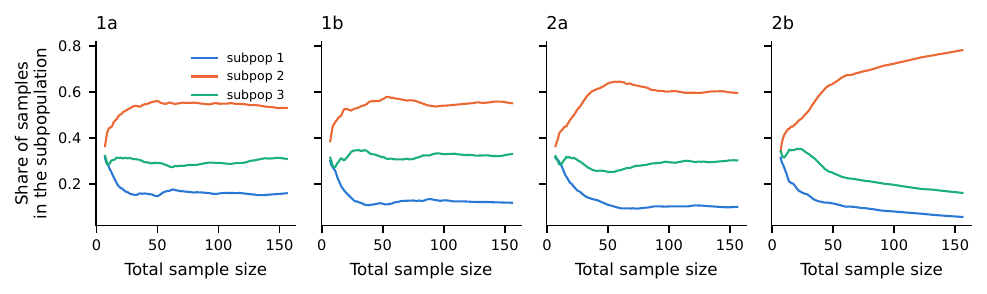}
\caption{Recruitment paths in the aligned $K = 3$ scenario: the share of
the accrued sample in each subpopulation as a function of the total sample
size, averaged over replications in one environment, for the four
subpopulation- and individual-level arms of the two families. With
borrowing fixed at $e_{k}$ the point estimators within a family and level
coincide on identical data, but the level-specific and family-specific
variance formulas in the
sensitivity index steer the arms to different data from the first episodes
on. This is the channel through which the four groups of
Section~\ref{sec:supp-degdiag} separate. The individual-level doubly robust arm
concentrates most strongly, which is the path effect of
Section~6.6.}
\label{fig:paths}
\end{figure}

With $\Dx = 2$, $\Tpre = 4$, and $\Dz = 2$, the number of hard constraints is
$m = 7$ and the noiseless rank is $1 + 2 + \min(4, 2) = 5$, so every
$K = 3$ scenario of Sections~6.1 and~6.2 is
degenerate, while the $K = 8$ scenarios of Section~6.3 are not.
Numerically, on the running means of a simulated trial the plain
constrained solve, run without the rank shortcut of the production
solvers, returns $\beta = e_{k}$ to numerical precision (departures below
$10^{-12}$) for all $K \le 7$ and departs from it by order one at $K = 8$,
so the transition is sharp and lies exactly at $m$; on the noiseless
means it lies at $K = 6$, which is the noiseless-rank prediction
(Figure~\ref{fig:degeneracy} shows both). The runs confirm the collapse of
Proposition~1. Since the production solvers return
exactly $e_{k}$ whenever the realized constraint matrix has rank $K$,
without a rank-deficient solve, the variants within a family and level
coincide on identical data,
the oracle bias term that distinguishes the MSE-optimal variants is
inactive, their sensitivity indices agree, and exact ties in the
look-ahead are broken by one rule, so the arms of a pair collect the same
data: the operating characteristics of \textbf{(1b)} and \textbf{(1c)},
and of \textbf{(2b)} and \textbf{(2c)}, coincide to the last digit in
every $K = 3$ scenario (Tables~\ref{tab:mainK3} and~\ref{tab:auc}).
Exact coincidence of the curves within a family is itself the recommended
diagnostic. Observing it in a pilot or a dry run indicates $K \le m$, in
which case the synthetic control machinery is inert and the design
comparison reduces to the estimators of the individual subpopulations,
while the optimal transport relaxation of Section~\ref{sec:ot} is the
documented escape (Section~\ref{sec:supp-ot-results}). In
the confounded configurations, where all arms share a single
potential outcomes pool, the observed structure is four visible
groups: $\{$\textbf{1b},\,\textbf{1c}$\}$ and $\{$\textbf{2b},\,\textbf{2c}$\}$
coincide, while \textbf{(1a)} and \textbf{(2a)} each stand apart from
their individual-level counterparts, with maximal gaps between the mean false positive rate trajectories of $1.4$ and $11.6$ percentage points in the aligned $K = 3$ scenario. The mechanism is the recruitment-path
channel. The subpopulation-level arms use level-specific variance inputs
in the sensitivity index (\textbf{(1a)} the homoscedastic $\sigma^{2}$,
\textbf{(2a)} influence curve estimates), and since they store the
centroid of the subpopulation as the patient covariate, the nuisance
models of \textbf{(2a)} cannot remove the confounding bias that those of
\textbf{(2b)} remove (on centroid covariates the targeting shift is zero,
so the point estimates of \textbf{(2a)} equal those of \textbf{(1a)};
Section~6.2), and both their paths and their operating points
differ from the individual-level arms. At $K > m$
the constraints leave $K - m$ free dimensions and borrowing becomes
active. The nondegenerate scenarios show the predicted separation within
a family with their single free dimension: \textbf{(1b)} and \textbf{(1c)}
separate by $-0.6 \pm 0.3$ points of FPR and $+0.005 \pm 0.002$ AUC in the
randomized scenario, and the maximal trajectory gaps within a pair, which
are zero in every $K = 3$ scenario, are $0.8$ and $0.6$ points in the
randomized $K = 8$ scenario and $0.9$ and $0.8$ points in the aligned one.
Figure~\ref{fig:degeneracy} shows the transition and the escape from it.
On the running means of a simulated trial the constrained solve returns
$e_{k}$ to numerical precision up to $K = 7$ and departs by order one from
$K = 8$ on, on the noiseless means from $K = 6$ on, while the
implementable optimal transport relaxation borrows at every $K$. Figure~\ref{fig:paths} shows the
recruitment-path channel through which the summary-level and
individual-level arms separate.

\FloatBarrier
\subsection{The Map Inside the Adaptive Design}
\label{sec:supp-map}
The middle and right panels of Figure~1 place the adaptive
grid of Section~6.5 on the map. At the design level,
averaging over environments, replications, and subpopulations, the doubly
robust family has the larger mean squared error in all $32$ grid configurations
(for example $2.67$ against $2.16$ in the randomized $K = 3$ configuration
at $H = 150$ and $4.00$ against $3.13$ at $K = 8$) and in all eight core
scenarios, by $+0.90 \pm 0.18$, $+1.72 \pm 0.27$, $+1.41 \pm 0.27$, and $+0.90 \pm 0.08$ in the randomized, aligned, random, and misspecified $K = 3$ scenarios and by $+0.76 \pm 0.10$, $+1.02 \pm 0.13$,
$+1.37 \pm 0.16$, and $+0.63 \pm 0.07$ at $K = 8$ (paired over
environments); the anchor configuration ($r = (-10, 0, +10)$, $K = 3$) is the
one adaptive configuration in which the doubly robust family is not behind
($-0.012 \pm 0.022$; six aligned $K = 3$ grid configurations at $H = 300$ and
$600$ are within two standard errors, at $+0.29$ to $+1.03$).
The source is the targeted fits on the few patients of starved
subpopulations: observations in which an arm holds at most three
patients are $27$--$74\%$ of the observations of \textbf{(2b)} in the
core scenarios and carry $87$--$97\%$ of its squared error; the targeted
cells among them (six or more patients, at least two per arm) carry
$91$--$122\%$ of the gap and the fallback cells $18$--$52\%$; the cells with more than three patients per arm offset the excess ($-24$ to $-43\%$); and fits on six to nine patients carry $82$--$100\%$ of the targeted cells' share. On the
randomized band at $3$--$5$ patients per arm its realized mean squared
error is $6.6$ and $3.8$ times the fixed-count formula at $K = 3$ and
$K = 8$ (against $1.7$ and $1.5$ for the sample mean), whereas at own
counts of $13$ or more per arm it has the smaller error in every core
scenario and in $28$ of the $32$ grid configurations.
Because the two families recruit along different paths, a subpopulation
ends a replication with different per-arm counts under the two families
(a median factor of $1.8$ between them), and a paired comparison placed at the mean of the two counts mixes two sample sizes (such a placement, with the same clusters, has $54$ decided tiles, $50$ agreeing with the prediction and the doubly robust family the better one in only $5$). Each
family's squared error in every replication and subpopulation is
therefore placed at the per-arm count that its own recruitment gave the
subpopulation by the horizon and at the subpopulation's bias magnitude,
the plane is tiled, and a tile compares the two families' mean squared
errors over the observations that fall into it, decided when the
difference exceeds two cluster-robust standard errors (clusters:
environment and subpopulation, pooled over the grid configurations of the
panel, which share the environments and the trial seeds, with the
small-cluster factor $G/(G - 1)$; the two families' standard errors are
combined as independent, which is conservative since the families share
their units) and blank when either family has fewer than $30$
observations in it. This comparison
conditions on outcome-dependent counts, for which the fixed-count
statements do not hold exactly (the remark after
Assumption~\ref{ass:counts}); below eight patients per arm the tile
magnitudes depart from the fixed-count formulas by factors of up to
$6.6$ for the doubly robust family. Of the $60$ tiles at $K = 3$ and $59$
at $K = 8$, $30$ and $30$ are decided, and the prediction
of~(11) has the sign of every one of them: the doubly robust
family is the better one beyond the boundary ($18$ and $14$ tiles, at
large counts or large bias) and the sample mean below it. The agreement
is consistent with the boundary rather than a sharp test of it, since
only $2$ of the $20$ tiles adjacent in the count direction to a change of
predicted sign are decided and cruder predictors agree nearly as well (a
constant $\kappa = 2$ in $25$ of $30$ and $27$ of $30$ tiles; the rule ``sample
mean if the own count is below $13$ or the bias below $0.25$'' in $27$
and $28$); and at $K = 8$ the families borrow with different weights, a
case to which Theorem~1 is not extended, so that panel
is an empirical extension. On the randomized band the paired comparison,
which sets the two families' estimates of the same subpopulation and
replication against each other, favors the sample mean at $8$ to $80$
patients per arm through the replications in which the doubly robust
path gave the subpopulation less than half the patients of the sample
mean path ($15\%$ and $25\%$ of them at $K = 3$ and $K = 8$, with a mean
squared error difference of $+10.6$ and $+4.6$ there against $-0.5$ and
$-0.8$ in the others); the reverse starvation, in which the sample mean
path gives less than half the patients of the doubly robust path, is
about as frequent ($19\%$ and $24\%$) and costs the sample mean $2.3$
and $2.5$ in the same units, so the asymmetry is the fragility of the
targeted fits at six to nine patients, which the path exposes. The configurations
at $H = 150$ and strengths $0$ and $2$ are the first ten environments of
the core scenarios. Two features of the fixed-count variance inflation curves of the left panel are recorded here. The raw bin medians are not monotone (rises above $0.05$ between neighboring sizes occur at strengths $1.5$, $2$, and $3$; smaller ones, of at most $0.015$, at strengths $0$ to $1$): at strength $2$ they rise by $0.29$ ($6$ to $8$ patients per arm), $0.16$ ($10$ to $13$), $0.04$ ($20$ to $26$), and $0.06$ ($52$ to $64$), from $2.13$ at six to $2.56$ at thirteen, at strength $1.5$ by $0.08$ ($6$ to $8$), and at strength $3$ the departure is largest ($\kappa$ rises from $2.9$ at four patients per arm to about $4$ at $26$--$40$). The isotonic fit imposes the monotone curve of Assumption~\ref{ass:inflation}; the single-crossing structure of the predicted difference is nevertheless observed with the raw curves: with the raw bin medians and each record's own constants and realized $s$ the predicted difference changes sign at most once in every record group at strengths $0$--$2$ (once in $743$ of the $990$ groups, never in $247$) and in $216$ of $219$ at strength $3$. A weaker sufficient condition than monotonicity is that $s(n)[\kappa(n)\sigma^{2}_{k,\mathrm{cond}} - E_{k}]$ decrease where it is positive on the grid; it holds in most groups but not in all (it fails in $40$ of $220$ groups at strength $2$ and $130$ of $219$ at strength $3$, in nearly all of which the single crossing holds nonetheless), so the structure is a property of these data rather than a consequence of the raw curves. The nonmonotone raw curves matter little for the agreement, $99.3$--$99.8\%$ at every strength, because the $\mathcal{O}(1)$ bias term decides most points.

\subsection{variance inflation Curves, Crossing Condition, and Selector}
\label{sec:supp-inflation}
\begin{figure}[!htbp]
\centering
\includegraphics[width=\linewidth]{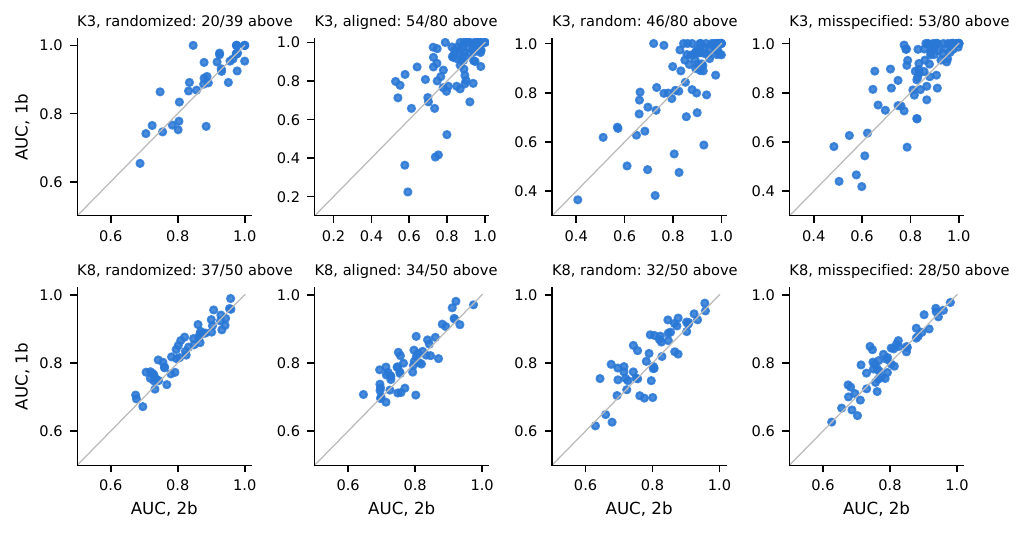}
\caption{AUC at the end of the trial for each environment, for the
individual-level sample mean arm \textbf{(1b)} against the doubly robust
arm \textbf{(2b)} in each scenario, one dot per environment ($39$ of $50$ in the randomized $K = 3$ scenario and $80$ of $100$ in the confounded ones, the others having no AUC; $50$ at $K = 8$), with the
identity line. The spread across environments that dominates the standard
deviations of Table~\ref{tab:mainK3} is the spread along the diagonal; the
family contrast is the displacement from it.}
\label{fig:scatter}
\end{figure}

Figure~\ref{fig:scatter} shows the pairing by environment behind the
standard errors of the family contrasts; the paragraphs that follow give
the full account of the simulation-based estimates and of the selector
record summarized
in Section~6.5.

\begin{figure}[!htbp]
\centering
\includegraphics[width=0.48\linewidth]{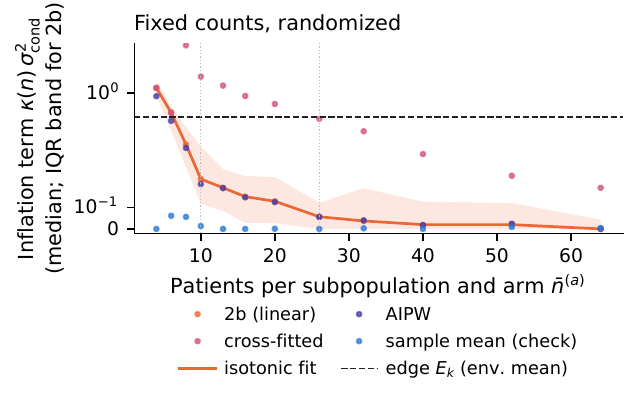}\hfill
\includegraphics[width=0.48\linewidth]{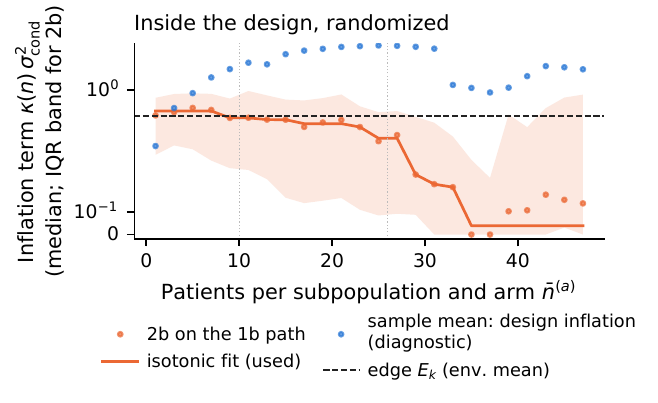}\\[2pt]
\includegraphics[width=0.48\linewidth]{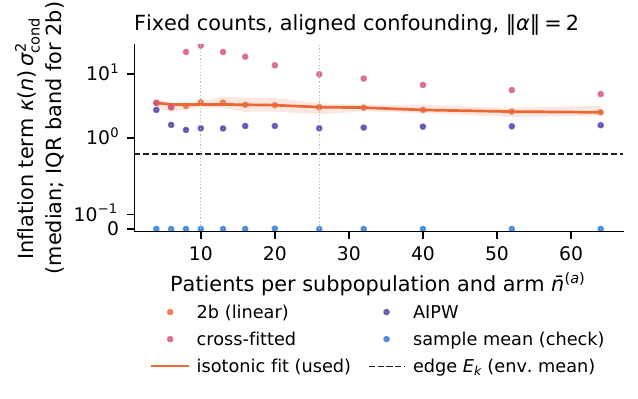}\hfill
\includegraphics[width=0.48\linewidth]{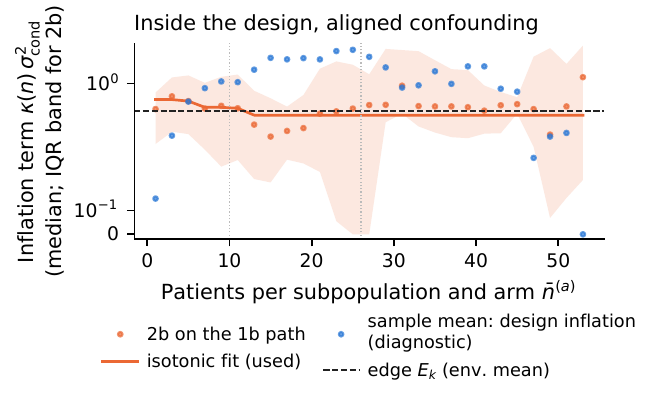}
\caption{Simulated variance inflation curves, plotted as the inflation term $\kappa(\bar n)\,\sigma^{2}_{k,\mathrm{cond}}$ against the number of patients per subpopulation and arm (vertical axis symmetric-log, linear below $0.5$), with the
explained-variance edge $E_{k}$ as a horizontal reference (environment
means) and the sampled sizes $10$ and $26$ marked. Left column: the
design stage estimate at fixed sample sizes for the linear-fluctuation
targeted estimator, AIPW, the cross-fitted estimator, and the sample mean
(a check, near zero), under randomized (top) and aligned confounded
assignment at the anticipated strength (bottom); medians over
environments and subpopulations with the interquartile band for
\textbf{(2b)}. Right column: the design-embedded estimate on ten
held-out environments, the ratio of the replication variances of the
doubly robust and sample mean analyses of the same trial recruited by the
sample mean family, binned by the shared sample size and pooled over
$K = 3$ and $K = 8$; the design inflation of the sample mean itself is
drawn as a diagnostic. The count-weighted isotonic fit of the bin medians
(line) is the curve that the family-2 recruitment uses and, under
randomization, the median-rule selector's step (ii); step (iii) uses the
design stage confounded curve of the left column. The band is the
interquartile range of the per-cell ratios
within each bin, and the isotonic fit of its upper edge is the curve of
the upper-quartile rule. The design stage randomized curve of the targeted estimator (top left) is used by no arm of the core scenarios or selector step (the AIPW curve of that panel recruits the AIPW arm and the curve-test arm of Section~\ref{sec:supp-sens} under randomization); its fixed-count replicate is the strength-$0$ boundary of Figure~1.}
\label{fig:kappa}
\end{figure}

\begin{figure}[!htbp]
\centering
\includegraphics[width=\linewidth]{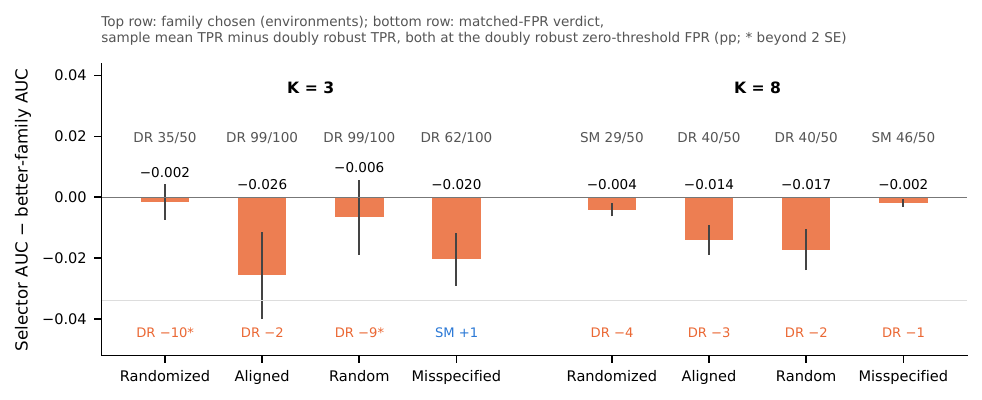}
\caption{Selector record (median rule): AUC of the pre-specified selector
at the end of the trial minus that of the better family, for each
scenario, with $\pm 1$ paired standard error, and the family it chose
with the environment vote (top row). The bottom row gives the co-primary
verdict at matched false positive rate, the true positive rate of
\textbf{(1b)} minus that of \textbf{(2b)}, both re-thresholded to the
zero-threshold false positive rate of \textbf{(2b)}, in percentage points
(SM: sample mean ahead;
DR: doubly robust ahead; asterisk beyond two standard errors). The
selector is within two paired standard errors of the better family in
five of the eight scenarios; under aligned and random confounding it
selects the doubly robust family in most environments, so its deficits
there are those of that family in AUC at small counts ($-0.014 \pm 0.005$
and $-0.017 \pm 0.007$ at $K = 8$), against a gain at matched
false positive rate under aligned confounding, within error, and in the misspecified $K = 3$ scenario ($-0.020 \pm 0.009$) the anticipated band overstates the shift.}
\label{fig:regret}
\end{figure}

\emph{Two variance inflation curves.} The inflation $\kappa_{k}(\bar n)$
of Proposition~4 was estimated by simulation in two
ways (Figure~\ref{fig:kappa}). The design stage estimate follows
the prescription of the selector. Samples of fixed size $2\bar n^{(a)}$
are drawn from the data generating process of the first ten $K = 3$
evaluation environments, regenerated from their seeds, the
pre-specified nuisance estimators are fitted, and the replication variance
of the effect estimate of each estimator is compared with the reference
$\sigma^{2}_{k,\mathrm{cond}}\,s_{k}$ at the realized arm counts (its
asymptotic value under randomization; under confounding the asymptotic
value is $\omega_{k}^{\dagger}\,\sigma^{2}_{k,\mathrm{cond}}\,s_{k}$, so
the confounded curve converges to $\omega_{k}^{\dagger} - 1$), with $200$
replications per environment, subpopulation, and grid point
($\bar n^{(a)} = 4$ to $64$) and, under confounding, $\alpha$ along the
mean outcome-weight direction of the environment as in the trials. The
sample mean, compared with $\sigma^{2}_{k,\mathrm{eff}}\,s_{k}$, returns
$\kappa$ between $-0.05$ and $0.04$ under randomization and between
$-0.09$ and $0.00$ under confounding, where the within-arm covariate
variance is $\bar E_{k}^{c} < E_{k}$, which checks the procedure. For
the linear-fluctuation targeted estimator under randomization this
inflation falls from $0.77$ at $\bar n^{(a)} = 4$ through $0.43$ at $6$, $0.27$ at
$8$ and $0.16$ at $10$ to $0.04$ at $26$, $0.01$ at $40$ and zero at $64$.
It reaches the environment-mean edge ($E_{k} = 0.56$ against
$\sigma^{2}_{k,\mathrm{cond}} = 1.45$) between the grid points $6$ and
$8$, at $n^{*}(0) \approx 7$ by interpolation; the inflation of AIPW is
slightly lower ($0.64$ at $4$, $0.26$ at $8$). Under aligned confounding at the
anticipated strength the same simulation returns $\kappa \approx 2.1$--$2.5$
for the targeted estimator up to $\bar n^{(a)} = 26$ and $1.7$ at $64$,
converging on $\omega_{k}^{\dagger} - 1 \approx 1.5$--$1.6$, one less
than the large-sample factor of the remark after
Proposition~4, and $0.92$ at $8$ and $0.97$--$1.09$
beyond for AIPW, whose limit $\omega_{k}^{\dagger} - 1$ is $1.1$; so $B(\bar n)$ never changes sign and a variance-only threshold is
infinite for both. The cross-fitted
estimator, which uses the full-sample fit below $8$ patients per arm
(Section~\ref{sec:supp-sim}), has inflation $2.1$ at $8$, $1.0$ at $10$,
and still $0.37$ at $26$ under randomization (threshold
$n^{*}(0) \approx 25$ on the isotonic fit), and $3$--$19$ under confounding: halving the nuisance sample, with the per-fold targeting, is expensive at every count the design reaches. The
design-embedded estimate measures the inflation inside the
adaptive trial. On ten environments drawn for this
purpose only, the trial is recruited by the sample mean family and
analyzed by both families on the identical patients, so at every
environment, episode, and subpopulation the two estimates share the sample
sizes and the recruitment path. The ratio $\chi$ of the replication
variance of the doubly robust analysis to that of the sample mean analysis
on the identical patients isolates the estimator's own variance
inflation from the inflation that the adaptive design induces in both; the replication
variance of the sample mean inside the design is $1.3$ to $2.8$ times its
nominal value (Figure~\ref{fig:kappa}). With the sample mean's realized
inflation as the reference,
$1 + \kappa_{\mathrm{emb}} = \chi\,(1 + E_{k}/\sigma^{2}_{k,\mathrm{cond}})$,
so the comparison of the inflation with the edge is the comparison of
$\chi$ with one. Pooled over the $K = 3$ and $K = 8$ held-out trials the
embedded inflation is $\kappa_{\mathrm{emb}} \approx 0.42$ for $\bar n^{(a)} \le 7$, $0.36$--$0.32$
for $9$--$23$, $0.28$ at $25$--$27$, $0.18$--$0.14$ at $29$--$33$, and
near zero beyond $35$, against an edge-to-variance ratio
$E_{k}/\sigma^{2}_{k,\mathrm{cond}} = 0.37$ ($E_{k} = 0.55$,
$\sigma^{2}_{k,\mathrm{cond}} = 1.49$). That is, $\chi$ is about $1.03$
below $\bar n^{(a)} = 9$, $0.98$ up to $23$, and $0.75$--$0.86$ beyond
$27$: on identical patients the doubly robust estimator matches the
sample mean below nine patients per arm, beats it by a few percent of
variance up to about $23$, and by $14$--$25\%$ beyond $27$. The median curve gives $n^{*}(0) \approx 9$ and
the isotonic fit of the upper quartile of the bins, which lies at
$0.55$--$0.61$ up to $21$ and at $0.31$--$0.46$ beyond, gives
$n^{*}(0) \approx 29$. Under aligned
confounding the embedded inflation is $0.48$--$0.39$ up to $\bar n^{(a)} = 11$
and $0.34$ beyond, so the median curve would reach the edge at
$n^{*} \approx 12$ and the upper quartile ($0.72$--$0.74$) never; this
curve tracks the edge because of the fallback cells discussed below and
is not used by step (iii) of the selector. At
$\sigma_{x} = 2$ the inflation is $1.45$ at $\bar n^{(a)} \le 7$ and
$1.44$ at $9$--$23$ against an edge-to-variance ratio of
$1.48$ ($\chi \approx 0.99$), so the two sides of the crossing condition nearly balance on identical patients at every sampled size and the median curve places the boundary at the smallest sample size. The design stage and the design-embedded
curves agree on the shape that Assumption~\ref{ass:inflation} requires, a
single sign change of $B(\bar n)$ once the raw bins are smoothed by a
count-weighted isotonic fit, and on the threshold under randomization
($7$ against $9$), and they differ in the level: inside the design the
doubly robust family pays about a third more variance than its asymptotic
value at every sample size up to $25$, about twice the fixed-size
inflation at
$\bar n^{(a)} = 10$. Two features of the embedded curve are not nuisance
estimation. In a subpopulation below the fallback threshold (fewer than
six patients, or fewer than two in an arm) the doubly robust analysis
is the sample mean, so the two replication variances coincide,
$\chi = 1$ exactly at $K = 3$, where both analyses use the weights
$e_{k}$, and up to the difference between the two families' weights at
$K = 8$, and the embedded inflation is $\kappa_{\mathrm{emb}} =
E_{k}/\sigma^{2}_{k,\mathrm{cond}}$ by construction (in general, under
confounding, $1 + \kappa_{\mathrm{emb}} =
(1 + \kappa)(\sigma^{2}_{k,\mathrm{cond}} + E_{k})/
(\sigma^{2}_{k,\mathrm{cond}} + \bar E_{k}^{c})$ with $\kappa$ the
inflation of Proposition~4, so the embedded curve also absorbs
the arm-conditional correction). Such cells are $54\%$ of the
cell-replications of the randomized $K = 8$ held-out runs, $37\%$ and
$62\%$ of the confounded ones at $K = 3$ and $K = 8$, and still
$25$--$33\%$ of the cells at $9$--$13$ patients per arm in the randomized
and $30$--$45\%$ in the confounded runs, which is why the embedded curves are
not measurements of the targeted estimator at small counts: without the
fallback replications the confounded inflation at $5$--$9$, $9$--$13$,
$13$--$17$, and $17$--$23$ patients per arm is $1.6$, $1.2$, $1.1$, and
$1.0$ at $K = 3$ and $1.2$, $0.8$, $0.6$, and $0.5$ at $K = 8$, against $0.27$--$0.64$ with them. Under randomization the same reading gives $0.40$, $0.35$, and $0.00$ at $K = 3$ and $0.52$, $0.47$, and $0.38$ at $K = 8$ at $5$--$9$, $9$--$13$, and $13$--$17$ patients per arm, against edge-to-variance ratios $E_{k}/\sigma^{2}_{k,\mathrm{cond}}$ of $0.34$--$0.43$ in those bins ($\kappa = 0.22$ at $17$--$23$ patients per arm at $K = 8$ against an edge ratio of $0.38$), so the targeted estimator on its own is already below the edge in the $5$--$9$ bin at $K = 3$ and crosses it between the $13$--$17$ and $17$--$23$ bins at $K = 8$; the threshold $n^{*}(0) \approx 9$ quoted for the embedded curve refers to the implemented estimator, fallback rule included. Further, the bins are indexed by the mean count of
a subpopulation along the adaptive path and pool subpopulations on both
sides of the fallback threshold and of the crossing, so the near-constancy
of the randomized curve up to $\bar n^{(a)} \approx 23$ reflects this
mixture rather than a flat nuisance inflation; the design stage curve,
which
has no fallback cells above $\bar n^{(a)} = 4$, falls monotonically. The
embedded curve under randomization is the one the family-2 recruitment
and step (ii) of the selector use, because it is the inflation that the
design realizes, fallback rule included; for step (iii) the design stage
confounded curve is used, since an inflation diluted to the edge by
fallback
cells would let any positive band route a trial to the doubly robust
family, although in a fallback cell that family does not remove the bias.

\emph{Crossing condition.} The sampled sizes lie between the median-rule
threshold and the upper-quartile one, and on a fixed recruitment path the
two families are within error of each other in discrimination. Under
uniform recruitment at $\bar n^{(a)} \approx 10$, just beyond
$n^{*}(0) \approx 9$, the doubly robust family is favored by $0.009 \pm 0.005$ AUC ($1.8$ standard errors, within error); at $\bar n^{(a)} \approx 26$
it is within $0.010 \pm 0.008$ of the sample mean at $K = 3$ and within
$0.011 \pm 0.006$ of it on the adaptive path at $K = 8$; and on the identical patients of the same-path runs the two
families are within $0.003$ AUC of each other at $K = 8$
(Section~6.6). The adaptive
$K = 8$ scenarios at $H = 150$, where the sample mean family is ahead by
$0.013$--$0.020$, are the exception, and Section~6.6 shows
that their gap is the recruitment path. This is the transition that the
crossing condition~(10) draws along the randomized axis
(Figures~\ref{fig:forest} and~1; the pairing by environment
behind the standard errors is shown in Figure~\ref{fig:scatter}). The
fixed-count map places $n^{*}(0)$ at about $7$ patients per arm, where
the design stage curve puts it (the design-embedded curve gives $9$), and
the same-path
variance ratio falls to $0.75$--$0.86$ beyond about $27$; the adaptive
$K = 3$ scenario at $\bar n^{(a)} \approx 26$ is within error of parity
in discrimination at the individual-level, which does not separate the
median-rule threshold from the upper-quartile one. Two remarks apply. The crossing condition compares
the estimators at fixed sample sizes, so it gives the sign of the mean
squared error difference per subpopulation and says nothing about the
recruitment path,
which at $K = 8$ and $H = 150$ dominates the family contrast. Further, the
outcome model is correctly specified and fitted at parametric rates (the
propensity model too, except in the misspecified scenario), with no
unmeasured confounding (Figure~\ref{fig:dag}), although
$20$--$48\%$ of the patients lie outside the truncation range of the propensity under confounding. Ensemble nuisances, model selection, or
misspecification would raise the inflation and move the boundary outward
(Section~4.4: the location of $\mathcal{T}_{k}$ moves with
$\kappa_{k}$), whereas an outcome model that also uses the observed
pre-treatment trajectory, which with $\Tpre > \Dz$ spans the directions
of the latent loadings and would act as a proxy for
$\mu_{T,k}^{\top} z_{kj}$, would enlarge the edge beyond $E_{k}$ but also
raise the inflation through its extra regressors; the net effect on the
boundary is not evaluated here.

\emph{Selector.} Run at the anticipated counts $\bar n = H/(2K) + 1$
($26$ at $K = 3$, $10.4$ at $K = 8$ and $H = 150$, $26$ at $H = 400$) with
the design-embedded curve in step (ii), the median rule routes the
randomized $K = 3$ scenario to the doubly robust family in $35$ of $50$
environments, the randomized $K = 8$ scenario at $H = 150$ to the
sample mean family in $29$ of $50$, the uniform-recruitment probe in seven
of ten, and at $H = 400$ eight of ten environments to the doubly robust
family ($-0.008 \pm 0.006$ against \textbf{(1b)}, $+0.002 \pm 0.002$
against \textbf{(2b)}). Under confounding, step (iii)
compares the squared anticipated worst case band with the net variance
term $s(\kappa\sigma^{2}_{k,\mathrm{cond}} - E_{k})$ of the design stage
confounded curve ($\kappa \approx 2.1$ at $\bar n = 26$ and $2.3$ at $10.4$, each median carrying a Monte Carlo uncertainty of about $\pm 0.1$ over its $30$ environment--subpopulation values; the term is about $0.20$ and $0.57$, so the band must exceed about $0.45$ and $0.75$ in outcome units; with the AIPW curve of the same simulation, $\kappa \approx 1.0$ at both sizes, the term would be about $0.08$ and $0.20$ on the same environments and the band would need to exceed only about $0.3$ and $0.45$): the squared band exceeds it in $99$ of the $100$ environments of the aligned and random scenarios at $K = 3$, in every environment of the aligned scenario at $H = 400$, in $40$ of $50$ at
$K = 8$ and $H = 150$ ($30$ and $27$ under the upper-quartile rule), and
in $62$ of $100$ ($56$) misspecified environments at $K = 3$ and $4$ ($2$) at $K = 8$, where the realized shift is smaller than the worst case band. Step (iii) takes the balanced $s = 2/\bar n$; under confounded
assignment the realized $s_{k}$ exceeds it by the factor
$1/\{4\pi_{k}(1-\pi_{k})\}$ (with each environment's median factor over
its subpopulations, $1.17$--$1.24$ under the linear-index mechanisms and $1.6$--$1.7$ under the quadratic one), so the balanced value favors the doubly robust family: with the realized factor the median rule would route $97$ instead of $99$ environments in the aligned and random $K = 3$ scenarios, $35$ instead of $62$ in the misspecified $K = 3$ scenario,
$26$ and $25$ instead of $40$ at $K = 8$ under aligned and random confounding, and none instead of $4$ in the misspecified $K = 8$ scenario. The record is also sensitive to the anticipation of the band itself, the quantity a protocol can least verify: with the anticipated worst case band halved at the true strength curve (a band error alone) the median rule routes $40$, $41$, and $8$ of $100$ environments in the aligned, random, and misspecified $K = 3$ scenarios, none at $K = 8$, and $2$ of $10$ at $H = 400$, and the selector arm's deficit against the better family vanishes at $K = 8$ and is $+0.006 \pm 0.008$, $+0.006 \pm 0.009$, and $-0.004 \pm 0.002$ AUC at $K = 3$; with the band doubled it routes $100$, $100$, and $89$ at $K = 3$ and $50$, $50$, and $44$ at $K = 8$, and the deficits grow to $-0.032 \pm 0.009$ in the misspecified $K = 3$ scenario and $-0.016 \pm 0.005$, $-0.020 \pm 0.007$, and $-0.010 \pm 0.004$ at $K = 8$; the anticipated edge matters little (halving it or raising it by half changes the routing by at most four environments and the deficits by at most $0.005$). A consistent error in the anticipated strength changes the band and the curve together: with the band and the curve at strength $1$ (the fixed-count curve of the regime map at that strength, simulated on held-out environments) the median rule routes $100$, $100$, and $92$ of $100$ environments at $K = 3$ and $41$, $41$, and $29$ of $50$ at $K = 8$, and at strength $3$ it routes $98$, $97$, and $45$ at $K = 3$ and $33$, $29$, and $1$ at $K = 8$, against $99$, $99$, $62$ and $40$, $40$, $4$ at the true strength $2$ with the in-sample curve of step (iii); with the map curve at the true strength the median rule routes $99$, $99$, and $57$ at $K = 3$ and $40$, $40$, and $4$ at $K = 8$, the like-for-like baseline of the two strength rows. Halving the strength does not halve the band, since $c_{k}$ rises as the propensity flattens: the band falls to about two thirds under the linear-index mechanisms ($0.65$--$0.66$ in median over the environments) and to about $0.85$ under the quadratic one ($0.84$--$0.86$ in median, range $0.30$--$1.61$), where it is not monotone in the strength. Under the linear-index mechanisms the routing at $K = 8$, where the band and the threshold are of the same size, is thus sensitive to a band error at a given curve and much less to a consistent strength error, and at $K = 3$ only under-anticipation of the band moves it; under the quadratic mechanism an under-anticipated strength routes $29$ instead of $4$ of $50$ environments at $K = 8$ and $92$ instead of $57$ of $100$ at $K = 3$, with the selector arm's deficit at $-0.033 \pm 0.009$ AUC ($3.5$ standard errors) in the latter. Under an anticipation error the record can be better or worse than under exact anticipation. Over the environments a rule routes to the doubly robust family, the selector arm, which copies that family's arrays and so follows its recruitment path, has that family's deficit there, largely the path cost of Section~\ref{sec:supp-mismatch}: under the median rule $-0.032 \pm 0.013$ AUC in the misspecified $K = 3$ scenario ($51$ of the $62$ routed environments have a defined AUC), $-0.010 \pm 0.005$ in the randomized $K = 8$ scenario ($21$
environments), $-0.017 \pm 0.006$ and $-0.022 \pm 0.008$ under aligned and
random confounding at $K = 8$ ($40$ each), and $-0.024 \pm 0.011$ in the
misspecified $K = 8$ scenario ($4$), against $-0.020 \pm 0.009$, $-0.004 \pm 0.002$, $-0.014 \pm 0.005$, $-0.017 \pm 0.007$, and
$-0.002 \pm 0.001$ over all environments (Table~3). The selector arm is therefore \textbf{(2b)} in all but one environment of the aligned and random $K = 3$ scenarios and in every environment of the aligned $H = 400$ probe, and a mixture elsewhere. Its record (Figure~\ref{fig:regret}) is within two paired
standard errors of the better family in five of the eight core
scenarios; the deficits are $-0.026 \pm 0.014$ (aligned $K = 3$, $1.8$ standard errors), $-0.006 \pm 0.012$ (random $K = 3$), $-0.020 \pm 0.009$ (misspecified $K = 3$, $2.3$), $-0.004 \pm 0.002$
(randomized $K = 8$, $2.0$), $-0.014 \pm 0.005$ and $-0.017 \pm 0.007$
(aligned and random $K = 8$, $2.8$ and $2.6$), and $-0.002 \pm 0.001$
(misspecified $K = 8$), offset at matched false positive rate under aligned confounding ($2.1 \pm 1.6$ points for the selector arm at $K = 8$, $2.7 \pm 1.8$ for the doubly robust arm). In mean squared
error the median-rule arm exceeds \textbf{(1b)} by $+0.66 \pm 0.19$, $+1.71 \pm 0.27$, $+1.40 \pm 0.27$, and $+0.48 \pm 0.07$ in the randomized, aligned, random, and misspecified $K = 3$ scenarios and by
$+0.27 \pm 0.08$, $+0.79 \pm 0.13$, $+1.10 \pm 0.17$, and
$+0.04 \pm 0.03$ at $K = 8$ (the upper-quartile arm by $+0.15 \pm 0.06$, $+1.71 \pm 0.27$, $+1.40 \pm 0.27$, and $+0.40 \pm 0.06$ at $K = 3$ and
by $0$, $+0.58 \pm 0.13$, $+0.77 \pm 0.17$, and $+0.00 \pm 0.01$ at
$K = 8$), the cost of routing environments to the family that is behind
in every core configuration. Its miss on the
uniform-recruitment probe is $-0.010 \pm 0.004$ ($2.5$ standard errors),
where it keeps the sample mean family in seven environments although the point estimate favors the doubly robust family, the sample size sitting on the threshold.
The upper-quartile rule, which requires the edge to exceed the upper
quartile of the inflation term, about half an interquartile range above its
median, keeps the sample mean family under randomization except in $12$
environments of the $K = 3$ scenario, one at $H = 400$, and three of the
ten anchor environments, at little cost
there ($-0.003 \pm 0.002$ at $K = 3$; $-0.009 \pm 0.005$ under uniform
recruitment, $1.8$ standard errors); under confounding it routes fewer
environments to the doubly robust family ($-0.009 \pm 0.004$ and
$-0.010 \pm 0.005$ against \textbf{(1b)} in the aligned and random $K = 8$ scenarios, $-0.019 \pm 0.008$ in the misspecified $K = 3$ scenario). Both rules see the confounded crossing
at $H = 400$ in the sense of the theorem, by selecting the family that
removes the bias, and the co-primary metric there shows the doubly robust
family ahead by $8.5 \pm 4.2$ points ($2.0$ standard errors) at matched
false positive rate while
the AUC contrast is $+0.007 \pm 0.013$ in favor of the sample mean. The
worst case band that enters the decision averages $0.87$ and $0.86$ over the environments of the aligned $K = 3$ and $K = 8$ scenarios, $0.86$ and $0.85$ under random orientation, whose $\|\alpha\|$ is the same, and $0.56$ and $0.57$ in the misspecified scenarios; the realized band reported alongside, $|c_{k}\,\alpha^{\top} w_{T,k}|$ averaged over the subpopulations, is $0.62$ and $0.56$ under aligned, $0.55$ and $0.55$ under random, and $0.39$ and $0.37$ under misspecified confounding. Two
sensitivities are recorded. A variance-only step (iii) with the
design stage curve has an infinite threshold in every environment; with
the design-embedded confounded curve it would have found finite
thresholds ($1$--$13$ patients per arm) in six of the first ten
environments at $K = 3$ and three at $K = 8$, because that curve is
diluted to the edge by the fallback cells. And the band rule itself, run with the design-embedded
confounded curve, would route every one of the first ten environments of
every confounded scenario to the doubly robust family under the median rule (the
upper-quartile rule all but two misspecified environments of the first
ten at $K = 3$ and one at $K = 8$), since against an inflation at the
edge any positive band decides; that routing is a property of the
fallback share of the held-out runs and not of the crossing condition,
and the selector arm
reported in the tables uses the design stage curve.

\emph{The selector with the recommended pairing.} The selector arm of this section recruits and analyzes with the family it selects, so where it routes to the doubly robust family it also follows that family's recruitment path. The protocol of the fourth recommendation recruits with the sample mean functional in every trial and routes the analysis only. Its arm is built environment by environment from the doubly robust analysis of the data recruited by the sample mean family (the mismatch arm recruit-1b/analyze-2b of Section~\ref{sec:supp-mismatch}, run on every environment of the eight core scenarios) where the median rule routes to the doubly robust family and from \textbf{(1b)} elsewhere; the mismatch arm follows the stored path of \textbf{(1b)} exactly (identical recruitment counts in every environment). Table~\ref{tab:recommended} compares it with \textbf{(1b)} and with the matched selector arm of Table~3 in AUC, at the zero threshold and at matched false positive rate, and in design-level mean squared error; every column is the arm minus the reference named. The arm is within two standard errors of the better family in AUC in every core scenario ($-0.005$ to $+0.012$ AUC; under the upper-quartile rule as well), above the matched arm by $0.001$ to $0.035$ AUC (beyond two standard errors in six of the eight scenarios), and removes the sample mean's false positive rate excess where the bias is aligned ($-13.3 \pm 2.0$ points against \textbf{(1b)} at $K = 3$, the whole excess, and $-3.8 \pm 0.9$ at $K = 8$, two thirds of the $+5.7 \pm 1.1$ of Table~2, leaving it $+1.9 \pm 0.8$ above \textbf{(2b)}). Its design-level error is within $0.09$ of the sample mean family's and $0.04$ to $1.76$ below the matched arm's (beyond two standard errors in seven of the eight scenarios), whose targeted fits are starved on the subpopulations of the doubly robust path; the pairing removes about nine tenths of the matched arm's excess ($90\%$ and $92\%$ in the aligned and random $K = 8$ scenarios, all or nearly all of it elsewhere), and the remainder, $+0.08 \pm 0.02$ and $+0.09 \pm 0.03$ against \textbf{(1b)} at $K = 8$ under aligned and random confounding ($3.3$ and $3.2$ standard errors), is the doubly robust analysis's own error on the starved subpopulations of the sample mean path ($+0.12 \pm 0.03$ and $+0.11 \pm 0.03$ for the mismatch arm against \textbf{(1b)} over all environments, Section~\ref{sec:supp-mismatch}). The discrimination cost of Table~3 is therefore a cost of the recruitment path, not of the routed analysis, and its error cost is largely one.

\begin{table}[!htbp]
\centering
\caption{The selector with the recommended pairing (median rule): recruit with the sample mean functional, analyze with the routed family.}
\label{tab:recommended}
\footnotesize
\setlength{\tabcolsep}{2pt}
\resizebox{\ifdim\width>\linewidth\linewidth\else\width\fi}{!}{%
\begin{tabular}{@{}lccccccc@{}}
\toprule
Scenario & DR & AUC$-$(1b) & AUC$-$matched & FPR$-$(1b) & TPR$_{\mathrm{m}}-$(1b) & MSE$-$(1b) & MSE$-$matched \\
\midrule
\multicolumn{8}{@{}l}{$K = 3$, $\bar n^{(a)} \approx 26$} \\
Randomized & 35 & $+0.005 \pm 0.003$ & $+0.006 \pm 0.007$ & $-0.4 \pm 0.3$ & $+0.2 \pm 1.1$ & $-0.02 \pm 0.01$ & $-0.69 \pm 0.19$ \\
Aligned & 99 & $+0.010 \pm 0.010$ & $+0.035 \pm 0.008$ & $-13.3 \pm 2.0$ & $+2.3 \pm 2.4$ & $-0.05 \pm 0.04$ & $-1.76 \pm 0.27$ \\
Random & 99 & $+0.012 \pm 0.010$ & $+0.019 \pm 0.007$ & $-3.2 \pm 1.9$ & $+3.8 \pm 1.9$ & $-0.07 \pm 0.03$ & $-1.47 \pm 0.27$ \\
Misspecified & 62 & $-0.005 \pm 0.005$ & $+0.015 \pm 0.006$ & $+0.6 \pm 1.0$ & $-0.9 \pm 1.3$ & $+0.02 \pm 0.02$ & $-0.46 \pm 0.07$ \\
\addlinespace[4pt]
\multicolumn{8}{@{}l}{$K = 8$, $\bar n^{(a)} \approx 10$} \\
Randomized & 21 & $+0.001 \pm 0.001$ & $+0.005 \pm 0.002$ & $+0.2 \pm 0.2$ & $+0.5 \pm 0.5$ & $+0.01 \pm 0.01$ & $-0.26 \pm 0.08$ \\
Aligned & 40 & $+0.001 \pm 0.003$ & $+0.015 \pm 0.004$ & $-3.8 \pm 0.9$ & $+2.7 \pm 1.4$ & $+0.08 \pm 0.02$ & $-0.71 \pm 0.13$ \\
Random & 40 & $-0.005 \pm 0.004$ & $+0.013 \pm 0.005$ & $-0.1 \pm 0.9$ & $+0.1 \pm 1.5$ & $+0.09 \pm 0.03$ & $-1.01 \pm 0.17$ \\
Misspecified & 4 & $-0.001 \pm 0.001$ & $+0.001 \pm 0.001$ & $+0.4 \pm 0.2$ & $-0.7 \pm 0.5$ & $0.00 \pm 0.00$ & $-0.04 \pm 0.03$ \\
\bottomrule
\end{tabular}}
\par\vspace{4pt}
\begin{minipage}{\linewidth}\footnotesize
NOTE: For each core scenario the number of environments routed to the doubly robust analysis (DR; of $50$, or $100$ in the confounded $K = 3$ scenarios), the AUC of the arm minus that of \textbf{(1b)} and minus that of the matched selector arm of Table~3, its zero-threshold false positive rate minus that of \textbf{(1b)}, its true positive rate minus that of \textbf{(1b)} at the arm's zero-threshold false positive rate (both arms re-thresholded by the quantile rule; TPR$_{\mathrm{m}}$; percentage points), and its design-level mean squared error minus that of \textbf{(1b)} and minus that of the matched arm; every column is the arm minus the reference named, so a positive AUC or TPR$_{\mathrm{m}}$ entry and a negative FPR or MSE entry favor the arm. Mean $\pm 1$ paired standard error across environments; the AUC and TPR$_{\mathrm{m}}$ columns use the environments with a defined AUC ($39$ of $50$ in the randomized $K = 3$ scenario, $80$ of $100$ in the confounded $K = 3$ scenarios, all $50$ at $K = 8$), the other columns all environments. In the routed environments the arm is the mismatch arm recruit-1b/analyze-2b of Section~\ref{sec:supp-mismatch}, run on every environment of the eight core scenarios; its recruitment counts equal those of \textbf{(1b)} in every environment.
\end{minipage}
\end{table}

\emph{Arm-conditional edges and overlap factors.} In the decomposition
$E_{k}^{c}(a) = E_{k} - \gamma_{k,a}\,(\alpha^{\top} w_{T,k})^{2}$ of the
remark before Proposition~4,
$\gamma_{k,a} = (\sigma_{x}^{2} - \tau_{k,a}^{2})/\|\alpha\|^{2}$ with
$\tau_{k,a}^{2}$ the arm-conditional variance of the standardized index
$\alpha^{\top} x_{kj}/\|\alpha\|$; when the selection weight is
log-concave in the index the tilted Gaussian law of the index has
$\tau_{k,a}^{2} \le \sigma_{x}^{2}$, so $\gamma_{k,a} \ge 0$, whereas
under the quadratic index $\expit(\alpha^{\top}x + (\alpha^{\top}x)^{2})$
of the misspecified mechanism only the complement of the treated-arm
weight is log-concave, so $\gamma_{k,0} \ge 0$ while $\gamma_{k,1} < 0$
in most subpopulations. In the theorem the term $\bar\gamma_{k}\, s(\bar n)\, u^{2}$ that the
arm-conditional edge adds to $\Delta\mathrm{MSE}_{k}$ is absorbed into the edge:
for $\bar\gamma_{k} \ge 0$ the hypothesis on the small-sample inflation is
unaffected, the proof of (ii) uses a connectivity argument that does not
need monotonicity in $|u|$, and the comparative statics in $|u|$ of part
(iv) hold with $c_{k}^{2}$ replaced by $c_{k}^{2} - \bar\gamma_{k} s(\bar n)$
wherever that difference is positive, where $c_{k}$ is the between-arm
shift coefficient $\{\E[\alpha^{\top} x_{kj} \mid A_{kj} = 1] -
\E[\alpha^{\top} x_{kj} \mid A_{kj} = 0]\}/\|\alpha\|^{2}$, equal
to~(5) under a linear index; under the linear-index
mechanisms the condition reads $\bar n > 2\bar\gamma_{k}/c_{k}^{2}$. The
quantities quoted in the remarks after Proposition~4
and Theorem~1 were computed on the first ten environments
of the confounded scenarios. The ratio $E_{k}^{c}(a)/E_{k}$ has medians
$0.81$--$0.84$ (aligned) and $0.85$--$0.93$ (randomly oriented) in both
arms; in the misspecified scenario it is $1.02$--$1.06$ in the treated and
$0.49$--$0.57$ in the control arm. Under the linear-index mechanisms
$\bar\gamma_{k} s(\bar n) \le 0.02$ against $c_{k}^{2} \ge 0.36$ at
$\bar n \ge 10$; under the quadratic index the shift coefficient is
smaller ($c_{k}^{2}$ has median $0.09$ at $K = 3$ and $0.14$ at $K = 8$)
and $c_{k}^{2} - \bar\gamma_{k} s(\bar n)$ is nonpositive in $5$ of $30$
and $15$ of $80$ subpopulations, in which the mean squared error
difference increases
with $|u|$. In these scenarios $20$--$48\%$ of the patients of a subpopulation have a true propensity outside $[0.05, 0.95]$; $\omega_{k}$
has median $6$--$8$ (it is infinite in the misspecified scenario),
$\omega_{k,\mathrm{lin}}$ has median $1.5$ ($1.1$ in the misspecified
scenario), the truncated weights with a known outcome regression would
give a factor of $1.7$--$1.8$ under the linear-index mechanisms, and
$\omega_{k}^{\dagger}$ has median $2.5$--$2.6$ (quartiles $2.4$--$2.8$)
under aligned and randomly oriented confounding and $1.1$--$1.2$ in the
misspecified scenario, in which the working logistic model reaches the
truncation bounds for few patients in most subpopulations (median share
$1\%$ at $K = 3$ and $7\%$ at $K = 8$, but above $90\%$ in a few). Hence
$\kappa_{k,\infty} \approx 1.5$--$1.6$ exceeds
$\bar E_{k}^{c}/\sigma_{k,\mathrm{cond}}^{2} \approx 0.3$ in every
subpopulation under aligned and randomly oriented confounding, whereas in
the misspecified scenario $\kappa_{k,\infty}$ (medians $0.11$ at $K = 3$
and $0.23$ at $K = 8$) is of the order of the edge (medians $0.23$ and
$0.18$) and lies below it in $57\%$ and $45\%$ of the subpopulations.

% ===========================================================================

\FloatBarrier
\subsection{Sensitivity Runs}
\label{sec:supp-sens}
Five sensitivity runs vary implementation choices of the main analysis (the fallback threshold, the outcome model, the member of the doubly robust family and its recruitment curve, and the environments behind the design stage curve), and a sixth simulates the inflation curve of the earlier targeting step. Each run is compared with the stored cell on the same environments and seeds (paired standard errors across environments; the sample mean arm is the stored one wherever the change does not touch it); the settings, the commands, and the contrast tool are documented in the README of the code.

\emph{Fallback threshold.} With the fallback of the doubly robust arm
means raised from six to ten and to fourteen patients per subpopulation
(two per arm unchanged), on the first $50$ environments of the aligned
$K = 3$ scenario the design-level mean squared error gap
$\mathrm{MSE}_{2} - \mathrm{MSE}_{1}$ falls from $+1.78 \pm 0.35$ to
$+0.12 \pm 0.10$ and $+0.12 \pm 0.11$, the AUC contrast from
$+0.021 \pm 0.022$ to $-0.003 \pm 0.019$ and $-0.006 \pm 0.020$, the
false positive rate excess of the sample mean stays at $15.5 \pm 3.0$ and
$14.3 \pm 3.0$ points (against $14.2 \pm 3.1$), and the matched-FPR
verdict at $-5.4 \pm 3.9$ and $-5.8 \pm 4.0$ (against $-4.1 \pm 4.0$),
with $22\%$ and $27\%$ of the cells on the fallback at the horizon (against $16.5\%$). In the random $K = 8$ scenario the gap falls from
$+1.37 \pm 0.16$ to $+0.16 \pm 0.04$ and $+0.05 \pm 0.04$, the AUC
contrast from $+0.020 \pm 0.007$ to $+0.002 \pm 0.005$ and
$-0.006 \pm 0.005$, the false positive rate excess rises from
$+0.4 \pm 1.1$ to $+2.3 \pm 1.0$ and $+2.2 \pm 1.0$ points, the matched-FPR
verdict moves to $-5.1 \pm 2.1$ and $-7.9 \pm 2.4$ (against $-1.7 \pm 1.6$),
and the fallback share rises to $49\%$ and $57\%$ (against $37\%$). The mean squared error deficit of the doubly robust family is thus mostly the cost of the targeted fits on six to thirteen patients: a threshold of ten removes it in the aligned $K = 3$ scenario and about nine tenths of it in the random $K = 8$ scenario ($+0.16 \pm 0.04$ remains, $4.0$ standard errors; $+0.05 \pm 0.04$ at fourteen), and the bias removal in the remaining targeted cells keeps the false positive rate advantage at $K = 3$ (in the random $K = 8$ scenario the sample mean had no excess under the six-patient rule and shows one of about $2$ points at ten and fourteen, with fewer targeted cells). These runs recruit with the design-embedded curve estimated under the six-patient rule. The design-level statements of Sections~6.5 and~7 are therefore largely statements about the six-patient rule.

\emph{Misspecified outcome model.} With the mean-zero quadratic term $0.25\,(\|\eta_{kj}\|^{2} - \Dx\sigma_{x}^{2})$ added to the post-treatment outcome of every patient (identically in both arms, so the effects are unchanged; standard deviation about $0.5$ against the outcome noise of $1$; the pre-treatment outcomes, the oracle recruitment inputs, and the design-embedded curve estimated under the linear outcome are unchanged), the linear outcome model of family 2 is wrong. Under confounded assignment the term also adds to the sample mean's bias, since $\E[\|\eta_{kj}\|^{2} \mid A_{kj}]$ depends on the arm: by $0.17$ on average in magnitude against $0.62$ and $0.57$ for the linear part $c_{k}\alpha^{\top} w_{T,k}$ (by quadrature on the first $50$ environments; the realized band of Section~\ref{sec:supp-inflation}, $0.62$ and $0.56$, is the same quantity by Monte Carlo over all environments, $100$ at $K = 3$), with the sign of $-\alpha^{\top} x_{k}$, so that the total bias changes sign in $6$ of $150$ subpopulations at $K = 3$ and $35$ of $400$ at $K = 8$. In the aligned
scenarios the sample mean's false positive rate excess remains
$15.2 \pm 3.5$ points at $K = 3$ and $7.3 \pm 1.4$ at $K = 8$ (against
$14.2 \pm 3.1$ and $5.7 \pm 1.1$ under the linear outcome), the AUC
contrasts are $+0.009 \pm 0.024$ and $+0.006 \pm 0.006$, the matched-FPR
verdicts $-7.8 \pm 5.0$ and $-7.3 \pm 2.0$, and the mean squared error gaps
$+1.48 \pm 0.18$ and $+1.17 \pm 0.15$; the AIPW comparator at the same
weights exceeds the targeted estimator by $+0.017 \pm 0.004$ and
$+0.010 \pm 0.002$ AUC and has the smaller error by $0.55 \pm 0.05$ and
$0.34 \pm 0.02$. The false positive rate advantage persists because the linear part of the bias, which dominates, is removed by any adjustment that is linear in the covariates, whatever the propensity; the probe is not a demonstration of double robustness. With the truncation of the fitted propensity binding for about a quarter of the patients, the propensity side cannot carry the consistency: the large-sample bias of the targeted estimator under this outcome has median magnitude $0.07$ and $0.08$ at $K = 3$ and $K = 8$ over the $50$ environments of the probe (zero with the linear outcome, and zero without the truncation; the quadratic part averages $0.17$ in magnitude), with the sign opposite to that part in $94\%$ and $97\%$ of the subpopulations, so the estimator over-corrects it: in median over the subpopulations its limit is $0.62$ and $0.61$ of the quadratic part (a least-squares slope on that part of $-0.21$ and $-0.24$), whereas the AIPW comparator's limit is $0.25$ and $0.22$ of it in median, with the opposite sign in $63\%$ and $60\%$ of the subpopulations but a positive slope ($+0.49$ and $+0.42$), so it under-corrects where the quadratic part is large. The crossing condition and the selector's band, which assume the linear outcome, were not re-derived for this process, and the probe is exploratory.

\emph{Other members of the family.} Run as recruiting arms in place of the targeted estimator, the one-step AIPW and the regression-adjusted (plug-in) estimators change the family contrasts. Each recruits with its own design stage inflation curve at the confounded or randomized strength (the AIPW curve of Figure~\ref{fig:kappa} and a curve simulated for the regression-adjusted estimator), whereas the stored targeted arm recruits with the design-embedded curve; the design stage curves are much steeper at $4$--$6$ patients per arm (isotonic medians $1.9$ and $1.1$ at $4$ and $6$ for AIPW under confounding against $0.47$ and $0.43$ for the embedded targeted curve), which pushes recruitment toward the small subpopulations and lowers the fallback share at the horizon (from $0.39$ to $0.36$ and $0.33$ under aligned confounding at $K = 8$, from $0.35$ to $0.31$ under randomization). In
the aligned $K = 3$ scenario the AUC contrast \textbf{(1b)} minus the
family-2 arm is $-0.018 \pm 0.020$ (AIPW) and $-0.021 \pm 0.021$
(regression) against $+0.021 \pm 0.022$ for the targeted arm on the same
$50$ environments, the false positive rate excess of the sample mean
$17.2 \pm 3.2$ and $18.1 \pm 3.3$ points (against $14.2 \pm 3.1$), the
matched-FPR verdict $-11.6 \pm 4.9$ and $-15.6 \pm 4.8$ (against
$-4.1 \pm 4.0$), and the mean squared error gap $+1.01 \pm 0.37$ and
$+0.71 \pm 0.26$ (against $+1.78 \pm 0.35$). In the aligned $K = 8$ scenario the AUC contrast is $-0.006 \pm 0.005$ and $-0.013 \pm 0.005$ (against $+0.016 \pm 0.005$), the false positive rate excess
$8.0 \pm 1.1$ and $9.8 \pm 1.2$ (against $5.7 \pm 1.1$), the matched-FPR
verdict $-9.9 \pm 2.3$ and $-15.0 \pm 2.6$ (against $-2.7 \pm 1.8$), and
the error gap $+0.73 \pm 0.14$ and $+0.56 \pm 0.12$ (against
$+1.02 \pm 0.13$); under randomization at $K = 8$ the sample mean's margin falls from $+0.017 \pm 0.004$ to $+0.005 \pm 0.003$ and $+0.006 \pm 0.003$ AUC, with the matched-FPR verdict at $-6.0 \pm 2.2$ and $-6.8 \pm 2.4$ (against $-3.9 \pm 2.0$) and the error gap at $+0.66 \pm 0.10$ and $+0.58 \pm 0.10$ (against $+0.76 \pm 0.10$). To separate the curve from the estimator, the targeted arm was rerun with the AIPW design stage curve on the same environments and seeds. Under randomization the sample mean's margin over it is $+0.008 \pm 0.004$: of the $0.0116$ by which the AIPW arm narrows the margin, the curve accounts for $0.0095 \pm 0.0023$ and the estimator for $0.0021 \pm 0.0016$ (in mean squared error $-0.08 \pm 0.04$ and $-0.02 \pm 0.03$; the split takes the curve first, so an interaction between curve and estimator falls into the estimator term). Under aligned confounding the margin over it is $+0.011 \pm 0.005$: of the $0.021$, the curve accounts for $0.005 \pm 0.004$ and the estimator for $0.016 \pm 0.004$ (in mean squared error $-0.05 \pm 0.12$ and $-0.24 \pm 0.13$). The member of the family thus matters for the discrimination verdict under confounding, where the AIPW arm closes the targeted arm's gap at $K = 8$ mainly through the estimator, as the comparators at fixed weights (Table~2) suggested; under randomization its gain is mainly the recruitment curve. The regression-adjusted arm was not decomposed; its design stage curve is as steep as the AIPW curve at $4$--$6$ patients per arm and lower beyond, so the AIPW split does not transfer to it. All three remain behind in mean squared error through the same starved cells (their fallback rule is the six-patient one).

\emph{Held-out design stage curve.} The design stage confounded curve
re-estimated on the held-out environments (offset $100$) routes $99$,
$99$, and $61$ of the $100$ confounded $K = 3$ environments, $40$, $39$,
and $3$ of the $50$ at $K = 8$, and $10$ of $10$ at $H = 400$ under the median rule, against $99$, $99$, $62$, $40$, $40$, $4$, and $10$ with the in-sample curve of step (iii) (under the upper-quartile rule $99$, $99$, $56$, $29$, $25$, $1$, and $10$ against $99$, $99$, $56$, $30$, $27$, $2$, and $10$); the selector arm's deficits over all environments change by at most $0.002$ AUC ($-0.022 \pm 0.009$ against $-0.020 \pm 0.009$ in the misspecified $K = 3$ scenario). The in-sample estimation of that curve thus changes the routing by one environment per scenario (two under the upper-quartile rule) and the record by less than its resolution.

\emph{Bounded-outcome fluctuation.} Simulated at fixed counts with $60$ replications per size, the design stage randomized inflation of the bounded-outcome (logistic) fluctuation used by the preliminary runs behind the pre-specified predictions has raw bin medians $1.57$, $0.77$, and $0.11$ at $4$, $10$, and $26$ patients per arm against $0.77$, $0.16$, and $0.04$ for the linear fluctuation (two to five times larger at $4$--$10$; at $26$ the two are within the Monte Carlo noise of $60$ replications, and the raw medians are not monotone, $1.57$, $2.11$, and $1.37$ at $4$, $6$, and $8$), and its threshold $n^{*}(0)$ on the isotonic fit is about $15$ against $7$: the location of the randomized boundary is specific to the targeting step, as the remark after Theorem~1 says. The earlier basis $n^{*}(0) > 26$ of Table~\ref{tab:predictions} was not a curve estimate but an inference from the locked adaptive runs of that estimator, which reflected the recruitment path as well as the targeting step; the estimator's own fixed-count threshold is the $15$ found here.

\section{PROOFS}
\label{app:proofs}
% ===========================================================================

\subsection{Assumptions}
\label{app:assumptions}
The results of Sections~3 and~4 are
stated under the following conditions, referenced from the proofs below.
\begin{assumption}[Data generating process]
\label{ass:dgp}
The data follow the linear-factor Gaussian process of
Section~5.1, with Gaussian covariate and latent dispersions,
global pre-treatment weights, subpopulation-specific post-treatment
weights, an additive treatment effect $r_{k}$ constant in $(x, z)$ within
subpopulation, and mutually independent $\eta_{kj}$, $\zeta_{kj}$,
$\varepsilon_{kjt}$, $\varepsilon_{kjT}^{(a)}$.
\end{assumption}
\begin{assumption}[Nuisance estimation]
\label{ass:nuisance}
The outcome model $Q$ lies in a correctly specified linear class
(Section~5.1) and is fitted by least squares at the parametric
$n_{k}^{-1/2}$ rate; the propensity $g$ is fitted by a logistic working
model (Section~2), correctly specified except in the
misspecified scenario, and truncated to $[0.05, 0.95]$. Where the
truncation binds, or the working model is misspecified, the limit
$g^{\dagger}$ of the fitted propensity differs from $g$ and the product
remainder of the doubly robust expansion is of first order rather than
$o_{P}(n_{k}^{-1/2})$; its conditional mean given the covariates and the
arms is zero under the correct outcome model, so it contributes to the
variance, through the factor $\omega_{k}^{\dagger}$ of the remark after
Proposition~4, and not to the bias; the estimator is
then regular in the model with a linear outcome regression, not in the
nonparametric model.
\end{assumption}
\begin{assumption}[Counts]
\label{ass:counts}
All fixed-count statements (Propositions~2, 3, and~4 and Corollaries~1 and~2) are stated conditionally on the arm assignments, for subpopulation
sizes that are fixed in advance or chosen independently of the outcomes
and, under design-controlled assignment, for an allocation of the arms
within each subpopulation that uses the counts only (a fixed schedule, or
the variance-reduction step of the look-ahead at $\beta = e_{k}$ with
oracle variances), with both arms of each subpopulation nonempty and
$\pi_{k} \in (0,1)$; conditionally on such assignments the patients whose
arms the mechanism draws are i.i.d.\ draws from the arm-conditional law,
which is what the proofs use (the forced burn-in patients of the
simulations are not; Section~\ref{sec:supp-sim}). Under
outcome-dependent recruitment or allocation, as in the estimated-variance
variants and \textbf{(2a)}, whose look-ahead uses influence curve
variances, and in the arms whose borrowing weights depend on the running
pre-treatment means (the hard-constrained arms at $K > m$ and the
transport arms at every $K$), the statements hold as approximations
(remark below).
\end{assumption}
\begin{assumption}[variance inflation curve]
\label{ass:inflation}
$\kappa_{k} : (0,\infty) \to [0,\infty)$, a function of the per-arm count
$\bar n$, is continuous and nonincreasing with limit
$\kappa_{k,\infty} \ge 0$ as $\bar n \to \infty$. Then
$B(\bar n) = \kappa_{k}(\bar n)\,\sigma_{k,\mathrm{cond}}^{2} - E_{k}$ is
nonincreasing, has at most one sign change, and $s B$ is strictly
decreasing wherever $B > 0$ because $s$ is strictly decreasing. The proof
of Theorem~1(i) uses these properties; a single sign
change of $B$ alone suffices at $u = 0$ but not for $u \neq 0$. Strict monotonicity of $\kappa_{k}$, assumed in the theorem,
is sufficient for them but not necessary. When $B$ has a plateau at zero,
part (i) holds with $\Delta\mathrm{MSE}_{k} \le 0$ on the plateau at $u = 0$, and
$n_{k}^{*}(0)$ is taken as the smallest root, which is the operational
threshold $n^{*} = \inf\{\bar n : \kappa_{k}(\bar n)\,
\sigma_{k,\mathrm{cond}}^{2} \le E_{k}\}$ of Section~4.4.
A terminal plateau of $B$ at zero arises only when $E_{k}$ equals the
terminal value
$\kappa_{k,\infty}\sigma_{k,\mathrm{cond}}^{2}$ of the inflation term,
where the families have the same variance at large counts and the
crossing is not strict; the selector's step (ii)
uses the strict inequality $E_{k} > \kappa(\bar n)\,
\sigma_{k,\mathrm{cond}}^{2}$, so a trial exactly on the boundary is routed
to the sample mean family.
\end{assumption}
\begin{assumption}[Confounding orientation]
\label{ass:orient}
For the symmetric-orientation case of Corollary~2:
conditionally on the alignment magnitude and the remaining geometry,
$\sgn(\alpha^{\top} w_{T,k})$ is $\pm 1$ with equal probability across
environments. The random orientation scenarios of the simulation satisfy this by
construction, drawing $\alpha$ uniformly on the sphere at the confounding
scale.
\end{assumption}
\noindent
A remark on Assumption~\ref{ass:counts} is needed. The sample sizes of
the design depend on the outcomes through the sensitivity index, and so
does the allocation of the arms whenever the look-ahead uses estimated
variances or borrowing weights that depend on the running means, so the
fixed-count statements are exact for subpopulation sizes fixed in advance
or chosen independently of the outcomes and for allocations that use the
counts only, conditionally on the arm assignments (under confounded assignment the arm counts themselves are
random and correlated with the outcomes through the covariates, which is
why the conditioning is on the assignments rather than on counts drawn
independently of everything), and under adaptive recruitment they do not
hold conditionally on the realized counts or assignments, because the conditional law of
the data given the counts differs from the unconditional one
(adaptive-sampling bias). That distortion is of order $n^{-1/2}$ in the
bias and of order $n^{-1}$ in the mean squared error, the same order as
the variance terms that locate the $u = 0$ boundary, and we do not bound
it; the $\mathcal{O}(1)$ confounding bias that separates the families at
large sample sizes is unaffected. Two features of the study limit its
effect on the comparisons: the design-embedded inflation is measured as
a
variance ratio of the two analyses on identical adaptive paths, so it
carries the component common to both, and the paired contrasts on common
seeds difference out the part shared by arms with similar paths (the
families follow different paths, so the removal is partial). A formal
unconditional treatment is beyond the scope of this article.

\subsection{Proof of the Asymptotic-Variance Proposition}
This subsection proves Proposition~2.
Fix subpopulation $k$ and condition throughout on the arm assignments,
with the subpopulation size fixed in advance or chosen independently of
the outcomes and an allocation of the arms that uses the counts only
(Assumption~\ref{ass:counts}); such assignments are independent of the
patients' covariates, loadings, and noise, so the conditioning leaves
their law unchanged. \emph{Sample mean.} The outcome of a
recruited patient decomposes into independent components
$w_{T,k}^{\top}\eta_{kj}$, $\mu_{T,k}^{\top}\zeta_{kj}$, and
$\varepsilon_{kjT}^{(a)}$ around the arm mean of the subpopulation, so the arm-specific running mean
has variance $\Var(\Ybar_{k,a}) = \sigma_{k,\mathrm{eff}}^{2}/n_{k}^{(a)}$
with $\sigma_{k,\mathrm{eff}}^{2} = \sigma^{2}
+ \sigma_{x}^{2}\|w_{T,k}\|^{2} + \sigma_{z}^{2}\|\mu_{T,k}\|^{2}$.

\emph{(i) Doubly robust arm mean.} The estimand is the marginal arm mean
$\theta_{k}^{(a)}$, whose efficient influence curve given the observed
covariate is
$\psi = \ind\{A=a\}\,\{Y - Q(a,x)\}/\Pr(A=a \mid x) + Q(a,x) - \theta_{k}^{(a)}$
(Hahn 1998). Under randomized assignment
($\Pr(A=a \mid x) = n_{k}^{(a)}/n_{k}$ by design) its variance is
$\Var(\psi) = (n_{k}/n_{k}^{(a)})\,\sigma_{k,\mathrm{cond}}^{2}
+ \Var_{x}(Q(a,x))$, with $\sigma_{k,\mathrm{cond}}^{2}
= \sigma^{2} + \sigma_{z}^{2}\|\mu_{T,k}\|^{2}$ the conditional outcome
variance and, from the linear outcome model,
$\Var_{x}(Q(a,x)) = \sigma_{x}^{2}\|w_{T,k}\|^{2} = E_{k}$. Under a
correctly specified outcome model (Assumption~\ref{ass:nuisance}) the
targeted estimate is asymptotically linear with influence curve $\psi$
over the $n_{k}$ observations of the subpopulation (Hahn 1998; van~der Laan and Rubin 2006), giving
asymptotic variance
$\Var(\psi)/n_{k} = \sigma_{k,\mathrm{cond}}^{2}/n_{k}^{(a)} + E_{k}/n_{k}$.
Subtracting this from the sample mean's variance
$(\sigma_{k,\mathrm{cond}}^{2} + E_{k})/n_{k}^{(a)}$ gives the arm-level
gap $E_{k}(1/n_{k}^{(a)} - 1/n_{k}) \ge 0$.

\emph{(ii) Effect within the subpopulation.} At $\beta = e_{k}$ the doubly robust
effect estimate is the difference of the two targeted arm means, both
averaging $Q^{*}(a,\cdot)$ over the same empirical covariate distribution,
with influence curve
$\{\ind\{A=1\}/g(x) - \ind\{A=0\}/(1-g(x))\}\{Y - Q(A,x)\}
+ Q(1,x) - Q(0,x) - r_{k}$. The regression difference
$Q(1,x) - Q(0,x) = r_{k}$ is constant in $x$ (Assumption~\ref{ass:dgp}),
so its variance contribution is zero: the two covariate sampling
components have canceled exactly. The remaining term has variance
$\sigma_{k,\mathrm{cond}}^{2}\,\E[1/\{g(x)(1-g(x))\}]/n_{k}$, which under
randomized assignment with $g \equiv n_{k}^{(1)}/n_{k}$ equals
$\sigma_{k,\mathrm{cond}}^{2}\,s_{k}$, against the sample mean
difference's $\sigma_{k,\mathrm{eff}}^{2}\,s_{k}$; the effect-level gap is
$E_{k}\,s_{k}$. Both gaps vanish if and only if the covariates explain
none of the post-treatment outcome variance. \qed

\subsection{Variance of the Borrowing Decomposition at Frozen Weights on
the Individual-Level Pool}
\label{app:crossterm}
The corrected penalty rate of~(3) is
\begin{equation}
\tilde\lambda_{k}
= \lambda_{k}
+ \frac{\sigma_{x}^{2}\,\|W_{\neg T} v_{k}\|^{2}
      + \sigma_{z}^{2}\,\|M_{\neg T} v_{k}\|^{2}}{\sigma^{2}},
\qquad v_{k} = M_{\neg T}^{+}\mu_{T,k},\quad \lambda_{k} = \|v_{k}\|^{2},
\label{eq:lambda}
\end{equation}
with $W_{\neg T} = [w_{1}\,\cdots\,w_{\Tpre}]$ and
$M_{\neg T} = [\mu_{1}\,\cdots\,\mu_{\Tpre}]$ the global pre-treatment
weight matrices of the factor model of Section~5.1,
$\sigma_{x}$ and $\sigma_{z}$ the covariate and latent dispersions within
a subpopulation, and $M_{\neg T}^{+}$ the Moore--Penrose pseudoinverse.
Away from $e_{k}$ on the
individual-level pool the variance of the error representation with the
weights frozen at $\beta$ contains in addition the cross-covariance
\begin{equation}
-\,2 \sum_{l} \frac{(\beta_{l} - \ind\{l=k\})^{2}}{n_{l}}\, C_{kl},
\qquad
C_{kl} = \sigma_{x}^{2}\, w_{T,l}^{\top} W_{\neg T} v_{k}
+ \sigma_{z}^{2}\, \mu_{T,l}^{\top} M_{\neg T} v_{k},
\qquad v_{k} = M_{\neg T}^{+}\mu_{T,k},
\label{eq:crossterm}
\end{equation}
for both families, which the individual-level variants include in their
working variance.
Consider homogeneous post-treatment weights and a weight vector $\beta$
that satisfies the sum-to-one, covariate, and trajectory constraints of
the realized matrix $\Gamma$; with subpopulation-specific post-treatment
weights the decomposition carries in addition the deterministic term
$-d_{k}^{\top}\beta$ of~(4), which the numerical check
includes. Write $\xi_{l,a}$ for the sampling error of $\Ybar_{l,a}$,
$\Xi_{\neg T}$ for the $\Tpre \times K$ matrix of sampling errors of the
running pre-treatment means, $Z \in \R^{K \times \Dz}$ for the matrix
with rows $z_{l}^{\top}$, $\vartheta = \beta - e_{k}$ for the weight
deviation (the symbol $u$ is reserved for the alignment), and
$v_{k} = M_{\neg T}^{+}\mu_{T,k}$. The trajectory constraint
$\widehat{Y}_{\neg T} \vartheta = 0$ reads
$M_{\neg T}^{\top} Z^{\top} \vartheta = -\Xi_{\neg T} \vartheta$ once the covariate and
sum-to-one constraints have removed the intercepts and the covariate rows,
so that, when $M_{\neg T}$ has full row rank,
$z_{k} - \sum_{l}\beta_{l} z_{l} = -Z^{\top} \vartheta = (M_{\neg T}^{+})^{\top}
\Xi_{\neg T} \vartheta$ and $\mu_{T,k}^{\top}(z_{k} - \sum_{l}\beta_{l} z_{l})
= v_{k}^{\top} \Xi_{\neg T} \vartheta$. Hence
\begin{equation}
r_{k} - \hat r_{k}(\beta)
= -\xi_{k,1} + \beta^{\top} \xi_{\cdot,0} - v_{k}^{\top} \Xi_{\neg T}\,(\beta - e_{k}),
\label{eq:decomp}
\end{equation}
with a negative sign in front of the representation term. On the
individual-level pool patient $j$ of subpopulation $l$ contributes
$w_{T,l}^{\top}\eta_{lj} + \mu_{T,l}^{\top}\zeta_{lj} + \varepsilon_{ljT}^{(a)}$
to the post-treatment error of its arm $a$ and
$w_{t}^{\top}\eta_{lj} + \mu_{t}^{\top}\zeta_{lj} + \varepsilon_{ljt}$ to
the pre-treatment error at period $t$, so the two are correlated through
$\eta_{lj}, \zeta_{lj}$. Each of the $n_{l}^{(a)}$ patients of an arm
enters the arm mean with weight $1/n_{l}^{(a)}$ and the running
pre-treatment mean with weight $1/n_{l}$, so with
$C_{kl} = \sigma_{x}^{2} w_{T,l}^{\top} W_{\neg T} v_{k}
+ \sigma_{z}^{2} \mu_{T,l}^{\top} M_{\neg T} v_{k}$,
\begin{equation*}
\Cov\big(\beta_{l} \xi_{l,0},\, -v_{k}^{\top} \Xi_{\neg T, \cdot l}\, \vartheta_{l}\big)
= -\frac{\beta_{l} \vartheta_{l} C_{kl}}{n_{l}},
\qquad
\Cov\big(-\xi_{k,1},\, -v_{k}^{\top} \Xi_{\neg T, \cdot k}\, \vartheta_{k}\big)
= \frac{\vartheta_{k} C_{kk}}{n_{k}}.
\end{equation*}
Since $\beta_{l} \vartheta_{l} = \vartheta_{l}^{2}$ for $l \neq k$ and
$\vartheta_{k} - \beta_{k} \vartheta_{k} = -\vartheta_{k}^{2}$, the total cross-covariance is
$-\sum_{l} \vartheta_{l}^{2} C_{kl}/n_{l}$, and the variance of~\eqref{eq:decomp} at frozen weights is $V_{k}(\beta)$ of~(3)
with the marginal variances plus twice this sum, which is~\eqref{eq:crossterm}. The term
vanishes at $\beta = e_{k}$ and on the subpopulation-level pool, where the
pre- and post-treatment draws are independent. Each coefficient of
$\beta_{l}^{2}$ in this variance is itself a variance, that of
$\xi_{l,0} - v_{k}^{\top} \Xi_{\neg T,\cdot l}$, so the quadratic form stays
positive semidefinite.

For the doubly robust family under design-controlled assignment the
targeted arm mean of subpopulation $l$ equals the arm mean plus the
fitted outcome model evaluated at the difference between the covariate
mean of all $n_{l}$ patients and that of the arm,
$\hat\theta_{l}^{(a)} = \Ybar_{l,a} + \hat w^{\top}(\bar x_{l} - \bar x_{l,a})
+ \mathcal{O}_{P}(1/n_{l})$ (under confounded assignment the remainder is
$\mathcal{O}_{P}(n_{l}^{-1/2})$ and the functional below is a working
approximation), so it carries the covariate sampling error
$w_{T,l}^{\top}\bar\eta_{l}$ of the
whole subpopulation in place of that of the arm. Writing $\tilde\xi_{l,a}$
and $\widetilde \Xi_{\neg T}$ for the errors without their $\eta$ components,
\begin{equation}
r_{k} - \hat r_{k}(\beta)
= -\tilde\xi_{k,1} + \beta^{\top}\tilde\xi_{\cdot,0}
+ \sum_{l} \vartheta_{l}\,(w_{T,l} - W_{\neg T} v_{k})^{\top}\bar\eta_{l}
- v_{k}^{\top}\widetilde \Xi_{\neg T}\, \vartheta + \mathcal{O}_{P}(1/n),
\label{eq:decompdr}
\end{equation}
whose variance is $V_{k}(\beta)$ with the conditional variances, plus the
covariate sampling terms $\sum_{l} \vartheta_{l}^{2} E_{l}/n_{l}$
discussed after Proposition~2, plus~\eqref{eq:crossterm} with the full
constant $C_{kl}$: the covariate block collapses to
$\sum_{l} \vartheta_{l}^{2}\,\sigma_{x}^{2}\|w_{T,l} - W_{\neg T} v_{k}\|^{2}/n_{l}$.
In the first ten $K = 8$ environments of the simulation study the ratio
$2|C_{kl}| / (E_{l} + \tilde\lambda_{k}\sigma^{2})$, taken over all $K^{2}$
pairs $(k, l)$ of each environment since the sum runs over $l = k$ as
well, has median $0.52$ (quartiles $0.22$ and $0.82$), so the term is of the same order as the
coefficients it corrects, and $C_{kl} > 0$ in $57\%$ of the pairs.

Both identities were checked numerically with the estimators as
implemented at $K = 9 > m$, with $\beta$ drawn
from the feasible set of the realized constraints in each of $2{,}000$
draws at $15$ patients per arm and $10{,}000$ at $200$, in two configurations:
one in which the cross term is small
($C_{kl} = 0.08$, the $\sigma_{x}^{2}$ and $\sigma_{z}^{2}$ parts nearly
canceling) and one in which $2|C_{kl}|/(E_{l} + \tilde\lambda_{k}\sigma^{2})
= 0.54$, close to the median $0.52$ of the $K = 8$ environments. The sample mean
identity~\eqref{eq:decomp} holds to below $10^{-14}$ with the negative sign in both and fails by up to $3.2$ with a positive sign. The
doubly robust identity~\eqref{eq:decompdr} has largest residuals of
$0.90$ and $1.5$ with $15$ patients per arm and $0.075$ and $0.083$ with
$200$, the $\mathcal{O}_{P}(1/n)$ nuisance-estimation error. The
empirical variance at the minimizer of each functional is compared with
the functional's value: at $15$ patients per arm the
error of the sample mean has variance $0.335 \pm 0.011$ against a functional
value of $0.334$ in the first configuration and $0.329 \pm 0.010$ against
$0.322$ in the second, and the doubly robust error has variance
$0.228 \pm 0.007$ against $0.211$ and $0.233 \pm 0.007$ against $0.204$,
the excess at $15$ patients per arm being in part the finite-sample
inflation of the nuisance fits that the variance inflation curve of
Section~6.5 measures. At $200$ patients per arm
both functionals agree with the empirical variances within $2\%$ in both
configurations ($0.0255 \pm 0.0004$ against $0.0250$ and $0.0238 \pm
0.0003$ against $0.0241$ for the sample mean, $0.0158 \pm 0.0002$ against
$0.0158$ and $0.0151 \pm 0.0002$ against $0.0153$ for the doubly robust
estimator; with $500$ draws the gaps in the second configuration were $8$ and $17\%$, Monte Carlo noise at that size), so the
doubly robust excess at $15$ patients per arm shrinks with the count as
nuisance inflation does. The qualifier of the frozen-weight statement
remains an argument rather than a measured gap: at a large cross term the
minimizer $\beta^{*}$ depends on the same pre-treatment noise that enters
the variance, and $V_{k}(\beta^{*})$ is then not exactly the variance of
$\hat r_{k}(\beta^{*})$; at these sizes the discrepancy is below the
$2\%$ resolution of the check.
With subpopulation-specific post-treatment weights, the setting of the
simulations, the sample mean identity holds to $4 \times 10^{-15}$ as well
($2{,}000$ draws at $15$ patients per arm), and the empirical variances
match the functionals within $1.5$ standard errors ($0.311 \pm 0.010$
against $0.316$ for the sample mean, $0.211 \pm 0.007$ against $0.202$
for the doubly robust estimator). This comparison checks the level of the
functionals and
not the cross term on its own: with seven hard constraints at $K = 9$ the
minimizers of the frozen-weight functional with and without the term
differ little, and their empirical variances ($0.329$ against
$0.332$ in the second configuration) lie inside one standard error of
each other, so the sign and size of the term rest on the pointwise
identity. Finally, the coefficient of $\vartheta_{l}^{2}$ alone,
$(\tilde\lambda_{k}\sigma^{2} + E_{l} - 2C_{kl})/n_{l}$, is not a variance
and is negative for $59$ of the $640$ (target, donor) pairs of those ten
$K = 8$ environments ($163$ of $640$ for the coefficient of the sample
mean without
$E_{l}$); the coefficient of $\beta_{l}^{2}$ stated above is what stays
nonnegative. One consequence for the recruitment is that the look-ahead
value of the individual-level arms need not fall when a treated patient
is added to such a donor, since that raises $n_{l}$ but not
$n_{l}^{(0)}$; the look-ahead therefore ranks the $2K$ candidates by
value and does not rely on monotonicity. \qed

\subsection{Proof of the Directional-Bias Proposition}
This subsection proves Proposition~3.
\label{app:proof-bias}
Throughout, expectations are over the patient draws and the outcome
noise, conditionally on the environment and on the arm assignments, for
subpopulation sizes fixed in advance or chosen independently of the
outcomes, with both arms of each subpopulation nonempty
(Assumption~\ref{ass:counts}); the remark after Assumption~\ref{ass:counts} describes the distortion that outcome-dependent recruitment induces, which these statements do not cover.

\emph{Step 1 (Stein's identity).} For $x \sim \mathcal{N}(\mathsf{m}, \sigma_{x}^{2}
I_{\Dx})$ and differentiable $f$ with integrable gradient, Gaussian
integration by parts gives $\E[(x - \mathsf{m})\,f(x)] = \sigma_{x}^{2}\,
\E[\nabla f(x)]$ (Stein 1981). With $f = g$, the logistic propensity of
Section~2, $\nabla g(x) = g(x)\big(1-g(x)\big)\,\alpha$, so
\begin{equation}
\Cov\big(x_{kj},\, g(x_{kj})\big)
= \sigma_{x}^{2}\, \E\big[g(x_{kj})\big(1-g(x_{kj})\big)\big]\, \alpha .
\label{eq:steincov}
\end{equation}

\emph{Step 2 (arm-conditional covariate means).} With
$\pi_{k} = \E[g(x_{kj})]$, the tower property over $A_{kj} \mid x_{kj}$ gives
$\pi_{k}\,\E[x_{kj} \mid A_{kj}=1] = \E[x_{kj}\, g(x_{kj})]$ and its
control-arm analogue, hence
\begin{equation}
\E[x_{kj} \mid A_{kj}=1] = x_{k} + \tfrac{1}{\pi_{k}}
\Cov\big(x_{kj}, g(x_{kj})\big),
\qquad
\E[x_{kj} \mid A_{kj}=0] = x_{k} - \tfrac{1}{1-\pi_{k}}
\Cov\big(x_{kj}, g(x_{kj})\big),
\end{equation}
so the between-arm separation is $\Cov/\{\pi_{k}(1-\pi_{k})\} = c_{k}\alpha$
with $c_{k}$ as in~(5), and the arm-specific shifts are
$(1-\pi_{k})\,c_{k}\alpha$ (treated) and $-\pi_{k}\,c_{k}\alpha$ (control).
Since $\E[g(1-g)] = \pi_{k}(1-\pi_{k}) - \Var(g(x_{kj}))$ and
$0 < g < 1$ strictly, $c_{k} = \sigma_{x}^{2}\{1 -
\Var(g)/[\pi_{k}(1-\pi_{k})]\} \in (0, \sigma_{x}^{2}]$, with
$c_{k} \to \sigma_{x}^{2}$ as $\Var(g) \to 0$.

\emph{Step 3 (latent loadings do not shift).} $A_{kj}$ is a function of
$x_{kj}$ and exogenous randomness, and $\zeta_{kj} \perp \eta_{kj}$, so
$z_{kj} \perp A_{kj}$ and $\E[z_{kj} \mid A_{kj}=a] = z_{k}$ for both arms.

\emph{Step 4 (outcome bias).} Conditionally on the arm assignments the
patients of an arm are i.i.d.\ draws from the arm-conditional law, because
the pairs $(x_{kj}, A_{kj})$ are independent across patients and $A_{kj}$
depends on $x_{kj}$ and an independent uniform draw only
($A_{kj} = \ind\{U_{kj} < g(x_{kj})\}$), so the arm mean satisfies
$\E[\Ybar_{k,a}] = \E[Y_{kjT}^{(a)} \mid A_{kj}=a]$ for every realized
count $n_{k}^{(a)} \ge 1$ (all expectations here are conditional on the
assignments), and the within-arm independence gives the variance
$(\sigma_{k,\mathrm{cond}}^{2} + E_{k}^{c}(a))/n_{k}^{(a)}$ of each arm
mean in Proposition~4; hence
\begin{equation}
\E[\Ybar_{k,1}] = \bar y_{kT} + r_{k}
+ (1-\pi_{k})\,c_{k}\,\alpha^{\top} w_{T,k},
\qquad
\E[\Ybar_{l,0}] = \bar y_{lT} - \pi_{l}\,c_{l}\,\alpha^{\top} w_{T,l},
\end{equation}
with $\bar y_{kT}$ the unconditional mean of the subpopulation. Substituting into
$\hat r_{k}(\beta)$ and separating off the terms that vanish under
homogeneous weights (the representation bias of~(4)) yields
the confounding component~(6); at $\beta = e_{k}$ the two shifts add to
$c_{k}\,\alpha^{\top} w_{T,k}$, which is~(5) and is
exact for such counts.

\emph{Step 5 (removal by the doubly robust family).} The targeted arm mean
solves the efficient-influence curve estimating equation, so its bias is
the second-order remainder; for the treated mean,
\begin{equation}
\E[\hat\theta_{k}^{(1)}] - \theta_{k}^{(1)}
= \E\!\left[\frac{g(x) - \hat g(x)}{\hat g(x)}
\Big(Q(1, x) - \hat Q(1, x)\Big)\right] + o\big(n_{k}^{-1/2}\big),
\end{equation}
and symmetrically for the control mean (Bang and Robins 2005; van~der Laan and Rubin 2006);
conditionally on the arm assignments each targeted arm mean carries in
addition the term $c_{k}\alpha^{\top}w_{T,k}\,(n_{k}^{(1)}/n_{k} - \pi_{k})$,
the covariate shift of the whole subpopulation relative to its
expectation, which is common to both arms and cancels in the contrast.
With the outcome model correctly specified and fitted by least squares,
$\E[\hat Q(1, x) - Q(1, x) \mid x, A] = 0$, so the remainder has mean zero
whatever the limit $g^{\dagger}$ of the truncated propensity fit and the
confounding bias is removed to leading order; when $g^{\dagger} \neq g$
the remainder is $\mathcal{O}_{P}(n_{k}^{-1/2})$ rather than of smaller
order and contributes to the variance (the factor $\omega_{k}^{\dagger}$),
not to the bias. With the propensity correctly specified, untruncated, and
fitted at the parametric rate, the symmetric argument applies. In the misspecified
scenario of Section~5 the outcome model carries the
consistency. \qed

\subsection{Proof of the Corollary on Removal by Borrowing}
This subsection proves Corollary~1.
Write $u_{l} = \alpha^{\top} w_{T,l}$; $\sgn(u_{l})$ is constant in $l$ by
assumption. From~(6),
\begin{equation*}
b_{k}(\beta)
= \sgn(u_{k})\,\Big[(1-\pi_{k})\,c_{k}\,|u_{k}|
+ \sum_{l} \beta_{l}\,\pi_{l}\,c_{l}\,|u_{l}|\Big],
\end{equation*}
where every summand is nonnegative because $\beta_{l} \ge 0$ on the
simplex, $\pi_{l} \in (0,1)$, and $c_{l} > 0$
(Proposition~3). The bracket is a strictly positive constant
plus a convex combination of the strictly positive terms
$\pi_{l} c_{l} |u_{l}|$, hence bounded below by
$(1-\pi_{k})c_{k}|u_{k}| + \min_{l}\pi_{l}c_{l}|u_{l}|$, which is~(7); the bound involves no sample size. \qed

\subsection{Proof of the Corollary on the Visibility of Bias Correction}
This subsection proves Corollary~2.
By asymptotic normality of $\hat r_{k}$ around $r_{k} + b_{k}$,
$\Pr(\hat r_{k} > 0) = \Phi\big((r_{k} + b_{k})/\mathrm{se}_{k}\big)$ up to
lower-order terms, and a first-order expansion in $b_{k}$ gives
$\Phi(r_{k}/\mathrm{se}_{k}) + \phi(r_{k}/\mathrm{se}_{k})\,
b_{k}/\mathrm{se}_{k} + \mathcal{O}(b_{k}^{2})$. Averaging the increment over
environments within the sets $\{r_{k} > 0\}$ and $\{r_{k} \le 0\}$ yields~(8). Under the symmetric-orientation condition, conditioning
on the magnitude $|b_{k}|$ and the remaining geometry leaves
$\E[\sgn(b_{k}) \mid \cdot\,] = 0$ (this conditioning requires that $\mathrm{se}_{k}$ carry no information on $\sgn(b_{k})$, which holds for fixed counts; under adaptive recruitment the standard error depends on the path and the path on the running estimates, so inside the design the statement is an approximation, as the remark after Assumption~\ref{ass:counts} says), so both expectations in~(8) vanish and the first-order aggregate shift is zero,
which leaves the variance difference as the gap between the families;
under aligned orientation $\sgn(b_{k})$ is positive in most
subpopulations (Section~5.2), so the increments of those
subpopulations share a sign and the gap, conditionally on $r_{k}$, equals
$\E_{\mathrm{env}}[c_{k}\,|\alpha^{\top} w_{T,k}|\,
\phi(r_{k}/\mathrm{se}_{k})/\mathrm{se}_{k} \mid r_{k}]$ in absolute value
up to the minority of opposite sign. Per-environment
error measures that are even in $b_{k}$, such as mean absolute deviation, are
first-order unaffected by the orientation and expose the bias in either case.
\qed

\subsection{Proof of the Crossing-Condition Proposition}
This subsection proves Proposition~4.
Fix the arm assignments (Assumption~\ref{ass:counts}). At $\beta = e_{k}$ the
sample mean estimate is $\Ybar_{k,1} - \Ybar_{k,0}$ with independent arms
and bias $b_{k} = c_{k}\alpha^{\top}w_{T,k}$ by Proposition~3.
Given $A_{kj} = a$, the outcome $Y_{kj}(a) = \delta_{T} + w_{T,k}^{\top}
x_{kj} + \mu_{T,k}^{\top} z_{kj} + r_{k} a + \varepsilon_{kjT}^{(a)}$ has
variance $w_{T,k}^{\top}\Cov(x_{kj} \mid A_{kj} = a)\, w_{T,k}
+ \sigma_{z}^{2}\|\mu_{T,k}\|^{2} + \sigma^{2} = E_{k}^{c}(a)
+ \sigma_{k,\mathrm{cond}}^{2}$, since $z_{kj} \perp A_{kj}$ (Step 3 of
the proof of Proposition~3), so
$\mathrm{MSE}_{1} = b_{k}^{2} + \sum_{a}[\sigma_{k,\mathrm{cond}}^{2}
+ E_{k}^{c}(a)]/n_{k}^{(a)} = b_{k}^{2} + (\sigma_{k,\mathrm{cond}}^{2}
+ \bar E_{k}^{c})\, s_{k}$ exactly. For Gaussian covariates the
components of $x_{kj}$ orthogonal to $\alpha$ are independent of
$\alpha^{\top} x_{kj}$ and hence of $A_{kj}$, so
$\Cov(x_{kj} \mid A_{kj} = a) = \sigma_{x}^{2} I - (\sigma_{x}^{2}
- \tau_{k,a}^{2})\,\alpha\alpha^{\top}/\|\alpha\|^{2}$ (with
$\tau_{k,a}^{2}$ the arm-conditional variance of the standardized index,
as in Section~\ref{sec:supp-inflation}), which gives
$E_{k}^{c}(a) = E_{k} - \gamma_{k,a}(\alpha^{\top} w_{T,k})^{2}$. The
doubly robust estimate has bias of smaller order (Step 5 above) and
variance $(1 + \kappa_{k})\,\sigma_{k,\mathrm{cond}}^{2} s_{k}$ by the
definition of $\kappa_{k}$. In the model with an unrestricted outcome regression the efficient
variance of the contrast under confounding is
$\sigma_{k,\mathrm{cond}}^{2}\,\E[1/\{g(x)(1-g(x))\}]/n_{k}
= \omega_{k}\,\sigma_{k,\mathrm{cond}}^{2} s_{k}$ (the
heterogeneous-effect term vanishes because $Q(1,x) - Q(0,x) = r_{k}$ is
constant), with $\omega_{k} = \pi_{k}(1-\pi_{k})\,\E[1/\{g(1-g)\}] \ge 1$
by Jensen's inequality and $\omega_{k} = 1$ under randomized assignment.
The estimator as implemented is regular and asymptotically linear in the
model with a linear outcome regression (Assumption~\ref{ass:nuisance}),
with influence curve $\breve\varepsilon\,(h + \ell^{\top} \breve x)$, where
$\breve x = (1, x, A)$ is the regressor vector of the outcome model,
$\breve\varepsilon$ the outcome residual (of variance
$\sigma_{k,\mathrm{cond}}^{2}$, not the noise $\varepsilon$ of
Section~5.1), $g^{\dagger}$ the limit of the
truncated propensity fit (Assumption~\ref{ass:nuisance}),
$h = \iota_{1} A/g^{\dagger} - \iota_{0}(1-A)/(1-g^{\dagger})$ the truncated
inverse probability weight rescaled by the fluctuation constants of the
targeted estimator, $\iota_{1} = \E[1/g^{\dagger}]/\E[A/(g^{\dagger})^{2}]$
and $\iota_{0} = \E[1/(1-g^{\dagger})]/\E[(1-A)/(1-g^{\dagger})^{2}]$,
the large-sample ratios of the mean clever covariate to its mean square
in the linear fluctuation ($\iota_{a} = 1$ for AIPW), $e_{A}$
the coordinate vector that selects the coefficient of $A$, and
$\ell = \E[\breve x \breve x^{\top}]^{-1}(e_{A} - \E[h \breve x])$ the
term that carries the least-squares estimation of the outcome model; its
large-sample variance is $\omega_{k}^{\dagger}\,\sigma_{k,\mathrm{cond}}^{2}
s_{k}$ with $\omega_{k}^{\dagger} = \pi_{k}(1-\pi_{k})\,\E[(h +
\ell^{\top}\breve x)^{2}]$. Write $L(f \mid \breve x)$ for the population
least-squares projection of $f$ on $\breve x$ and $L(f \mid 1, x)$ for
that on $(1, x)$. The orthogonal decomposition of $h + \ell^{\top}\breve x$
into $h - L(h \mid \breve x)$ and
$e_{A}^{\top}\E[\breve x \breve x^{\top}]^{-1}\breve x$ gives
$\omega_{k}^{\dagger} = \omega_{k,\mathrm{lin}}
+ \pi_{k}(1-\pi_{k})\,\E[\{h - L(h \mid \breve x)\}^{2}] \ge \omega_{k,\mathrm{lin}}$
with $\omega_{k,\mathrm{lin}} = \pi_{k}(1-\pi_{k})\,[\E(\breve x \breve x^{\top})^{-1}]_{AA}
= \pi_{k}(1-\pi_{k})/\E[\{A - L(A \mid 1, x)\}^{2}]$, the factor of the
least-squares coefficient of $A$; since $\E[\{A - L(A \mid 1,x)\}^{2}]
= \E[g(1-g)] + \E[\{g - L(g \mid 1,x)\}^{2}] \le \pi_{k}(1-\pi_{k})$,
$1 \le \omega_{k,\mathrm{lin}} \le \pi_{k}(1-\pi_{k})/\E[g(1-g)] \le
\omega_{k}$ by Jensen's inequality. Hence $\omega_{k}^{\dagger} \ge 1$,
no ordering between $\omega_{k}^{\dagger}$ and $\omega_{k}$ is implied
(the simulated configurations have $\omega_{k}^{\dagger} < \omega_{k}$
in every subpopulation), and $\kappa_{k} \to
\omega_{k}^{\dagger} - 1 \ge 0$ as the finite-sample inflation vanishes.
The condition $\mathrm{MSE}_{2} < \mathrm{MSE}_{1}$ then reads
$(1+\kappa_{k})\sigma_{k,\mathrm{cond}}^{2} s_{k}
< b_{k}^{2} + (\sigma_{k,\mathrm{cond}}^{2} + \bar E_{k}^{c}) s_{k}$,
which rearranges to~(10). \qed

\subsection{Proof of the Equal-Error-Boundary Theorem}
This subsection proves Theorem~1.
Proposition~4 gives $\Delta\mathrm{MSE}_{k}
= s(\bar n)\,[\kappa_{k}(\bar n)\,\sigma_{k,\mathrm{cond}}^{2}
- \bar E_{k}^{c}] - (c_{k} u)^{2}$, which is~(11) with $E_{k}$
standing for $\bar E_{k}^{c}$ at its anticipated value; for Gaussian
covariates $\bar E_{k}^{c} = E_{k} - \bar\gamma_{k} u^{2}$ adds
$\bar\gamma_{k} s(\bar n) u^{2}$ to $\Delta\mathrm{MSE}_{k}$; absorbing it into the
edge, the steps below for parts (i)--(iii) go through with $E_{k}$
replaced by $E_{k} - \bar\gamma_{k} u^{2}$ (the hypothesis
$\lim_{n \downarrow 0} \kappa_{k}(n)\,\sigma_{k,\mathrm{cond}}^{2}
> E_{k} - \bar\gamma_{k} u^{2}$ follows from the theorem's when
$\bar\gamma_{k} \ge 0$, and the argument for (ii) does not use
monotonicity in $|u|$, which the $u$-dependent edge can break at small
$n$), and the monotonicity in $|u|$ of part (iv) with
$c_{k}^{2}$ replaced by $c_{k}^{2} - \bar\gamma_{k} s(\bar n)$ wherever
this difference is positive (the whole sampled range under the
linear-index mechanisms and most, not all, subpopulations under the
quadratic index, as stated after the theorem). Fix $u$ and write
$B(n) = \kappa_{k}(n)\,\sigma_{k,\mathrm{cond}}^{2} - E_{k}$, so
$\Delta\mathrm{MSE}_{k}(n, u) = s(n)\,B(n) - (c_{k} u)^{2}$ with $s$ positive,
continuous, strictly decreasing, $s(n) \to \infty$ as $n \downarrow 0$ and
$s(n) \to 0$ as $n \to \infty$, as recorded below~(11). By the assumptions on
$\kappa_{k}$, $B$ is continuous and strictly decreasing with
$\lim_{n \downarrow 0} B(n) > 0$ and $\lim_{n \to \infty} B(n)
= B_{\infty} = \kappa_{k,\infty}\,\sigma_{k,\mathrm{cond}}^{2} - E_{k}$.
Throughout this proof $n$ denotes the per-arm count $\bar n$ of the
statement; any strictly increasing reparameterization leaves the argument
unchanged.

\emph{Case $B_{\infty} < 0$.} $B$ has a unique root $\nu_{0}$ and is
positive before it, negative after. On $(0, \nu_{0}]$ the product $s B$
is strictly decreasing (for $\nu_{1} < \nu_{2} \le \nu_{0}$: if
$B(\nu_{2}) > 0$ then $s(\nu_{1}) B(\nu_{1}) \ge s(\nu_{1}) B(\nu_{2}) >
s(\nu_{2}) B(\nu_{2})$, and if $B(\nu_{2}) = 0$ then
$s(\nu_{1}) B(\nu_{1}) > 0 = s(\nu_{2}) B(\nu_{2})$), with
$s B \to +\infty$ as $n \downarrow 0$ and $s(\nu_{0}) B(\nu_{0}) = 0$;
for $n > \nu_{0}$, $s B < 0$. Hence $\Delta\mathrm{MSE}_{k}(\cdot, u)$ decreases
strictly on $(0, \nu_{0}]$ from $+\infty$ to $-(c_{k} u)^{2}$ and
satisfies $\Delta\mathrm{MSE}_{k} < -(c_{k} u)^{2} \le 0$ on $(\nu_{0}, \infty)$.
If $c_{k} u \neq 0$ the unique zero $n_{k}^{*}(u)$ lies in
$(0, \nu_{0})$; if $c_{k} u = 0$ the zero is $n_{k}^{*} = \nu_{0}$; in
both cases $\Delta\mathrm{MSE}_{k} > 0$ exactly on $(0, n_{k}^{*})$ and
$\Delta\mathrm{MSE}_{k} < 0$ on $(n_{k}^{*}, \infty)$.

\emph{Case $B_{\infty} \ge 0$.} Then $B > B_{\infty} \ge 0$ at every
finite $n$ (a strictly decreasing function exceeds its limit), and $s B$
is strictly decreasing from $+\infty$ to $0$ (as $n \to \infty$, $s \to 0$
while $B$ is bounded on $[1, \infty)$). If $c_{k} u \neq 0$ there is a
unique root as before; if $c_{k} u = 0$, $\Delta\mathrm{MSE}_{k} = s B > 0$ for
every finite $n$, that is, $n_{k}^{*} = \infty$. Together the two cases
prove (i) and the finiteness criterion, which reduces to ``$|c_{k} u| > 0$
or $E_{k} > 0$'' when $\kappa_{k,\infty} = 0$.

\emph{(ii).} $u$ enters~(11) only through $u^{2}$, so on
the quadrant with $\upsilon = |u|$ the regions are
$\mathcal{R}_{1} = \{(n, \upsilon) : n < n_{k}^{*}(\upsilon)\}$ and
$\mathcal{R}_{2} = \{(n, \upsilon) : n > n_{k}^{*}(\upsilon)\}$, and by (i)
every horizontal slice of either region is an interval. The argument
below does not use monotonicity of $n_{k}^{*}$ in $\upsilon$, so it also
covers the $u$-dependent edge $E_{k} - \bar\gamma_{k} u^{2}$ of the remark
after the theorem, under which $n_{k}^{*}$ can increase with $\upsilon$
at small $n$. If $c_{k} = 0$, $\Delta\mathrm{MSE}_{k}$ does not depend on
$\upsilon$ and both regions are products of an interval with the
$\upsilon$-axis (with the $u$-dependent edge the ordinates with a finite
boundary form an interval, on any compact part of which the argument for
$\mathcal{R}_{2}$ below applies with $B(n) + \bar\gamma_{k}\upsilon^{2}$ in
place of $B(n)$), so let $c_{k} \ne 0$. $\mathcal{R}_{1}$ is path-connected:
for two of its points $(\nu_{1}, \upsilon_{1})$ and
$(\nu_{2}, \upsilon_{2})$ there is a count
$\nu_{\flat} < \min(\nu_{1}, \nu_{2})$ with $\Delta\mathrm{MSE}_{k}(n, \upsilon) > 0$
for all $n \le \nu_{\flat}$ and all $\upsilon$ between $\upsilon_{1}$ and
$\upsilon_{2}$, because $s(n)B(n) \to +\infty$ as $n \downarrow 0$ while
$(c_{k}\upsilon)^{2}$ is bounded on that range (the added term
$s(n)\bar\gamma_{k}\upsilon^{2}$ of the $u$-dependent edge is
nonnegative when $\bar\gamma_{k} \ge 0$; when $\bar\gamma_{k} < 0$ the
hypothesis on the small-sample inflation at the larger of the two
ordinates
gives the same bound); the path $(\nu_{1}, \upsilon_{1}) \to
(\nu_{\flat}, \upsilon_{1}) \to (\nu_{\flat}, \upsilon_{2}) \to
(\nu_{2}, \upsilon_{2})$ stays in $\mathcal{R}_{1}$, its horizontal legs by
(i) and its vertical leg by the choice of $\nu_{\flat}$. $\mathcal{R}_{2}$ is
path-connected: given two of its points, let
$\upsilon_{\min} = \min(\upsilon_{1}, \upsilon_{2})$; every ordinate
between $\upsilon_{1}$ and $\upsilon_{2}$ has a finite boundary, since by
the finiteness criterion $n_{k}^{*}(\upsilon) < \infty$ for every
$\upsilon > 0$, and $n_{k}^{*}(0) < \infty$ exactly when $B_{\infty} < 0$,
which holds if $\upsilon_{\min} = 0$ because a point of $\mathcal{R}_{2}$ at
ordinate $0$ certifies it. There is a count
$\nu^{\sharp} > \max(\nu_{1}, \nu_{2})$ with $\Delta\mathrm{MSE}_{k}(n, \upsilon) < 0$
for all $n \ge \nu^{\sharp}$ and all such $\upsilon$: if
$\upsilon_{\min} > 0$ this follows from $s(n)B(n) \to 0$ and
$(c_{k}\upsilon)^{2} \ge (c_{k}\upsilon_{\min})^{2} > 0$ (the term
$s(n)\bar\gamma_{k}\upsilon^{2}$ vanishes with $s$), and if
$\upsilon_{\min} = 0$ then $\Delta\mathrm{MSE}_{k} \le s(n)B(n) < 0$ for all
$n > \nu_{0}$ and every $\upsilon$ (with the $u$-dependent edge,
$\Delta\mathrm{MSE}_{k} = s(n)B(n) - (c_{k}^{2} - \bar\gamma_{k}s(n))\upsilon^{2}
< 0$ once also $s(n)\bar\gamma_{k} \le c_{k}^{2}$, that is,
$n \ge 2\bar\gamma_{k}/c_{k}^{2}$ at $s = 2/n$). The path
$(\nu_{1}, \upsilon_{1}) \to (\nu^{\sharp}, \upsilon_{1}) \to
(\nu^{\sharp}, \upsilon_{2}) \to (\nu_{2}, \upsilon_{2})$ stays in
$\mathcal{R}_{2}$, its horizontal legs by (i) and its vertical leg by the
choice of $\nu^{\sharp}$. Hence there are exactly two connected regions,
separated by the graph of $n_{k}^{*}(\cdot)$.

\emph{(iii)} is the statement that~(11) contains no other
quantities.

\emph{(iv).} $\Delta\mathrm{MSE}_{k}$ is pointwise strictly decreasing in
$(c_{k} u)^{2}$ and in $E_{k}$ (coefficients $-1$ and $-s(n) < 0$) and
pointwise strictly increasing in $\sigma_{k,\mathrm{cond}}^{2}$
(coefficient $s(n)\,\kappa_{k}(n) > 0$ at every finite $n$). If a
parameter change within the assumptions lowers $\Delta\mathrm{MSE}_{k}$ pointwise
and the old boundary $n_{k}^{*}$ is finite, then
$\Delta\mathrm{MSE}_{k}^{\mathrm{new}}(n_{k}^{*}) <
\Delta\mathrm{MSE}_{k}^{\mathrm{old}}(n_{k}^{*}) = 0$, and since
$\Delta\mathrm{MSE}_{k}^{\mathrm{new}} > 0$ precisely on
$(0, n_{k}^{*,\mathrm{new}})$ by (i), it follows that
$n_{k}^{*,\mathrm{new}} < n_{k}^{*}$; symmetrically for pointwise
increases. Finally, when $B_{\infty} < 0$ the $u = 0$ boundary satisfies
$s(\nu_{0})\,B(\nu_{0}) = 0$ with $s > 0$, that is,
$\kappa_{k}(\nu_{0})\,\sigma_{k,\mathrm{cond}}^{2} = E_{k}$, and the root
is unique because $B$ is strictly decreasing. \qed

\subsection{Proof of the Borrowing-Collapse Proposition}
\label{app:proof-collapse}
This subsection proves Proposition~1.
The constraint rows are $\one^{\top}$, the $\Dx$ rows of $X$, and the $\Tpre$
rows of $\widehat Y_{\neg T}$, whose entries are running means
$\hat y_{l,t} = y^{\circ}_{l,t} + \Xi_{t,l}$ with the noiseless values
$y^{\circ}_{l,t} = \delta_{t} + w_{t}^{\top} x_{l} + \mu_{t}^{\top} z_{l}$
(the entries of the noiseless matrix $\Gamma^{\circ}$; under
outcome-dependent recruitment they are not the expectations of the
running means, which the argument does not need) and sampling errors
$\Xi_{t,l}$, the entries of $\Xi_{\neg T}$, that have an absolutely
continuous joint law: the recruitment history takes finitely many values,
conditional on any one of positive probability the covariate, latent,
and outcome noise draws $(\eta, \zeta, \varepsilon)$ of the recruited
patients have a joint density (the Gaussian density restricted to an
event of positive probability and, under confounded assignment,
reweighted by the assignment probabilities; on the subpopulation-level
pool, where every patient sits at the centroid, the outcome noise alone),
and the errors
$\Xi_{t,l} = w_{t}^{\top}\bar\eta_{l} + \mu_{t}^{\top}\bar\zeta_{l}
+ \bar\varepsilon_{l,t}$, with bars for the means over the patients of
subpopulation $l$, are a surjective linear image of those draws
(Assumption~\ref{ass:dgp}). Modulo the
span of $\one^{\top}$ and the rows of $X$, the noiseless row $t$ is
$\mu_{t}^{\top} Z^{\top}$ with $Z \in \R^{K \times \Dz}$ the matrix with rows
$z_{l}^{\top}$, so the noiseless trajectory block has the row space of
$M_{\neg T}^{\top} Z^{\top}$, of rank at most $\min(\Tpre, \Dz)$ and equal
to it when $M_{\neg T}$ and $Z$ have full rank, which holds almost surely
under the continuous laws of the pre-treatment latent weights and the
loadings; hence $\rank(\Gamma^{\circ}) = \min(K, 1 + \Dx + \min(\Tpre, \Dz))$
for centroids, loadings, and latent weights in general position, and the
rank condition of the statement, $\rank([\one^{\top}; X]) \ge
\min(K, m) - \Tpre$, is what the realized rows need from the centroids
once the $\Tpre$ noisy rows contribute their full rank; under the
environment law the centroids are absolutely continuous, so
$\rank([\one^{\top}; X]) = \min(K, 1 + \Dx) \ge \min(K, m) - \Tpre$
almost surely, which is the almost-sure clause of the statement. The
realized matrix
is $\Gamma = \Gamma^{\circ} + [0; 0; \Xi_{\neg T}]$ with $\Xi_{\neg T}$ the
$\Tpre \times K$ error block. For $K \ge m$ the set of $\Xi_{\neg T}$ at
which $\rank(\Gamma) < m$ is the common zero set of the $m \times m$
minors of $\Gamma$, polynomials in the entries of $\Xi_{\neg T}$, and
under the rank condition of the statement one of them is not identically
zero: the condition gives $\rank([\one^{\top}; X]) = 1 + \Dx$, so some
$1 + \Dx$ columns carry a nonsingular block of $[\one^{\top}; X]$, and
choosing $\Xi_{\neg T}$ so that the $\Tpre$ trajectory rows equal unit
vectors on $\Tpre$ of the remaining $K - 1 - \Dx \ge \Tpre$ columns makes
the minor on these $m$ columns block-triangular with a nonzero
determinant. The zero set of a nonzero polynomial has Lebesgue measure
zero and $\Xi_{\neg T}$ has an absolutely continuous joint law (the
errors of one column share patients and are dependent, which the
argument does not need), so the set has probability zero and
$\rank(\Gamma) = m$ almost surely. For $K < m$ the condition gives
$\mathrm{rk} = \rank([\one^{\top}; X]) \ge K - \Tpre$: a row basis of
$[\one^{\top}; X]$ is nonsingular on some $\mathrm{rk}$ columns, and
$\Xi_{\neg T}$ can be chosen so that $K - \mathrm{rk} \le \Tpre$
trajectory rows are unit vectors on the other $K - \mathrm{rk}$ columns,
which gives a nonzero
$K \times K$ minor, so $\rank(\Gamma) = K$ almost surely by the same
argument. Together, $\rank(\Gamma) = \min(K, m)$. The
feasible set $\{\beta : \Gamma\beta = q\}$ is $e_{k} + \operatorname{null}(\Gamma)$,
since $\beta = e_{k}$ is feasible ($q$ is the $k$th column of $\Gamma$: the target reproduces its own covariates and realized trajectory exactly), of
dimension $K - \rank(\Gamma) = \max(K - m, 0)$. If $K \le m$ it is the single
point $e_{k}$ and every family-specific objective returns the same
weights; the quadratic objectives become active, and the hard-constrained
variants can separate, only when $K > m$. \qed

\subsection{Proof of the Mismatch-Penalty Proposition}
This subsection proves Proposition~5.
In this proof $n_{k}$ denotes the control count $n_{k}^{(0)}$ of
subpopulation $k$. Consider minimizing $\max_{k} \sigma_{k}^{2}/n_{k}$ over
allocations with $\sum_{k} n_{k} = N$. At any allocation where some subpopulation is strictly below
the maximum, budget can be moved from it to the maximizing subpopulations
(to all of them, in proportion to their $\sigma_{k}^{2}$, when several attain
the maximum) to reduce the objective,
so the optimum equalizes $\sigma_{k}^{2}/n_{k} = \sum_{l}\sigma_{l}^{2}/N$
with $n_{k} \propto \sigma_{k}^{2}$. Allocating by profile $F$ and evaluating
under profile $G$ gives the achieved worst case 
$\max_{k} \sigma_{k,G}^{2}/n_{k}^{F}
= (\sum_{l}\sigma_{l,F}^{2}/N) \max_{k}(\sigma_{k,G}^{2}/\sigma_{k,F}^{2})$,
and dividing by the matched value $\sum_{l}\sigma_{l,G}^{2}/N$ yields~(12). Since $\sum_{l}\sigma_{l,G}^{2}
= \sum_{l}\sigma_{l,F}^{2}\,(\sigma_{l,G}^{2}/\sigma_{l,F}^{2})
\le \max_{k}(\sigma_{k,G}^{2}/\sigma_{k,F}^{2}) \sum_{l}\sigma_{l,F}^{2}$,
the factor is at least one and equals one if and only if the ratio is
constant across subpopulations, that is, the profiles are proportional. Instantiating with
$\sigma_{k,\mathrm{eff}}^{2}$ and $\sigma_{k,\mathrm{cond}}^{2}$ gives the
statement in the text. \qed